\documentclass{ws-ijmpd}
\usepackage[super,compress]{cite}
\usepackage[dvips,breaklinks]{hyperref}
\hypersetup{colorlinks,urlcolor=black,citecolor=black,linkcolor=black,filecolor=black}
\usepackage{url}

\newcommand{\beq}{\begin{equation}}
\newcommand{\beqa}{\begin{eqnarray}}
\newcommand{\eeq}{\end{equation}}
\newcommand{\eeqa}{\end{eqnarray}}
\newcommand{\be}{\begin{equation}}
\newcommand{\ba}{\begin{eqnarray}}
\newcommand{\ee}{\end{equation}}
\newcommand{\ea}{\end{eqnarray}}
\newcommand{\et}{{\it et al.}}

\newcommand{\lmk}{\left(}
\newcommand{\rmk}{\right)}

\newcommand{\lla}{\left\langle}

\newcommand{\rra}{\right\rangle}
\newcommand{\so}{M_\odot}
\newcommand{\mch}{{\cal M}}

\newcommand{\fch}{{\cal F}}

\newcommand{\mrm}{\mathrm }

\newcommand{\mc}{\mathcal{M}}

\newcommand{\nn}{\nonumber}

\begin{document}

\markboth{Kent Yagi}
{Scientific Potential of DPF and  Testing GR with Space-Borne GW Interferometers}

%%%%%%%%%%%%%%%%%%%%% Publisher's Area please ignore %%%%%%%%%%%%%%%
%
\catchline{}{}{}{}{}
%
%%%%%%%%%%%%%%%%%%%%%%%%%%%%%%%%%%%%%%%%%%%%%%%%%%%%%%%%%%%%%%%%%%%%

\title{Scientific Potential of DECIGO Pathfinder and \\ Testing GR with Space-Borne Gravitational Wave Interferometers}

\author{Kent Yagi}

\address{Department of Physics, Montana State University, Bozeman, MT 59717, USA.\\
kyagi@physics.montana.edu}

%\author{SECOND AUTHOR}

%\address{Group, Laboratory, Address\\
%City, State ZIP/Zone, Country\\
%second\_author@group.com}

\maketitle

\begin{history}
\received{Day Month Year}
\revised{Day Month Year}
\end{history}

\begin{abstract}
DECIGO Pathfinder (DPF) has an ability to detect gravitational waves from galactic intermediate-mass black hole binaries. If the signal is detected, it would be possible to determine parameters of the binary components. Furthermore, by using future space-borne gravitational wave interferometers, it would be possible to test alternative theories of gravity in the strong field regime.  In this review article, we first explain how the detectors like DPF and DECIGO/BBO work and discuss the expected event rates. Then, we review how the observed gravitational waveforms from precessing compact binaries with slightly eccentric orbits can be calculated both in general relativity and in alternative theories of gravity. For the latter, we focus on Brans-Dicke and massive gravity theories. After reviewing these theories, we show the results of the parameter estimation with DPF using the Fisher analysis. We also discuss a possible joint search of DPF and ground-based interferometers. Then, we show the results of testing alternative theories of gravity using future space-borne interferometers. DECIGO/BBO would be able to place 4--5 orders of magnitude stronger constraint on Brans-Dicke theory than the solar system experiment. This is still 1--2 orders of magnitude stronger than the future solar system mission such as ASTROD I. On the other hand, LISA should be able to put 4 orders of magnitude more stringent constraint on the mass of the graviton than the current solar system bound. DPF may be able to place comparable constraint on the massive gravity theories as the solar system bound. We also discuss the prospects of using eLISA and ASTROD-GW in testing alternative theories of gravity. The bounds using eLISA are similar to the LISA ones, but ASTROD-GW performs the best in constraining massive gravity theories among all the gravitational wave detectors considered in this article.
\end{abstract}


\ccode{PACS numbers:}

\tableofcontents

\section{Introduction}

\subsection{Gravitational Waves in General}

In general relativity (GR), gravitational fluctuations propagate as ``ripples'' in the spacetime at the speed of light, known as \textit{gravitational waves} (GWs).
Since gravity is weak compared to other fundamental forces, the interaction of GWs with other matters is very faint.
This means that GWs can escape from highly dense or optically thick regions that  cannot be probed by electromagnetic (EM) waves (e.g. very early universe before the last scattering surface). 
Therefore, GWs have potentials to open up a novel astronomy and cosmology.

The existence of GWs has been proved indirectly by measuring the orbital decay rate of binary pulsars, caused by the energy loss through gravitational radiation.
% the GWs taking away the energy from the binary system.
This can be calculated analytically within the framework of GR, and this matches with the measurements beautifully~{\cite{weisbergtaylor,stairs}}.
However, this indirect GW measurement only detects the radiated energy via GWs and not the perturbation of the spacetime itself.
Therefore, the direct detection of GWs has been awaited for a long time to detect this latter effect.
Currently, several ground-based detectors that are aiming for the direct detection are shifting from initial phases (first generation) to advanced phases (second-generation) such as adv.~LIGO~{\cite{adv-LIGO1,adv-LIGO2}}, adv.~VIRGO~{\cite{adv-VIRGO1,adv-VIRGO2}}, GEO-HF~\cite{GEO1,GEO2} and KAGRA (formerly called LCGT)~{\cite{lcgt1,lcgt2}}.  
These are GW interferometers that consist of two arms.
When GWs pass through this detector, the length of each arm changes, modifying the interferometric pattern.
KAGRA is novel and sometimes called 2.5th-generation detector since it will be cryogenically-cooled and is buried underground to reduce the thermal and seismic noises.
These detectors have their best sensitivities at around 100--1000Hz.
Basically, the high-frequency parts of these detectors are limited by the shot noises while the low-frequency parts are limited by the radiation pressure and the seismic noises.

Sources and sciences from GW interferometers have been summarized in Refs.~{\cite{thorne-source,schutz-source}}.
Promising sources for these second-generation detectors are signals from compact binaries (see Secs. 3.4 and 3.5 of Ref.~\cite{schutz-source} for a theoretical overview and Refs.~\cite{ligo-compact1,ligo-compact2,ligo-compact3,compact4} for observational results) whose event rates are estimated as $\mathcal{O}(10)$/year~{\cite{abadie}}.
(See Ref.~\cite{kamble} for the review on the electro-magnetic counterparts of GW sources.)
Binary signals are divided into three phases, inspiral, merger  and ringdown.
During the inspiral phase, the binary separation gradually shrinks due to the energy loss via GW radiation.
The inspiral waveform has been studied well under the post-Newtonian (PN) formalism~{\cite{blanchet-living, willwiseman,futamase-living}}.
Merger phase is highly non-linear but there has been a great progress thanks to numerical relativity~{\cite{pretorius-review,centrella-review,shibata-living}}.
Ringdown phases are also well-studied using black hole (BH) perturbation method~{\cite{sasaki-living}}.
%Hence, the frequency gets higher and higher 
GWs from a merging compact binary is often called \textit{chirp signals} since their frequencies get higher and higher as they head towards merger.
Since the inspiral waveform modeling has been performed well using PN approach~{\cite{blanchet-living}}, it can be used to estimate binary parameters~\cite{cutlerflanagan}
%test alternative theories of gravity~\cite{will-living,mishra,cornish-ppe} and 
and to probe cosmology as standard sirens~{\cite{schutz-standard-siren}}.
Also, neutron-star (NS) binaries can be used to constrain equation of state (EoS) of NSs~{\cite{flanaganhinderer-tidal,read-love,hinderer-tidal,kiuchi,kyutoku,hotokezaka,vines1,vines2,lackey,damour-tidal-EOB}}.
Other interesting sources are burst signals (see Sec. 3.2 of Ref.~\cite{schutz-source} for a theoretical review and Refs.~\cite{ligo-burst1,ligo-burst2,ligo-magnetars,ligo-cosmic} for the current observational results) including supernovae~{\cite{kotake}}, magnetars and cosmic strings~{\cite{damourvilenkin1,damourvilenkin2}}, and continuous signals (see Sec. 3.3 of Ref.~\cite{schutz-source} for a theoretical overview and Refs.~\cite{ligo-continuous1,ligo-continuous2,ligo-crab} for observational results) from e.g. (newly-born) neutron stars.
There are also stochastic GW backgrounds (see Sec. 3.6 of Ref.~\cite{schutz-source} for a theoretical review and Ref.~\cite{nature-gwb} for observational results) as potential candidates for GW sources such as the ones from cosmic-strings~{\cite{siemens}}, pre-big-bang models~\cite{pbb} and preheatings~{\cite{easther}}.
GW backgrounds associated with inflation may be detected by ground-based detectors if the spectrum has a sufficient blue tilt.
There is also a proposed third-generation project called Einstein Telescope (ET)~{\cite{ET2,ET3}}.
The sensitivity is increased roughly by one order of magnitude compared to the second-generation ones and ET has interesting sciences which have been summarized in Ref.~{\cite{ET-science}}.

\subsection{Space-Borne Interferometers}

Ground-based interferometers have difficulty in detecting GW signals lower than 1--10Hz due to the seismic noise. To overcome this problem,
%These detectors have their best sensitivity around 100-1000Hz and their lower frequency sides are limited by the seismic noises.
%There are also several projects for space-borne interferometers.
%To overcome the low-frequency limit due to the seismic noise, 
space-borne interferometers have been proposed.
Among them, the Laser Interferometer Space Antenna (LISA) has been proposed by ESA and NASA with its optimal sensitivity around GW frequency of 1mHz~{\cite{danzmann,lisa2}}.
Classic LISA consists of three spacecrafts from which laser is emitted to one another.
This triangular-shaped interferometer has armlengths of $5\times 10^6$km.
Due to this extremely long armlength, LISA is a transponder-type (rather than a Michelson or a Fabry-Perot-type) interferometer.
(I.e. Instead of reflecting the incoming laser directly, each spacecraft once receives it and emits the laser with corresponding phase.)
The triangular cluster follows the same orbit as the Earth, keeping its position 20$^{\circ}$ behind the latter.
The expected targets are supermassive black hole (SMBH) binaries and white dwarf (WD) binaries.
The former detection may help in clarifying the mechanism of how SMBHs are formed.
A system consisting of a small compact object orbiting a SMBH is called an extreme mass ratio inspiral (EMRI)~{\cite{EMRI-rev}}, and GWs from this system encodes the information of the structure of the strong-field spacetime~{\cite{ryan1,ryan2,collins,barackcutler,glampedakis,gairbumpy,barausse,li,sopuertayunes,Panietal,prisgair,sarah,sarahleo,pani-gravastar,rodriguez,gairyunes,apostolatos-bumpy,gerakopoulos,contopoulos}}. 
Unfortunately, GWs from WD binaries may mask~\cite{nelemans,farmer} other signals including primordial GW background~{\cite{allen, maggiore-review}}.
As a prototype mission, LISA path finder (LPF) will be launched before LISA~{\cite{lpf}}.

Although, the classic LISA was planned and developed by NASA and ESA, the collaboration between these 2 agencies terminated in 2011. 
The new version of LISA (called eLISA) led by ESA alone is now under consideration. 
The arm-length would be shortened by 5 times compared to the classic one and the number of the arms would be reduced to 2.
The satellites are slowly drifting away from Earth to save propellant.
The expected operation period is 2-5 yrs.
See Refs.~\cite{elisa-science,elisa-science2,GWB-elisa1,GWB-elisa2,sopuerta-elisa,gair-elisa} for expected sources and sciences with eLISA.

\begin{figure}[tbp]
%\begin{center}
  \centerline{\includegraphics[scale=.5,clip]{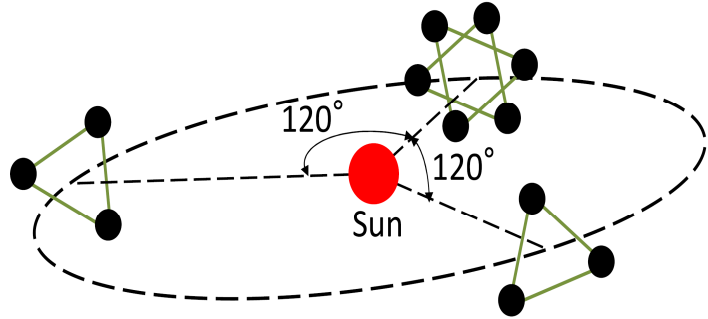}} 
\vspace*{8pt} 
\caption{\label{default} The configuration of DECIGO and BBO~{\cite{yagi:brane}}. There are eight effective interferometers in total. }
%\end{center}
\end{figure}

\begin{figure}[thbp]
  \centerline{\includegraphics[scale=1.5,clip]{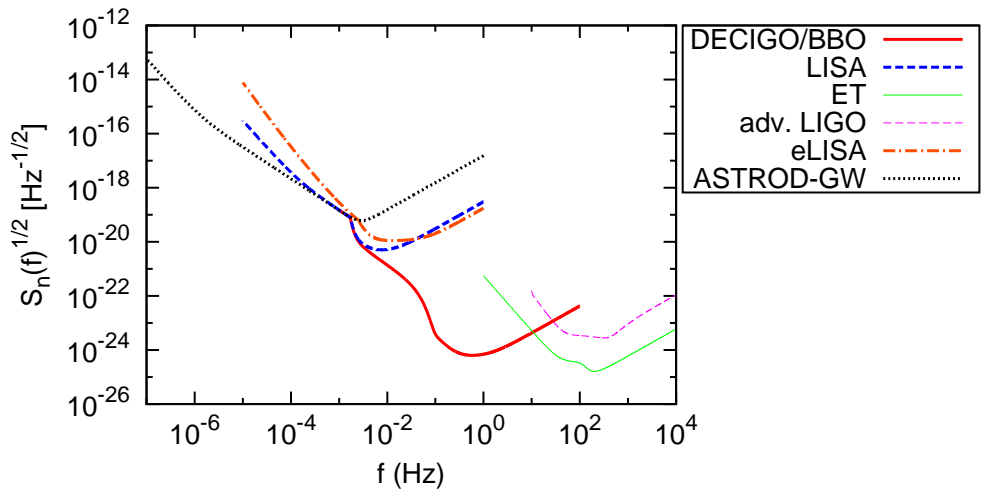}} 
\vspace*{8pt}
 \caption{\label{noise} The square root of the noise spectral densities $\sqrt{S_n(f)}$ of various detectors.
 (The noise spectral density roughly corresponds to the lower limit of the GW amplitude that each interferometer can detect.)
The (blue) thick dashed and the (red) thick solid curves represent the LISA and DECIGO/BBO ones, while the (green) thin solid and the (purple) thin dashed curves show the ones for ET and adv.~LIGO, respectively. The (orange) dotted-dashed and the (black) solid curves correspond to the eLISA and ASTROD-GW ones, respectively. }
\end{figure}

Another space-borne interferometer, the Deci-Hertz Interferometer Gravitational Wave Observatory (DECIGO) has been proposed by Seto, Kawamura and Nakamura~{\cite{setoDECIGO}}.
It consists of four clusters, each having a triangular interferometer.
The major differences between DECIGO and LISA are that the armlengths are shortened to $10^3$km and the former is a Fabry-Perot-type interferometer whereas the latter is a transponder-type as previously mentioned.
Two out of four clusters are situated on the same site so that the correlation analysis can be performed to detect primordial GW background, while the other two are placed far apart so as to increase the angular resolutions of the sources~\cite{kawamura2006,ando2010,kawamura2011}
(see Fig.~\ref{default}).
It is most sensitive at around 0.1--1Hz.
Similar interferometer, the Big Bang Observatory (BBO) has been suggested as a follow-on mission to LISA~{\cite{phinneybbo,cutlerholz}}. 
Since it is a transponder-type interferometer, it can be considered as a mini-LISA with its configuration similar to the one of DECIGO.
(The noise curves of adv.~LIGO, ET, LISA and DECIGO/BBO are shown in Fig.~\ref{noise}~\footnote{DECIGO has been proposed with a few times less sensitivity than BBO, but this is not a fixed design and here, we have assumed that it has the same sensitivity as BBO.}.)
Since the WD/WD binary signals has a high cutoff frequency at $f\sim 0.2$Hz~{\cite{farmer}}, the main target of DECIGO/BBO is the primordial GW background (PGWB)~{\cite{setoDECIGO,kuroyanagi1,kuroyanagi2}}.
NS/NS foreground signals might mask PGWB instead of WD/WD signals, but it is likely that BBO would have enough sensitivity to subtract sufficient amount~{\cite{cutlerharms,harms}}.
%DECIGO may have to improve its sensitivity by 2--2.5 times in order to achieve this goal~\cite{yagiseto}. 
See Refs.~\cite{setoDECIGO,cutlerholz,yagi:inhom,yagi:inhom2,nishizawayagi1,nishizawayagi2,nakayama1,nakayama2,kuroyanagi-thermal,grojean,caprini,saito1,bugaev,saito2,falta1,falta2,gair-SMBH,kuroyanagiCS} for other cosmological and astrophysical scientific potentials of DECIGO and BBO. 
%They have other interesting scientific objectives including direct detection of the accelerating expansion of the universe~\cite{setoDECIGO}, performing the precision cosmology~\footnote{The sensitivity of DECIGO/BBO required for performing precision cosmology and detecting primordial GW background is discussed in Ref.~\cite{nishizawa}.}~\cite{cutlerholz} including the test of the inhomogeneous universe~\cite{yagi:inhom}, revealing the thermal history of the early universe~\cite{nakayama1,nakayama2,kuroyanagi-thermal}, physics at cosmological phase transitions~\cite{grojean,caprini}, progenitors for the type Ia supernovae~\cite{falta1,falta2} and the formation process of SMBH~\cite{gair-SMBH}, and verifying the primordial BH as the dark matter candidate~\cite{saito1,bugaev,saito2}.
% and probing alternative theories of gravity~\cite{kent2} and the size of extra dimension~\cite{kentbrane}.

As for the prototype mission, DECIGO Pathfinder (DPF)~\cite{ando2009,ando2010,ando-ASTROD} is hoped to be launched in 2016--2017.
The main goals are to test the key technologies for the space mission and to carry out observations of GWs and Earth gravity. 
The first space-borne GW detector, SWIM, has already been launched in 2009 \ {\cite{toba,ando-ASTROD}}.
It is a torsion-bar type space antenna.
The ground-based torsion-bar antennae have been used to place upper limits on the energy density of stochastic GWs at 0.1--1Hz \ {\cite{ishidoshiro,shoda}}.
%As for the first space-borne GW detector, torsion-bar type space antenna called SWIM has already been launched in 2009
% and successfully performed its observation run.
%It has completed its mission and is currently under the phase of data analysis.
The improved version of the torsion-bar antenna (TOBA) has been proposed by Ando \textit{et al}.~{\cite{toba}}.
%In this review, we focus on the scientific significances of GW observations with DPF.

DPF has a sensitivity of $\sqrt{S_n(f)} \sim 2 \times 10^{-15} \ \mrm{Hz}^{-1/2}$ at $f \sim 1$--$100\mrm{Hz}$.
The main GW sources for DPF are the intermediate mass BH (IMBH) binaries.
There are evidences that stellar-mass BHs and SMBHs exist, but there is no unambiguous detection of individual IMBHs.
IMBH detection may reveal the formation mechanism of SMBHs.
% and also it may lead to the identification of the baryonic dark matter~\cite{van}. 
DPF has enough sensitivity to detect GWs from {\emph{galactic}} IMBH binaries that might exist  at the centers of globular clusters and massive young clusters (see Refs.~\cite{colbert,pasquato} for reviews on IMBHs and Ref.~\cite{gurkan} for the possibility of IMBH binary formation).
%DPF has an ability to detect the GW background with the energy density of $\Omega_\mrm{GW} \ge O(1)$.
%However, stronger bound has already been set from the Big Bang Nucleosynthesis (BBN)~\cite{allen}.
For the observation of the Earth gravity, DPF has an ability to perform complementary operation compared to other missions such as CHAMP~{\cite{champ1}}, GRACE~\cite{grace} and GOCE~{\cite{goce}}.
%~\cite{andolisa}.
%(DPF can still observe GWs since geogravity noise dominates other noises only below $f$=0.03Hz.)
%However, we do not consider these issues further in this review and stick to the observation of GWs from IMBH binaries. 

ASTROD-GW (Astrodynamical Space Test of Relativity using Optical Devices optimized for GW detection) is another space-based GW interferometer mission~\cite{ASTROD-GW}. 
It consists of 3 satellites that are placed at near Lagrange points L3, L4 and L5, respectively, forming a large triangluar interferometer similar to LISA. 
Since its armlength is 52 times larger than that of LISA, it is more sensitive than LISA at lower frequency.

%%%%%%%%%%%%%%%%%%%%%%%%%%%%%%%%%%%%%%%%%%%%%%%%%%%%%
\subsection{Testing Alternative Theories of Gravity using GWs}

Among all the sciences that have been mentioned above, there is yet another very interesting  and fundamentally important science that can be best probed by using GWs.
It is the \textit{tests on alternative theories of gravity}~{\cite{will-book,will-living, arun-ASTROD}}. 
In order to solve problems like dark energy~{\cite{tsujikawa-book}}, dark matter and inflation~\cite{bassett,mukhanov} within the context of GR, usually we need to introduce unknown matters or fields, but there are possibilities that these problems can be explained naturally by modifying gravitational theories.
Also, if the classical gravitational theory is to be realized at the low energy limit of more fundamental theory like superstring theory~{\cite{polchinski1,polchinski2}}, the classical one does not necessarily reduce to GR due to the additional fields (e.g. dilatons) that couple to gravity.

\begin{figure}[t]
 \begin{center}
  \centerline{\includegraphics[scale=.5,clip, angle=-90]{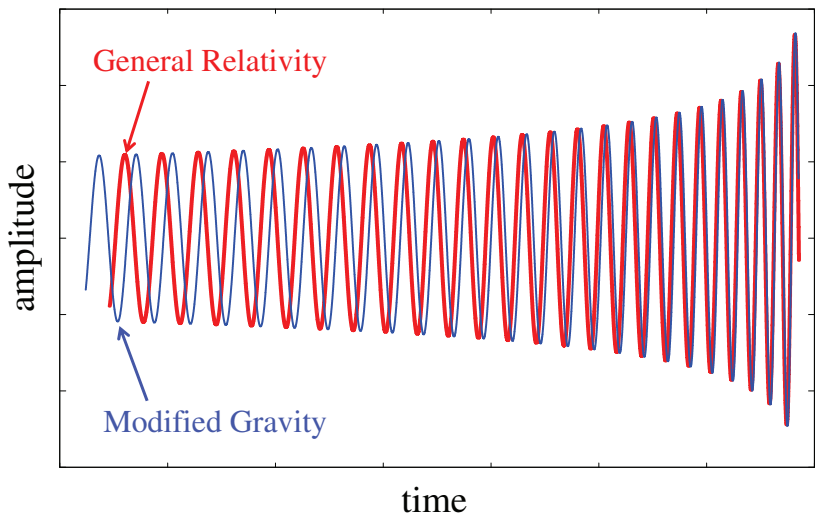}} 
   \end{center}
\vspace*{8pt}
 \caption{A schematic picture of gravitational waveforms from a compact binary in GR (red thick) and in alternative theories of gravity (blue thin).}
\label{fig:waveform}
\end{figure}

The tests on alternative theories of gravity have been performed with great accuracies under the solar system experiments and binary pulsar observations~\cite{will-living,stairs}  and no deviation from GR has been reported so far.
The former can probe only the weak field limit of the theory, while the latter can probe the strong field gravity~\cite{damour-review} such as the effacement principle and the scalarization~\cite{scalarization1,scalarization2,harada-scalarization} in scalar-tensor theory.
However, since the typical velocity of the binary component is $v/c \sim 10^{-3}$, the system is not so dynamical.
We are here interested in testing GR in the \emph{strong} and \emph{dynamical} field regime.
The best way to perform these tests in the near future is to use GWs (especially GWs from compact binaries) since they directly contain gravitational information in the strong-field regions.
Also, GWs lead to remarkably high precision GR tests in the following sense.
Usually, GWs are so weak that they are buried under the detector noises.
In order to dig them out from the noises, we need to perform the \textit{matched filtering analysis}~{\cite{data-analysis-liv-rev,cutlerflanagan}}.
This is nothing but taking correlations between the observed signals and the template waveforms so that if the former match with the latter, we can extract the signals out of the noises.
Therefore, parameter estimations using this matched filtering technique are sensitive to deviations in GW phase since the correlations are remarkably reduced even if the phases between GWs and the templates are different only slightly.
Now, it is often the case that alternative theories of gravity modify the phases of GWs from compact binaries (see Fig.~\ref{fig:waveform}).
Let us say that we observe a signal at a frequency $f=0.1$Hz for observation period $t_\mathrm{obs}=1$yr.
Then, the number of GW cycles (the number of phases) that we see is roughly $f\times t_\mrm{obs} \sim 3\times 10^6$.
This shows that we would be able to detect the deviations from GR if they modify the phase at least by $(3\times 10^6)^{-1} \sim 3\times 10^{-7}$! 
There are many works calculating how accurately we can probe alternative theories of gravity using GWs. (See e.g. Ref.~\cite{ppE} and references therein, and see Refs.~\cite{blanchet-tail1,blanchet-tail2,arun-model-indep,mishra,ppE,cornish-PPE,li-testing-GR,huwyler,katerina,hayama,cannella-maggiore,cannella-PhD,vallisneri-testingGR} for model-independent tests. See Refs.~\cite{dreyer,hughes-menou,berti-BHspectroscopy,kamaretsos,gossan} for possible tests on GR using ringdown signals.)

Two of the most important characteristics of GR are (i) it has only two tensor gravitational degrees of freedom and (ii) the graviton is massless (in other words, GWs propagate at the speed of light).
In this review, we focus on simple extensions of each of the above points by (i) introducing an additional scalar field and (ii) giving a finite mass to graviton.

As for the first point, we consider \textit{Brans-Dicke} (BD) theory~\cite{brans} which is the simplest representative of the scalar-tensor theory~{\cite{fujii}}.
This theory is parameterized by the so-called \textit{Brans-Dicke} parameter $\omega_\mrm{BD}$ which gives the (inverse) coupling between the additional (massless) scalar field and the matter fields.
If we take the decoupling limit of $\omega_\mrm{BD} \rightarrow \infty$, this theory reduces to GR.
The current strongest constraint has been obtained from the solar system experiment by Shapiro time delay measurements using the Cassini satellite~{\cite{cassini}} as $\omega_\mrm{BD} > 4\times 10^4$.
There is a future space mission called ASTROD which is expected to be launched in 2021~{\cite{ASTROD-I}}.
It would perform very precise tests of GR in the solar system and the constraint on $\omega_\mrm{BD}$ would be 3 orders of magnitude stronger than the Cassini bound. 
One of the remarkable points about considering gravitational radiation in this theory is that due to the additional scalar field, there exists scalar dipole radiation~\cite{eardley,will1977,zaglauer}!
Because of this additional energy release, the binary evolution is modified compared to GR, which further makes a change in GW phase.
The orbital decay of the binary pulsar PSR J1738-0333 has put a slightly weaker constraint than the solar system one~{\cite{freire}}. 
There have been several works calculating the possible bounds on $\omega_\mrm{BD}$ using future GW interferometers~{\cite{will1994,scharre,willyunes,bertibuonanno,yagiLISA}}.
(See Refs.~\cite{bertialsing,yunes-massiveBD} for the constraint on massive BD theory, Ref.~\cite{damour-GW-ST} for the one on generic scalar-tensor theories and Ref.~\cite{arun-dipole} for the bound on generic dipolar gravitational radiation.)
Unfortunately, these results show that the bounds from future GW observations using adv.LIGO and LISA would not be able to exceed the one from the solar system experiment. 
In this review, we will show that DECIGO/BBO would be able to place 4 orders of magnitude stronger constraint than the Cassini bound~{\cite{yagiDECIGO}}.

For the second point, there exist many theories in which graviton acquires a finite mass $m_g$ \ {\cite{fierz,dvali,rubakov2,dubovsky,chamseddine,derham1,derham2}}.
(See e.g. Refs.~\cite{rubakov,hinterbichler} for recent reviews.)
These theories are called \textit{massive gravity} or \textit{massive graviton} (MG) theories.
From a pioneering work by Fierz and Pauli~\cite{fierz} in 1939, most of massive gravity theories have suffered from problems like Boulware-Deser ghost modes appearing in the curved background~\cite{boulware} or violating the Lorentz invariance~{\cite{rubakov2,dubovsky}}.
Only recently a self-consistent massive gravity is proposed~{\cite{derham1,derham2}}.
Independent of these specific massive gravity theories, the gravitational potential is modified to the one of Yukawa type.
This means that the effective gravitational constant depends on the distance from the source, which modifies the Kepler's third law.
This has been tested in the solar system and the constraint has been put on the graviton Compton wavelength as $\lambda_g > 2.8\times 10^{17}$cm \ {\cite{talmadge}}.
As for GWs, the propagation speed is modified from the speed of light which again makes a change in GW phase.
GWs can also perform a model-independent tests on the graviton mass.
Compared to the solar system experiment, ground-based detectors can only put comparable (or slightly stronger) constraints~\cite{will1998,arunwill,keppel} whereas LISA bounds should be stronger by four orders of magnitude~\cite{will1998,willyunes,bertibuonanno,arunwill,stavridis,yagiLISA,keppel,delpozzo,cornish-PPE,bertisesana,huwyler,mirshekari}!
This is because the deviation in the propagation speed of GWs is larger for lower frequency GWs. 
DECIGO/BBO would be able to place slightly weaker constraint than LISA~\cite{yagiDECIGO} and DPF may be able to place comparable constraint to the solar system bound~{\cite{yagiDPF}}.

\subsection{Organization}

In this review, we will first explain how accurately DPF would be able to determine binary parameters by detecting GWs from galactic IMBH binaries. Then, we will describe the ability of testing GR using space-borne GW interferometers, especially DECIGO/BBO and DPF. 
This article is organized as follows. 
In Secs.~\ref{sec:DPF} and~\ref{sec:DECIGO}, we explain the basic concepts, noise curves and event rates of DPF and DECIGO/BBO, respectively. 
In Sec.~\ref{sec:binGW}, we derive GWs from compact binaries, starting from the quadrupole, leading contribution for the binary with a circular orbit, and extending it to include higher PN terms and a slightly eccentric orbit. 
Then, in Sec.~\ref{sec:obsGW}, we explain how to construct the observed gravitational waveforms for the spin-aligned and precessing binaries. 
We also introduce the inspiral-merger-ringdown hybrid waveform.
In Secs.~\ref{sec:BD} and~\ref{sec:MG}, we consider BD and MG theories, respectively. 
We explain how the GW phase is modified from GR, and review current constraints.
In Sec.~\ref{sec:results-DPF}, we show the results of the binary parameter estimation using DPF, and in Sec.~\ref{sec:testing-GR}, we show the proposed constraints on BD and MG theories (and also on other theories) using future GW interferometers. 
Finally, we conclude in Sec.~\ref{sec:conclusions}.

%%%%%%%%%%%%%%%%%%%%%%%%%%%%%%%%%%%%%%%%%%%%%%%%%%%%%%%%%%
\section{DECIGO Pathfinder}
\label{sec:DPF}

\subsection{Basic Designs}

DPF~\cite{ando2009,ando2010} is a prototype mission of DECIGO to test the advanced technologies of GW space mission such as (i) a precise position measuring system with Fabry-Perot (FP) cavity, (ii) a highly stabilized laser source and (iii) the drag-free control system which shields external forces due to solar radiation and residual atmosphere.
The weight of the satellite is about 350kg and it will be orbiting the earth at the Keplerian velocity with an altitude of 500km.
The expected observation period is 1 year.
It consists of 1-arm interferometer with armlength 30cm and the laser power of 100mW.
%This interferometer includes a FP cavity with a finesse of about 100 and the laser is emitted from a highly stabilized source with an output power of 100mW at a wavelength of 1030nm.
FP cavity has not been tested in space up to now and DPF is expected to have better sensitivity than LPF which uses a Mach-Zender interferometer~{\cite{lpf}}.
DPF provides new possibilities for a precise measurement of position and high-stabilized laser in space. 

Each mirror is placed inside a module called housing.
The relative positions of these mirrors and the frame will be measured by the local sensor and will be fed back to the satellite position using thrusters. 
%There are electrostatic-type local sensors on the frame of the housing, which are used for measuring the relative positions of the mirrors and the frame.
%The common motion signals of 2 mirrors are fed back to the satellite position using thrusters (drag-free control) while the differential motion signals are sent to the actuators on the frame of the housing to stabilize the FP cavity.
%The drag-free controls have been performed by several satellites such as TRIAD-I and Gravity Probe-B satellites.
LPF will operate at the Lagrange 1 (L1) point where the gravitational environment is stable, while DPF will demonstrate it in an earth orbit.
This will open a new window for future space missions.

%The main GW sources for DPF are IMBH binaries in our Galaxy.
DPF GW observations aim at the frequency of 0.1--1Hz which is important because the ground-based interferometers and LPF are not sensitive in this frequency range.
Also DPF data are expected to be more complicated than the one from ground-based interferometers due to the satellite orbital motion and the effects of the earth, and hence the development of data analysis technique for DPF has a significant meaning. 
%DPF can also measure the gravity of the earth with comparable sensitivity to other space missions currently operating and can provide complementary observation to others.
%~\cite{andolisa}.

%\if0%%%%%%%%%%%%%%%%%%%%%%%%%%%%%%%%%%%%%%%%%%
\begin{figure}[t]
  \centerline{\includegraphics[scale=1.4,clip]{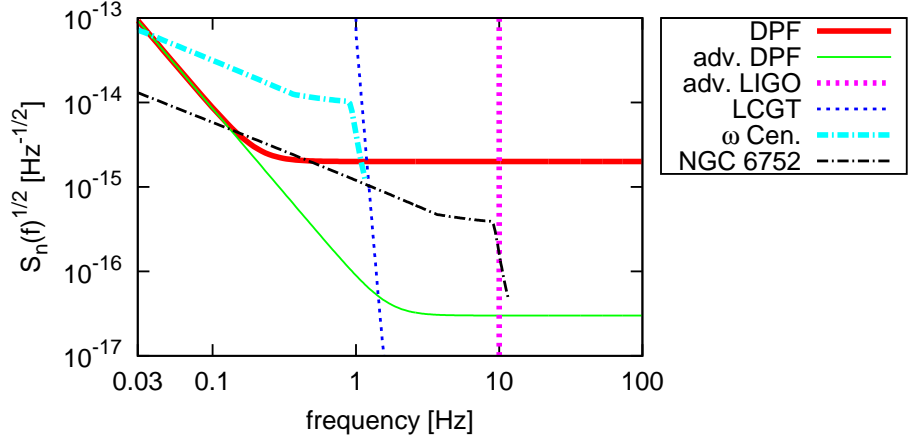} }
\vspace*{8pt}
 \caption{\label{noise-dpf}
The root noise spectral density $\sqrt{S_n(f)}$ of DPF is shown as the (red) thick solid curve, together with the one for adv.~DPF (high frequency part is limited by the shot noise rather than the laser frequency noise) as the (green) thin solid curve.
We also show the adv.~LIGO low frequency cutoff 10Hz as the (magenta) thick dotted line and the one for KAGRA down to 1Hz as (blue) thin dotted curve.
The sky-averaged amplitude of a $10^4\so$ equal-mass binary in $\omega$ Centauri is depicted as (light blue) thick dotted-dash curve and the one of a $10^3\so$ equal-mass binary in NGC 6752 is drawn as (black) thin dotted-dashed curve.
Here, we assumed that the observations are carried out from $f=0.03$Hz to mergers.
We also assumed that these BHs have dimensionless effective spin angular momentum $\chi=0.2$. (This figure is taken from Ref.~{\cite{yagiDPF}}.) }
\end{figure}
%\fi%%%%%%%%%%%%%%%%%%%%%%%%%%%%%%%%%%%%%%%%%%%%

\subsection{Noise Spectrum and Observable Range of DPF}
\label{dpf-noise}

The detected signal $s(t)=h(t)+n(t)$ can be expressed as the combination of the GW signal $h(t)$ and the noise $n(t)$.
If the noise is stationary, the noise spectral density $S_n(f)$ can be defined as
\beq
\left\langle \tilde{n}^*(f) \tilde{n}(f) \right\rangle = \delta (f-f') \frac{1}{2} S_n(f)\,,
\eeq
where $\tilde{n}(f)$ represents the Fourier component of the noise and the angle brackets denote the expectation value.
The root noise spectral density $\sqrt{S_n(f)}$ of DPF is given as%~\cite{dpf-noise}
\beq
S_n(f)=\frac{1.0\times 10^{-30}}{(2\pi )^4}\lmk \frac{f}{1\mrm{Hz}} \rmk^{-4} +4.0 \times 10^{-30} \ \mrm{Hz}^{-1},
\eeq
where the first term corresponds to the acceleration noises while the second term represents the laser frequency noise.
It is shown in Fig.~\ref{noise-dpf} as the (red) thick solid curve.
We set the lower frequency cutoff at $f=0.03$Hz due to the Earth gravity.
It may be possible to improve the sensitivity in which the higher frequency part is now limited by the shot noise.
We call this the ``adv.~DPF'' whose noise spectral density is given by~\cite{andoprivate}
\beq
S_n(f)^{(\mrm{adv})}=\frac{1.0\times 10^{-30}}{(2\pi )^4}\lmk \frac{f}{1\mrm{Hz}} \rmk^{-4} +9.0 \times 10^{-34} \ \mrm{Hz}^{-1}.
\eeq
It is shown in Fig.~\ref{noise-dpf} as the (green) thin solid curve.
We also show the lower cutoff frequencies of  the adv.~LIGO (magenta thick dotted) and KAGRA (blue thin dotted)~\footnote{It is likely that adv.~LIGO and adv.~VIRGO also have non-zero sensitivity at $f<10$Hz, but KAGRA seems to have better sensitivity at this frequency range. }.

\begin{figure}[t]
  \centerline{\includegraphics[scale=1.4,clip]{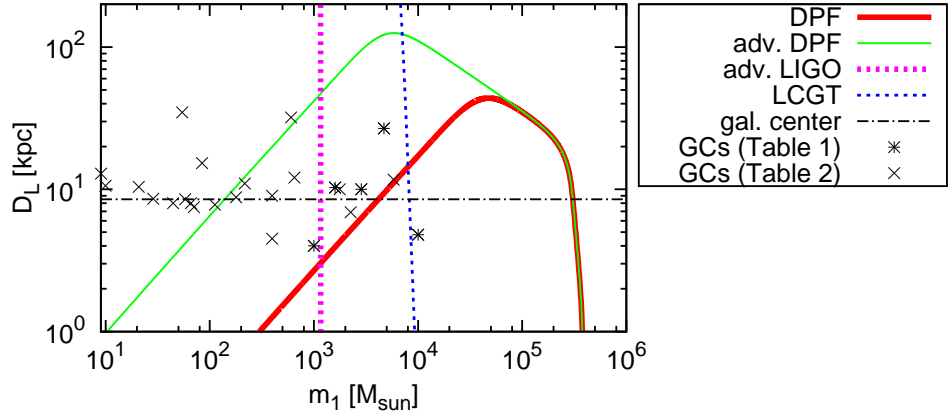} }
\vspace*{8pt}
 \caption{\label{range_shot}
This figure shows the observable range for each detector with SNR threshold $\rho_\mrm{thr}=5$ \ {\cite{yagiDPF}}.
The (red) thick solid curve represents the pattern-averaged one for DPF while the (green) thin solid curve shows the one with adv.~DPF.
Other 2 curves correspond to the same detectors as in Fig.~\ref{noise} \ {\cite{yagiDPF}}. 
``$*$'' shows the masses and distances for assumed equal-mass IMBH binaries in the galactic globular clusters shown in Table~\ref{table-mass1}, while ``$\times$'' shows the ones displayed in Table~2 of Ref.~{\cite{yagiDPF}}, whose masses have been determined by using $M-\sigma$ relation.} 
\end{figure}

\begin{table}
\tbl{The distances and possible IMBH masses at the centers of galactic globular clusters~{\cite{yagiDPF}}.}
%\begin{ruledtabular}
%\begin{center}
{\begin{tabular}{c||cc}  \hline\hline
 NGC & distance & (total) BH mass  \\ 
No. & (kpc) & ($\so$) \\ \hline\hline
5139 ($\omega$ Cen.) & 4.8 \ \cite{vandeven} & (3.0-4.75)$\times 10^4$ \ \cite{noyola} \\
 & & $\leq$1.2$\times 10^4$ \ \cite{anderson1-omega,anderson2-omega} \\
 & & (1.3-2.3) $\times 10^4$ \ \cite{miocchi} \\
6388 & 10.0 \ \cite{harris} & 5.7$\times 10^3$ \ \cite{lanzoni} \\
6715 (M54) & 26.8 \ \cite{harris} & 9.4$\times 10^3$ \ \cite{ibata} \\
6752 & 4.0 \ \cite{harris} & 2.0$\times 10^3$ \ \cite{ferraro} \\
7078 (M15) & 10.3 \ \cite{harris}  & 3.2$\times 10^3$ \ \cite{gerssen} \\ \hline\hline
\end{tabular} \label{table-mass1}}
%\end{center}
%\end{ruledtabular}
\end{table}

%\fi%%%%%%%%%%%%%%%%%%%%%%%%%%%%%%%%%%%%

In Fig.~{\ref{range_shot}}, we show the (sky-averaged) observable range of DPF and adv.~DPF as the (red) thick solid curve and the (green) thin solid curve, respectively. 
We also show the ones of adv.~LIGO and KAGRA with the same curves as in Fig.~{\ref{noise-dpf}}.
The (black) dashed horizontal line at $D_L=8.5$kpc corresponds to the galactic center.
We also plot the possible GW sources at the centers of the globular clusters shown in Table~{\ref{table-mass1}}, assuming that they consist of equal-mass IMBH binaries.

%%%%%%%%%%%%%%%%%%%%%%%%%%%%%%%%%%%%%%%%%%%%%%%%%%%%%%%%%%
\subsection{Beam Pattern Functions and the Effect of Detector Motion}
\label{app:1arm}

\begin{figure}[t]
  \centerline{\includegraphics[scale=.4,clip]{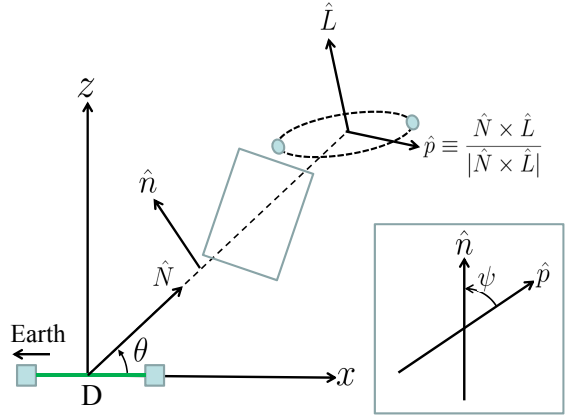}}
\vspace*{8pt}
 \caption{\label{polarization}
We introduce the detector coordinates $\{ x,y,z \}$ with its origin placed at the center of the DPF arm.
The $x$-axis coincides with the detector arm while $z$-axis is orthogonal to this $x$-axis and exists on $\hat{\bm{N}}$-$x$ plane, with $\hat{\bm{N}}$ denoting a unit vector pointing towards the source.
The angle of $\hat{\bm{N}}$ measured from $\hat{\bm{x}}$ is denoted as $\theta$ while the polarization angle $\psi$ is defined as the angle from $\hat{\bm{p}}\equiv \frac{\hat{\bm{N}}\times \hat{\bm{L}}}{ | \hat{\bm{N}}\times \hat{\bm{L}}|} $ to $\hat{\bm{n}}$ which is a unit vector made by projecting $\hat{\bm{z}}$ onto the plane perpendicular to $\hat{\bm{N}}$.
$\hat{\bm{n}}$ is orthogonal to $\hat{\bm{N}}$ and lies on $\hat{\bm{x}}$-$\hat{\bm{z}}$ plane. 
(This figure is taken from Ref.~{\cite{yagiDPF}}.)}
\end{figure}

%%%%%%%%%%%%%%%%%%%%%%%%%%%%%%%%%%%%%%%%%%%%%%%%%%%%%%%%%%%
%\section{1-Armed Detector}

The beam pattern functions of 1-armed interferometer are given by~\cite{yagiDPF}
\beqa
F^{+}(\theta,\psi) &=& \frac{1}{2}\sin^2\theta \cos 2\psi\,, 
\label{beamplus} \\
F^{\times}(\theta,\psi) &=& \frac{1}{2} \sin^2\theta \sin 2\psi\,.
\label{beam}
\eeqa
Here, $\theta$ is the angle between the arm and the incoming GW, and $\psi$ represents the polarization angle (see Fig.~\ref{polarization}).
The sky-averaged values of them are $\lla F^{+2} \rra = \lla F^{\times 2} \rra=2/15$.
In Fig.~\ref{noise-dpf}, we also show the sky-averaged GW amplitudes of a $10^4\so$ equal-mass IMBH binary in $\omega$ Centauri (light blue thick dotted-dashed) and a $10^3\so$ equal-mass one in NGC 6752 (black thin dotted-dashed).
We assume the dimensionless spin parameter $\chi$ of these BHs to be $\chi=0.2$ whose value is predicted to be the most probable~\cite{miller} (see also Ref.~\cite{buliga}).
%see also Ref.~\cite{buliga} for a recent discussion on the spins of the BHs at the centers of globular clusters).
We use the phenomenological inspiral-merger-ringdown hybrid waveforms~\cite{ajith,ajithspin} (see Sec.~\ref{hybrid}) to estimate these amplitudes.
One can see that GW frequency of IMBH binary in $\omega$ Centauri is too low for the ground-based interferometers, and hence its GW signals become unique sources for DPF.
%~\footnote{Higher harmonic signals of ringdown~\cite{bccc} may be detected with KAGRA if it is sensitive enough down to 1Hz.}.
On the other hand, the one in NGC 6752 can be detected with both DPF and the ground-based ones.
This suggests that it may be possible to perform joint searches between these detectors~{\cite{yagiDPF}}, which we explain in Sec.~\ref{sec:joint}.

\begin{figure}[t]
  \centerline{\includegraphics[scale=.45,clip]{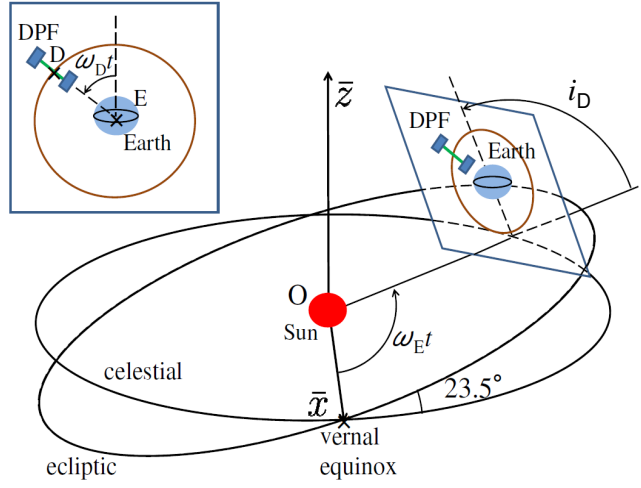}}
\vspace*{8pt}
 \caption{\label{orbit}
We introduce the celestial coordinates $\{ \bar{x},\bar{y},\bar{z} \}$ with $\bar{x}$-axis pointing the vernal equinox and the $\bar{z}$-axis is orthogonal to the celestial plane and is pointing the celestial north pole.
DPF orbits the earth with its plane always facing towards the sun and its arm pointing the earth.
It has an inclination angle $i_\mrm{D}=97.3^{\circ}$ against the celestial plane.
The magnified picture of the DPF orbital plane is shown at the top left panel.
The angular velocity of the earth $\omega_\mrm{E}$ is measured from the $\bar{x}$-axis while the one of the DPF $\omega_\mrm{D}$ is measured from the axis which is obtained by projecting the $\bar{z}$-axis onto the DPF orbital plane. 
(This figure is taken from Ref.~{\cite{yagiDPF}}.) }
\end{figure}

For later use, we introduce the celestial coordinates $\{ \bar{x},\bar{y},\bar{z} \}$ shown in Fig.~\ref{orbit}.
$\bar{x}$-axis points the vernal equinox, while $\bar{z}$-axis points the north celestial pole and is orthogonal to the celestial plane.
The source direction and the polarization angle $(\theta,\psi)$ can be re-expressed in terms of the source direction $(\bar{\theta}_\mrm{S},\bar{\phi}_\mrm{S})$ and the direction of the orbital angular momentum $(\bar{\theta}_\mrm{L},\bar{\phi}_\mrm{L})$ as~\cite{yagiDPF}
\beqa
\cos\theta &=&\cos\varphi_\mrm{D}(t) \sin i_\mrm{D} \cos{\bar{\theta}_\mrm{S}} \nonumber \\
& & -\left[ \cos\{ \varphi_\mrm{E}(t)-\bar{\phi}_\mrm{S}\} \cos\varphi_\mrm{D}(t) \cos i_\mrm{D}+\sin\{ \varphi_\mrm{E}(t)-\bar{\phi}_\mrm{S}\} \sin\varphi_\mrm{D}(t) \right] \sin{\bar{\theta}_\mrm{S}}.  \label{theta} \\
 \cos\psi  &=& \Bigl[ \cos\bar{\theta}_\mrm{S} \{ -\cos \{ \varphi_\mrm{E}(t)-\bar{\phi}_\mrm{L} \} \sin\varphi_\mrm{D}(t)+\sin \{ \varphi_\mrm{E}(t)-\bar{\phi}_\mrm{L} \} \cos\varphi_\mrm{D}(t)\cos i_\mrm{D} \} \sin\bar{\theta}_\mrm{L}  \nonumber \\
 & & \quad +\{ \cos \{ \varphi_\mrm{E}(t)-\bar{\phi}_\mrm{S} \} \sin\varphi_\mrm{D}(t) \cos\bar{\theta}_\mrm{L} \nonumber \\
 & & \quad - \cos\varphi_\mrm{D}(t) \{ \sin \{ \varphi_\mrm{E}(t)-\bar{\phi}_\mrm{S} \} \cos\bar{\theta}_\mrm{L}  \cos i_\mrm{D} %\nonumber \\
 + \sin \{ \bar{\phi}_\mrm{L}-\bar{\phi}_\mrm{S} \} \sin\bar{\theta}_\mrm{L}  \sin i_\mrm{D} \} \} \sin\bar{\theta}_\mrm{S} \Bigl] \nonumber \\
  & & \quad \Bigl[ (1-\cos^2\theta)\{ 1-(\sin\bar{\theta}_\mrm{S}\sin\bar{\theta}_\mrm{L}\cos(\bar{\phi}_\mrm{L}-\bar{\phi}_\mrm{S})+\cos\bar{\theta}_\mrm{S}\cos\bar{\theta}_\mrm{L})^2 \} \Bigl]^{-1/2}. \label{psi}
\eeqa

%%%%%%%%%%%%%%%%%%%%%%%%%%%%%%%%%%%%%%%%%%%%%%%%
\subsection{Event Rate}

In this subsection, we estimate the event rate for (I) Equal-mass BH binaries and (II) IMRIs using DPF.

\subsubsection{Equal-mass BH Binaries}

\if0%%%%%%%%%%%%%%%%%%%%%%%%%%%%%%%%%%%%%%

\begin{table}[t]
\caption{\label{table-mass} The distances and velocity dispersions of galactic globular clusters.
Possible masses of IMBHs, if they exit, are obtained from $M-\sigma$ relation~\cite{tremaine}}
%\begin{ruledtabular}
\begin{center}
\begin{tabular}{c||ccc}  \hline\hline
 NGC & distance & vel. disp. $\sigma$ & BH mass  \\ 
No. & (kpc)~\cite{harris} &  (km/s)~\cite{dubath} & ($\so$) \\ \hline\hline
104 & 4.5 & 10.0 & 794.7 \\
362 & 8.5 & 6.2 & 116.3 \\ 
1851 & 12.1 & 11.3 & 1299 \\
1904 & 12.9 & 3.9 & 18.04 \\
5272 & 10.4 & 4.8 & 41.57\\
5286 & 11.0 & 8.6 & 433.4 \\
5694 & 34.7 & 6.1 & 108.9 \\
5824 & 32.0 & 11.1 & 1209 \\
5904 & 7.5 & 6.5 & 140.6 \\
5946 & 10.6 & 4.0 & 19.97 \\
6093 & 10.0 & 14.5 & 3539 \\
6266 & 6.9 & 15.4 & 4508 \\
6284 & 15.3 & 6.8 & 168.6 \\
6293 & 8.8 & 8.2 & 357.9 \\
6325 & 8.0 & 6.4 & 132.4 \\
6342 & 8.6 & 5.2 & 57.35 \\
6441 & 11.7 & 19.5 & 11645 \\
6522 & 7.8 & 7.3 & 224.3 \\
6558 & 7.4 & 3.5 & 11.68 \\
6681 & 9.0 & 10.0 & 794.7 \\
7099 & 8.0 & 5.8 & 88.96  \\  \hline\hline
%6752 & 4.0 & 4.9 & 1000~\cite{ferraro} \\
%7078 (M15) & 10.3 & 15.1 & 3200~\cite{gerssen} \\
%5139 ($\omega$ Cen.) & 5.3 & 22~\cite{meylan} & 40000~\cite{noyola} \\
\end{tabular}
\end{center}
%\end{ruledtabular}
\end{table}

\fi%%%%%%%%%%%%%%%%%%%%%%%%%%%%%%%%%%%%%%%

DPF can detect GW signals from galactic IMBH binaries.
Two of the candidates where IMBH binaries might exist in our Galaxy are at the centers of globular clusters (GCs) and galactic massive young clusters (GMYCs). 

For the former case,
%IMBH binaries at the center of galactic globular clusters are also potential GW source candidates for DPF.
%Currently, about 150 galactic globular clusters have been discovered~\cite{harris}.
we select 21 globular clusters out of 150 globular clusters that have been found.
We determine the total mass of the possible IMBH binary by applying the $M-\sigma$ formula obtained by Tremaine \textit{et al.}~\cite{tremaine} as
\beq
M=1.35\times 10^8 \lmk \frac{\sigma}{200\mrm{km/s}} \rmk^4 \so\,.
\eeq
%
%o estimate the mass of the assumed IMBH in each globular cluster.
We assume the mass ratio of IMBH binary to be 1.
% with each BH having mass $M/2$. 
The results are listed in Table~2 of {Ref.~\cite{yagiDPF}}.
%, together with the distance to each globular cluster which is obtained from the catalogue of galactic globular clusters made by Harris (see the URL link shown in Ref.~\cite{harris}). 
%We also list the same data for NGC 6752, M15 and $\omega$ Centauri in the lower part of the table.
We plot IMBH binary of each globular cluster in Fig.~\ref{range_shot} and count how many of them lie within the observable range of DPF and Adv.~DPF.
%The results are shown in Fig.~\ref{gc}.
%The meaning of this figure is same as Fig.~\ref{range_shot}. 
One sees that 2 out of 26 globular clusters (5 shown in Table~\ref{table-mass1} + 21 shown in Table~2 of Ref.~\cite{yagiDPF}) might contain IMBH binaries detectable with DPF on average.
Then, the detection rate of the IMBH binaries in globular clusters $\dot{N}_\mrm{GC}$ is given by 
\beq
\dot{N}_\mrm{GC} \approx 2\times \frac{150}{26}\times\frac{1}{13.8 \ \mrm{Gyr}} = 8.4 \times 10^{-10} \mrm{yr}^{-1}.
\label{rate_GC}
\eeq
Here, following {Ref.~\cite{fregeau}}, we divide the number of globular clusters by the age of the universe.
This is because only one IMBH binary can be formed over its lifetime in each cluster.
% for the same reason discussed above Eq.~(\ref{rate_GMYC}).
%since only 1 IMBH binary is formed per cluster over its lifetime~\cite{fregeau}. 
When we use the adv.~DPF, 13 out of 26 IMBH binaries in the galactic globular clusters lie withing the observable reach of the detector.
This makes the event rate larger than Eq.~\eqref{rate_GC} by a factor $13/2=6.5$.

For the latter case, currently, more than 10 galactic massive young clusters (GMYCs) have been discovered~{\cite{gvaramadze}}. % in the galactic disk~\cite{gvaramadze}.
%At the galactic center, it is expected that there exist about 100 massive young clusters~\cite{gvaramadze} and more than half of them may contain IMBHs at their centers~\cite{port2002}.
G$\ddot{\mrm{u}}$rkan \et~\cite{gurkan} performed numerical simulations and found that  two IMBHs may form in GMYCs if the initial binary fraction is relatively large.
%at their centers with BH masses $10^3 \so$, which are likely to form binaries.
After IMBH binary formation, it shrinks due to the dynamical friction with the cluster stars.
The timescale of the dynamical encounters would be $\leq 1$Gyr \ {\cite{fregeau}}.
%The time scale of this process is typically $\leq$ 10 Myr which is independent of the local average stellar mass~\cite{gurkan,fregeau}.
%Then, binary shrinks via dynamical encounters with cluster stars, with the time scale of $\leq 1$Gyr~\cite{fregeau}.
%Finally, 2 IMBHs merge due to gravitational radiation within 1 Myr~\cite{gurkan}.
%If we assume that IMBH binaries are all situated at the galactic center, DPF would not be able to detect GW signals from IMBH binaries in GMYCs.
%We found that the expected event rate is more than one order of magnitude smaller than the GC case.
For simplicity, we assume that IMBH binaries are all located at the galactic center.
From Fig.~\ref{range_shot}, one sees that DPF is not sensitive enough to detect GW signals from IMBH binaries in GMYCs.
On the other hand, when we use adv.~DPF, it has ability to detect all of the IMBH binaries in GMYCs.
% leading to the event rate more than ten times larger than the one in Eq.~(\ref{rate_GMYC}).
Following Fregeau \et~{\cite{fregeau}}, we assume that the number of star clusters massive enough to form IMBH binaries is the same as the one of globular clusters, and 10$\%$ of them actually produce IMBH binaries.
In our galaxy, about 150 globular clusters~\cite{harris} have been observed.
This means that at least 15 IMBH binaries are expected to be within the reach of adv.~DPF.
%Since only one IMBH binary is formed over its lifetime for each cluster~\cite{fregeau}, 
The detection rate of the IMBH binaries in GMYCs with adv.~DPF can be roughly estimated as 
$\dot{N}_\mrm{GMYC}^\mrm{(adv.)} \approx 15/(13.8 \mrm{Gyr}) = 1.1 \times 10^{-9} \mrm{yr}^{-1}$.

\subsubsection{IMRIs}

\begin{figure}[t]
  \centerline{\includegraphics[scale=1.4,clip]{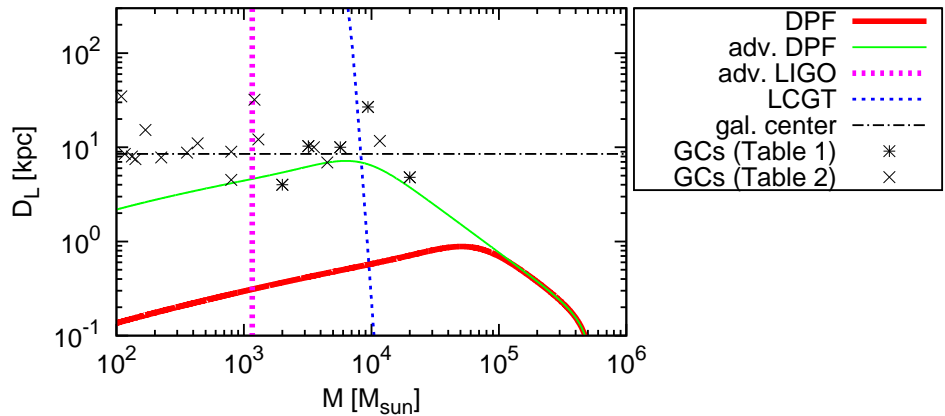} }
\vspace*{8pt}
 \caption{\label{range_shot_m2_10}
Observable ranges against total mass $M$ with $m_2=10\so$ \ {\cite{yagiDPF}}.
The meaning of each curve and plot is same as Fig.~\ref{range_shot}.}
\end{figure}

After supernova explosions, many stellar-mass BHs of $\sim 10\so$ would form. 
These would sink to the cluster cores due to the mass segregation, and their number density is expected to be comparable to the one of the main sequence stars, typically $n=10^6$pc$^{-3}$. 
In this subsection, we consider IMBH binaries with intermediate-mass ratio inspirals (IMRIs).
In Fig.~\ref{range_shot_m2_10}, we show observable ranges of DPF (red thick solid) and adv.~DPF (green thin solid) with $m_2=10\so$.
%Meanings of curves and plots are same as in Fig.~\ref{gc}.  
%We fix the total mass as $M=4\times 10^4\so$ and estimated the SNR for various $q$ with BH spin $\chi=0.2$.
%The result is shown in Fig.~\ref{centauri}.
%The horizontal dashed line represents the SNR threshold of 5.
One sees that DPF is not sensitive enough to detect any GWs from IMRIs with $m_2=10\so$, but adv.~DPF has ability to detect these signals from some of GCs and the galactic center.

For GMYCs, from Fig.~\ref{range_shot_m2_10}, one sees that even adv.~DPF is not sensitive, on average, to an IMRI signal of $(10+10^3)M_\odot$ at the galactic center.
%However, for some binaries with near optimal orientations of the angular momentum $\bm{L}$, SNRs go beyond $\rho=5$ with DPF. 
We found that 1.7$\%$ of these binaries have SNRs greater than 5 due to the optimal orientations~{\cite{yagiDPF}}.
 %binaries can be detected by adv.~DPF~\cite{yagiDPF}. 
On the other hand, the merger rate due to three-body interaction per cluster becomes $\dot{N}^\mrm{GMYC}=2\times10^{-8}\mrm{yr}^{-1}$ \ \cite{mapelli2010}~\footnote{For relatively large IMBH masses, the two-body interaction may give comparable contribution compared to the three-body interactions~{\cite{miller,mapelli2010}}.}. 
%Notice that this value does not exceed the limit we found in~\ref{sec_supply}.
%Gvaramadze \et~\cite{gvaramadze} proposed that there should be 70-100 massive young clusters in the galactic disk in total.
%They also suggested that about 50 young massive clusters exist in the galactic center.
%Following Ref.~\cite{mapelli2010}, we assume that 75$\%$ of GMYCs contain IMBHs at their centers, leading to about 50-75 GMYCs may contain IMBHs.
We make a conservative assumption that 50 GMYCs may contain IMBHs (see e.g. Refs.~\cite{gvaramadze,mapelli2010}).
In this case, the total detection rate becomes 
\beq
\dot{N}^\mrm{GMYC,tot}=0.017 \times 50 \times 2\times10^{-8}\mrm{yr}^{-1} =1.7\times 10^{-8} \mrm{yr}^{-1}.
\label{rate_imp}
\eeq
%
%This value is same as the one obtained in Ref.~\cite{mapelli2010}.

For GCs, 
%we use $\nu_\mrm{limit}^\mrm{GC}=\nu_\mrm{mass}^\mrm{GC}$ as our merger rate per cluster\footnote{Hopman~\cite{hopman} estimated the merger rate of $10\so$ BH into $3\times 10^4\so$ IMBH with Monte Carlo simulations, obtaining $\nu \approx 7\times 10^{-7}$ yr$^{-1}$. 
we found that the merger rate is limited by the requirement that the merger rate in a single cluster cannot exceed $\dot{N}_\mrm{mass} = (m_1/m_2) T_\mrm{age}^{-1}$ \ {\cite{miller,yagiDPF}}, where $T_\mrm{age}$ is the current age of the cluster.
Otherwise, the mass of the larger BH has been increased considerably. 
%If we assume that the order does not differ much for smaller IMBH mass cases, this exceeds the limit in Eq.~(\ref{limit}).}.
From Fig.~\ref{range_shot_m2_10}, one sees that IMRI signals from about 3 out of 26 globular clusters might be detected by adv.~DPF.
This leads to the estimate that it can detect IMRI signals from about $3\times 150/26=17.3$ globular clusters in total, and hence the detection rate of globular clusters in total becomes
\beq
\dot{N}^\mrm{GC,tot}  \lesssim 1.2\times 10^{-7} \mrm{yr}^{-1}.
\label{rate_imp}
\eeq
%

%%%%%%%%%%%%%%%%%%%%%%%%%%%%%%%%%%%%%%%%%%%%%%%%%%%%%%%%%%
\section{DECIGO/BBO}
\label{sec:DECIGO}

\subsection{Basic Designs}

DECIGO is first proposed by Seto, Kawamura and Nakamura~{\cite{setoDECIGO}}.
It consists of four triangular sets of detectors whose configuration is
shown in Fig.~\ref{default}.
This effectively corresponds to eight individual interferometers.
Each triangular detector has an arm-length of $10^3$km.
Its primary goal is to detect PGWB with the GW energy density of $\Omega_\mrm{GW}=10^{-16}$.
Since the
WD/WD confusion noise will have a cutoff frequency at around 0.2
Hz \ {\cite{farmer}}, DECIGO has an advantage on detecting this source over LISA (see Refs.~\cite{cutlerharms,harms,yagiseto} for the
discussions of the NS/NS confusion noise).
Therefore two of the four triangular detectors are located on the same site
forming a star of David so that correlation analysis~\cite{flanagan-corr,allenromano} can be performed to detect PGWB.
%In order to detect PGWB, it is necessary to perform correlation
%analysis~\cite{flanagan-corr,allenromano}.
The rest of the two detectors are placed far apart to increase the angular
resolutions of the source locations.

BBO has almost the same constellation as DECIGO.
The main difference is that while DECIGO is a Fabry-Perot type
interferometer, BBO is a transponder-type interferometer. 
BBO has arm-lengths of $5\times 10^4$km.

\subsection{Noise Spectrum}

The noise spectrum of BBO is given as follows.
The non-sky-averaged instrumental noise spectral density for BBO is
obtained from Ref.~\cite{cutlerholz} as~\cite{yagi:brane}
\beq
S_{n}^{\mathrm{inst}}(f)=
\left[
1.8\times 10^{-49} \left(\frac{f}{1 \mathrm{Hz}} \right)^2
                         +2.9\times 10^{-49}
                         +9.2\times 10^{-52}\left(\frac{f}{1 \mathrm{Hz}} \right)^{-4} \ \right]\mathrm{Hz^{-1}}\,.
\label{inst-BBO}
\eeq
It has 20/3 times better sensitivity than the one for the sky-averaged
sensitivity~{\cite{bertibuonanno}}.  
Apart from instrumental noise, there are astrophysical foreground confusion noises.
These confusion noise spectral densities and the energy densities of gravitational waves are related as~\cite{cutlerharms} 
\begin{equation}
S_n^{\mathrm{conf}}=\frac{3}{5\pi}f^{-3}\rho_c\Omega_{\mathrm{GW}}\,. 
\label{confusion}
\end{equation}
Here, $\rho_c\equiv \frac{3H_0^2}{8\pi}$ is the critical energy density of the Universe and 
\begin{equation}
\Omega_{\mathrm{GW}}\equiv\frac{1}{\rho_c}\frac{d \rho_{\mathrm{GW}}}{d \ln f}
\end{equation}
is the energy density of GWs per log frequency normalized by $\rho_c$.
The energy density of GWs that originate from extra-galactic WD binaries has been estimated as
$\Omega_{\mathrm{GW}}=3.6\times10^{-12}(f/10^{-3}\mathrm{Hz})^{2/3}$ which leads to the noise spectral density of~\cite{farmer}
\begin{equation}
S_n^{\mathrm{ex-gal}}(f)=4.2\times10^{-47}\left(\frac{f}{1\ \mathrm{Hz}}\right)^{-7/3} \ \mathrm{Hz^{-1}}\,.
\end{equation}
On the other hand, the one from galactic WD binaries has been calculated as~\cite{nelemans}
\begin{equation}
S_n^{\mathrm{gal}}(f)=2.1\times10^{-45}\left(\frac{f}{1\ \mathrm{Hz}}\right)^{-7/3} \ \mathrm{Hz^{-1}}.
\end{equation}
We multiply $S_n^\mrm{gal}$ and $S_n^\mrm{ex-gal}$ by a
factor $\fch \equiv \exp\{-2\left({f}/{0.05\mathrm{Hz}} \right)^2\}$,
which corresponds to the high frequency cutoff for the white dwarf
confusion noises.
We also have to take into account the confusion noise from NS binaries.
%If the merger rate today of these binaries is $\dot{n}_0 = 10^{-7} \mrm{Mpc}^{-3} \mrm{yr}^{-1}$, 
Its noise spectral density $S_n^{\mathrm{NS}}$ is estimated as~\cite{cutlerharms, yagiseto}
\beq
S_n^{\mathrm{NS}}(f) = 1.3\times10^{-48}\left(\frac{f}{1\ \mathrm{Hz}}\right)^{-7/3} \left(  \frac{\dot{n}_0}{10^{-7} \ \mathrm{Mpc}^{-3} \mathrm{yr}^{-1}} \right) \mathrm{Hz^{-1}}\,,
\eeq
where $\dot{n}_0$ denotes current merger rate density of NS/NS binaries.
Putting altogether, the total noise spectral density for BBO becomes
\ba
S_n(f) &=&  \min\left[ \frac{S_n^{\mathrm{inst}}(f)}{\exp(-\kappa T_\mrm{obs}^{-1}dN/df)},\ 
            S_n^{\mathrm{inst}}(f)+S_n^{\mathrm{gal}}(f) \fch (f) \right] \nn \\
& & +S_n^{\mathrm{ex-gal}}(f) \fch (f) + \mathcal{F}_\mrm{clean} S_n^{\mathrm{NS}}(f) . \label{noise-BBO}
\ea
Here $dN/df$ is the number density of galactic white dwarf binaries per unit frequency, which is given by~\cite{hughes}
\begin{equation}
\frac{dN}{df}=2\times10^{-3} \mathrm{Hz}^{-1}\left(\frac{f}{1 \ \mathrm{Hz}} \right)^{-11/3}\,.
\end{equation}
$\kappa\simeq 4.5$ is the average number of frequency bins that are lost when each galactic binary is fitted out.
The factor $\mathcal{F}_\mrm{clean}$ in front of $S_n^{\mathrm{NS}}(f)$ represents our assumption of the fraction of GWs that cannot be removed after foreground subtraction.
%~\cite{yagiDECIGO}. 
In this review, we assume that NS/NS foregrounds can be subtracted down to the level below the instrumental sensitivity.
The lower and higher frequency ends of the BBO sensitivity band are set as $f_{\mrm{low}}=10^{-3}$Hz and $f_{\mrm{high}}=100$Hz, respectively.
The noise spectrum of BBO is shown as a red thick solid curve in Fig.~\ref{noise}.
%We also show the amplitude of the GW signal from 
%a BH/NS of $(1.4+10)\so$ at $D_L=3$Gpc as a (red) thick dotted line.

DECIGO has been proposed with 3-4 times less sensitive spectrum than
BBO.
Its instrumental noise spectrum is given by
\be
S_{n,^{\mathrm{DECIGO}}}^\mathrm{inst}(f)=5.3\times 10^{-48}
\Biggl[(1+y^2)+\frac{2.3\times 10^{-7}}{y^4(1+y^2)}
+\frac{2.6\times 10^{-8}}{ y^4}\Biggr]\mathrm{Hz^{-1}},
\label{inst-DECIGO}
\ee
where $y\equiv f/f_p$ with $f_p \equiv 7.36$Hz.
However, this is not the fixed design sensitivity and there is a project
going on to improve the sensitivity to the same level as BBO.
%For the Fisher analyses in this review, we use Eq.~\eqref{inst-DECIGO} for Brans-Dicke and Massive gravity, while we use Eq.~\eqref{inst-BBO} for RS-II braneworld model and the effect of binary-disk interactions.
% assume that DECIGO has same sensitivity as BBO.
%BBO have the noise spectral densities shown in Eq.~(\ref{noise-BBO}).
The frequency range of DECIGO/BBO is given as
\be
(f_\mrm{low} ,f_\mrm{high}) = (10^{-3}\mrm{Hz}, 10^2\mrm{Hz})\,.
\ee
%

%%%%%%%%%%%%%%%%%%%%%%%%%%%%%%%%%%
\subsection{Event Rate}
\label{DECIGO-event}

The promising sources for DECIGO/BBO are inspirals of NS binaries.
In this subsection, following Cutler and Harms~{\cite{cutlerharms}}, we derive the detection rate of these sources.
Here, we adopt the model where NS/NS binaries only exist at redshift below $z=5$.
(This is inferred from observations. See the discussion below.)
Since DECIGO/BBO has enough sensitivity to detect the farthest NS/NS binaries considered here, the merger rate corresponds to the detection rate.
The NS/NS merger rate can be written as
\be
\dot{N}_\mrm{NSNS} = \int_0^{\infty} 4\pi [a_0 r_1(z)]^2 \dot{n}(z) \frac{d\tau_1}{dz}dz\,,
\ee
where 
\be
a_0 r(z)  = \int ^z_0 \frac{dz'}{H(z)}\,, \qquad 
\frac{d\tau}{dz} = \frac{1}{(1+z) H(z) }\,.
\label{a0rz}
\ee
Here, $H(z)$ represents the Hubble parameter at redshift $z$. 
For $\Lambda$-CDM cosmology, it is given by 
\be
H(z) \equiv H_0 \sqrt{(1+z)^3 \Omega_m + \Omega_\Lambda}\,.
\label{hz}
\ee
%
%We re-express the rate
$\dot{n}(z)$ is the merger rate of NS/NS binaries at redshift $z$ and it can be re-expressed   in terms of the current merger rate $\dot{n}_0$  and the redshift dependence $R(z)$  as $\dot{n}(z)=\dot{n}_0 \times R(z)$.
%In this review, we use  $\dot{n}_0=10^{-7}$ Mpc$^{-3}$ yr$^{-1}$ as a fiducial value.
%(This corresponds to the merger rate of $10^{-5} \mrm{yr}^{-1}$ per Milky Way equivalent galaxy.)
For $R(z)$, we adopt the following  piecewise linear fit ~\cite{cutlerharms, seto-lense}  based on observations discussed in Ref.~{\cite{schneider}};
\begin{eqnarray}
R(z)=\left\{ \begin{array}{ll}
1+2z & (z\leq 1) \\
\frac{3}{4}(5-z) & (1\leq z\leq 5) \\
0 & (z\geq 5). \\
\end{array} \right.
\label{Rz}
\end{eqnarray}
Then, we finally obtain the detection rate as
\be
\dot{N}_\mrm{NSNS} = 10^5 \left( \frac{\dot{n}_0}{10^{-7} \mrm{Mpc}^{-3} \mrm{yr}^{-1}} \right) \mrm{yr}^{-1}\,.
\ee
The detection rate for BH/NS binaries are expected to be roughly 1/10 of NS/NS detection rate.

We can also estimate the NS/NS foreground $S_n^{\mathrm{NS}}(f)$ as follows.
%For the total GW foreground by cosmological NS/NS binaries,  
First, we apply the convenient formula given by Phinney~\cite{phinney} as
%
%\begin{widetext}
\ba
\Omega_{\mathrm{GW}}^{\mathrm{NS}}&=&\frac{8\pi^{5/3}}{9}\frac{1}{H_0^2}\mathcal{M}^{5/3} f^{2/3}\int _{0}^{\infty} dz \frac{\dot{n}(z)}{(1+z)^{4/3}H(z)} \nn \\
&=& 3.74\times 10^{-12} h^{-3}_{72} \left( \frac{\mathcal{M}}{1.22 M_{\odot}} \right)^{5/3} 
                                                            \left( \frac{f}{1 \mathrm{Hz}}  \right)^{2/3} 
                                                             \left( \frac{\dot{n}_0}{10^{-7} \mathrm{Mpc}^{-3} \mathrm{yr}^{-1}}  \right). \label{nsb}
\ea
%%%%%%%%%%%%%%%%%%%%%%%%%%%%%%%%%%%%%%%%%%%%%%%%%%%
%
%
%The total GW foreground spectrum $S_n$ and the normalized energy density  $\Omega_{\mathrm{GW}}$ have the relation 
%
%\begin{equation}
%S_n=\frac{3}{5\pi}f^{-3}\rho_c\Omega_{\mathrm{GW}}, \label{confusion}
%\end{equation}
%
%where the critical energy density of the Universe $\rho_c$ is defined as $\rho_c \equiv (3/8\pi) H_0^2$.
By using Eqs.~\eqref{confusion} and~\eqref{nsb}, we obtain 
\begin{equation}
\sqrt{S_n^{\mathrm{NS}}}=1.76\times 10^{-24} h_{72} \left( \frac{\mathcal{M}}{1.22 M_{\odot}} \right)^{5/6} \left( \frac{\dot{n}_0}{10^{-7} \mathrm{Mpc}^{-3} \mathrm{yr}^{-1}}  \right)^{1/2}
                                       \left( \frac{f}{1 \mathrm{Hz}}  \right)^{-7/6}.
\end{equation}

%%%%%%%%%%%%%%%%%%%%%%%%%%%%%%%%%%%%%%%%%%%%%%%%%%%%%%%%%%
\section{Gravitational Waves from Compact Binaries}
\label{sec:binGW}

In this section, we focus on the GWs from inspiral binaries composed of compact stars such as BHs and NSs within the context of GR.
In Sec.~\ref{leading-circ}, we first derive the GWs from a circular binary at the leading order.
Then, In Sec.~\ref{higher-circ}, we extend the result to higher order terms.
Finally, in Sec.~\ref{sec:ecc}, we consider binaries with small eccentricity.

%%%%%%%%%%%%%%%%%%%%%%%%%%%%%%%%%%%%%%%%%%%%%%%%%%%%%%%%%%
\subsection{Quadrupole Gravitational Radiation from Binaries with Circular Orbit}
\label{leading-circ}

\begin{figure}[t]
  \centerline{\includegraphics[scale=.6,clip]{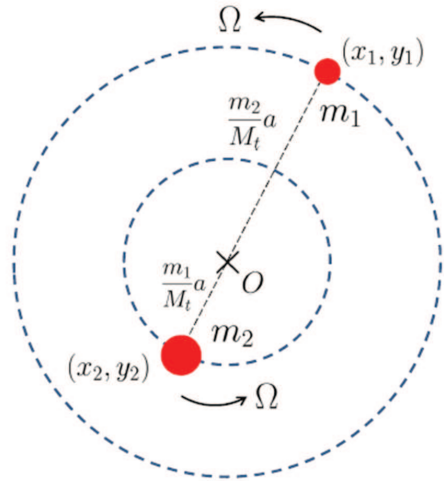} }
\vspace*{8pt}
 \caption{\label{binary}
A circular binary with masses $m_1$ and $m_2$.}
\end{figure}

In this section, we derive the leading order gravitational waves from compact binaries, namely the quadrupole radiation.
The binary circularizes quickly during its orbital decay due to the gravitational radiation reaction, and hence we restrict our attention to the binary with circular orbit in this section.
Also, we assume that the velocity of each body is sufficiently small compared to the speed of light.

Let us assume that we have two compact objects with masses $m_1$ and $m_2$ in the $x$-$y$ plane with orbital separation $a$ at a distance $D_L$ from the detector.
We define the total and the reduced mass as $M_t \equiv m_1 + m_2$ and $\mu \equiv m_1 m_2/M_t$, respectively.
The binary components obey the Kepler's Law at the leading order with orbital angular velocity defined as 
\be
\Omega \equiv \sqrt{\frac{M_t}{a^3}}\,.
\ee
As shown in Fig.~\ref{binary}, we denote the positions of these objects in the center of mass frame as $(x_1,y_1)$ and $(x_2, y_2)$, which can be expressed as
\ba
(x_1,y_1) & = &\left( \frac{m_2}{M_t}a \cos \left( \Omega t +\frac{\pi}{2} \right), 
                                              \frac{m_2}{M_t}a \sin \left( \Omega t +\frac{\pi}{2} \right) \right)\,, \\
(x_2,y_2) & = &\left( -\frac{m_1}{M_t}a \cos \left( \Omega t +\frac{\pi}{2} \right), 
                                              -\frac{m_1}{M_t}a \sin \left( \Omega t +\frac{\pi}{2} \right) \right)\,.
\ea
By taking the time derivative of the quadrupole moment $M_{ij} = \sum_{a=1,2} m_a x_{ai} x_{aj}$ twice,  we obtain 
\ba
h_+ &=& A_+ \cos 2\Omega t, \label{hplus}  \\
h_{\times} &=& A_{\times} \sin 2\Omega t\,. \label{hcross}
\ea
Here, we have defined
\ba
A_+ &\equiv & \frac{4\mu M_t}{a D_L} \left( \frac{1+(\hat{\bm{L}} \cdot \bm{n})^2}{2} \right), \label{Aplus}  \\
A_{\times} &\equiv & \frac{4\mu M_t}{a D_L} (\hat{\bm{L}} \cdot \bm{n})\,, \label{Across}
\ea
where $\bm{\hat{L}}=\bm{\hat{z}}$ is the unit orbital angular momentum vector and $\bm{n}$ is the unit vector pointing towards the direction of GW propagation.
%where we have redefined $\Omega t + \phi_S$ to $\Omega t$.
From this result, we see that the monochromatic GW frequency can be written as $f = \Omega/\pi$ at the leading order.

Next, we take the effect of radiation reaction into account and estimate the GW phase $\phi(t) \equiv \int 2\pi f dt$ \ {\cite{cutlerflanagan}}.
First, the total energy of the binary system is given as
\be
E = -\frac{\mu M_t}{2 a}\,.
\label{energy}
\ee
Next, by using the quadrupole formula
\be
\frac{dE_\mrm{GW}}{dt} = \frac{1}{5}\left\langle\dddot{M}_{ij}\dddot{M}_{ij}-\frac{1}{3}(\dddot{M}_{kk})^2\right\rangle\,,
\ee
we have the radiated energy due to GWs as
\be
\frac{dE_\mrm{GW}}{dt} = \frac{32}{5} \frac{\mu^2 M_t^3}{a^5}\,.
\ee
By using the balancing equation
\be
\frac{dE}{dt} = - \frac{dE_\mrm{GW}}{dt}\,,
\ee
the evolution of the binary separation can be expressed as 
\be
\dot{a} = -\frac{a}{E} \frac{dE}{dt} = -\frac{64}{5} \frac{\mu M_t^2}{a^3}\,,
\label{dotaGW}
\ee
which can be solved to yield
\be
a(t) = \left( \frac{256}{5} \mu M_t^2 (t_0-t) \right)^{1/4}\,,
\label{at-leading}
\ee
where $t_0$ is the coalescence time.
This can be turned into the frequency evolution as 
\ba
\dot{f}=\frac{\dot{\Omega}}{\pi} \! &=& \! -\frac{3}{2}\frac{M_t^{1/2}}{\pi}\frac{\dot{a} }{a^{5/2}} \notag \\
                                             \!  &=& \! \frac{96}{5}\pi ^{8/3} \mathcal{M} ^{5/3}f^{11/3}, \label{frequency}
\ea
where $\mc \equiv M_t^{2/5} \mu^{3/5}$ is the \textit{chirp mass}.
This equation can be solved as
\be
f=\left( \frac{5}{256} \right)^{3/8} \frac{1}{\pi \mathcal{M}^{5/8}} \frac{1}{(t_0-t)^{3/8}}\,. \label{f}
\ee
By integrating this by once, we get the phase as
\be
\phi(t)=\int 2\pi f dt = -2 \left( \frac{1}{5} \mathcal{M}^{-1} (t_0-t) \right)^{5/8} + \phi_0, \label{phi}
\ee
with $\phi_0$ representing the coalescence phase.

Next, we derive the waveform in the Fourier domain $\tilde{h}(f)$ defined as
\be
\tilde{h}(f) \equiv \int^{\infty}_{-\infty} e^{2\pi ift} h(t) dt\,.
\ee
Since we have assumed that the velocity of each binary component is much smaller than the speed of light, the amplitude $A(t)$ and the phase satisfy the following conditions: $d \ln A/dt \ll d \phi/dt$ and $d^2 \phi/dt^2 \ll (d \phi /dt)^2$. 
Under this situation, we can apply the \textit{stationary phase approximation}, where the Fourier component of a function $B(t)=A(t) \cos \phi(t)$ becomes~\cite{cutlerflanagan} 
\be
\tilde{B}(f) \approx \frac{1}{2} A(t) \left( \frac{df}{dt} \right)^{-1/2} \exp{i(2\pi ft-\phi(f)-\pi/4)}.
\ee
Here, $t$ is the time that satisfies $d\phi(t)/dt=2\pi f$ and $\phi(f)=\phi[t(f)]$.
From Eqs.~\eqref{f} and~\eqref{phi}, we have 
\ba
t(f) &=& t_0-5\mc(8\pi \mc f)^{-8/3} \,, \\
\phi(f) &=& \phi_0-2(8\pi \mathcal{M} f)^{-5/3}\,.
\ea
Therefore, we obtain
\ba
\tilde{h}_{+}(f) &=& A f^{-7/6} \left( \frac{1+(\hat{\bm{L}} \cdot \bm{n})^2}{2} \right) e^{i\Psi_{+}(f)}, \\
\tilde{h}_{\times}(f) &=& A f^{-7/6} (\hat{\bm{L}} \cdot \bm{n})  e^{i\Psi_{\times}(f)},
\ea
where the amplitude and phases are given as
\ba
A &=& \sqrt{\frac{5}{24}}\frac{1}{\pi^{2/3}} \frac{\mathcal{M}^{5/6}}{D_L}, \\ \label{amp}
\Psi_{+}(f) &=& 2\pi ft_0-\phi_0-\frac{\pi}{4}+\Psi_\mrm{0PN}(f), \\
\Psi_{\times}(f) &=& \Psi_{+}(f) + \frac{\pi}{2},
\ea
respectively.
Here, $\Psi_\mrm{0PN}(f)$ is defined as
\be
\Psi_\mrm{0PN}(f) \equiv \frac{3}{128}(\pi \mathcal{M}f)^{-5/3}\,,
\label{Psi-0PN}
\ee
and $D_L$ is the luminosity distance given as 
\be
D_L = (1+z) \int_0^z \frac{dz'}{H(z')}\,.
\ee
%
%For $\Lambda$CDM cosmology, the Hubble parameter at redshift $z$ becomes
%
%\be
%H(z) \equiv H_0 \sqrt{(1+z)^3 \Omega_m + \Omega_\Lambda}\,.
%\label{hz}
%\ee
%

%%%%%%%%%%%%%%%%%%%%%%%%%%%%%%%%%%%%%%%%%%%%%%%%%%%%%%%%%%
\subsection{Higher Post-Newtonian Corrections}
\label{higher-circ}

In the previous section, we derived the waveform at the leading order, where we used the quadrupole formula for the radiated energy flux and assumed that the binary orbit is of Newtonian.
In this section, we extend the previous analysis to higher orders.
Since we go beyond Newtonian, this expansion is called the \textit{post-Newtonian} (PN).
Our expansion parameter is $(v^2/c^2)$ where $v$ is the typical velocity of the binary component, and the terms that are proportional to $(v^2/c^2)^n$ are of $n$PN orders.
In this section, we consider the extension up to 2PN order~{\cite{blanchet2pn}}.

Furthermore, we neglect the PN corrections to the amplitude.
This is because when we take the correlation between two different waveforms, it is more sensitive to the deviation in the phase than the one in the amplitude. 
This type of waveform is called the \textit{restricted} PN waveform.

In Ref.~{\cite{blanchet2pn}}, expressions for $E$, $\dot{E}$ and $\Omega$ are shown to 2PN order.
By using them, up to this PN order, $\dot{f}$ becomes
\ba
\dot{f} &=& \frac{96}{5}\pi^{8/3} \mc^{5/3} f^{11/3} \biggl[ 1-\left( \frac{743}{336}+\frac{11}{4}\eta \right) x 
           +(4\pi-\beta)x^{1/2}  \nn \\
          & &+\left( \frac{34103}{18144}+\frac{13661}{2016}\eta +\frac{59}{18}\eta^2+\sigma \right) x^2  \biggr]\,, \label{fdot}
\ea
where $\eta \equiv \mu/M_t$ is the symmetric mass ratio 
with PN expansion parameter $x \equiv (\pi M_t f)^{2/3} = M_t/a = v^2$.
The spin-orbit and the spin-spin couplings are given as
\ba
\beta&=& \frac{1}{12}\sum_{i=1,2}\left( 113\frac{m_i^2}{M_t}+75\eta \right) \hat{\bm{L}}\cdot \bm{\chi_i},  \\
\sigma&=&\frac{1}{48}\eta\left\{ -247 \bm{\chi_1} \cdot \bm{\chi_2}
                 +721 \left(\hat{\bm{L}}\cdot \bm{\chi_1} \right) \left(\hat{\bm{L}}\cdot \bm{\chi_2} \right)\right\},
\ea
where the dimensionless spin parameter is defined as $\bm{\chi}_i = \bm{S}_i/m_i^2$ with $\bm{S}_i$ denoting the spin angular momentum of the $i$-th compact object.
The gravitational wave phase in the Fourier domain can be calculated as
\ba
\Psi_{+}(f) &= &2\pi ft_0-\phi_0-\frac{\pi}{4}+\Psi_\mrm{0PN}(f) \biggl[1   
                 + \left( \frac{3715}{756}+\frac{55}{9}\eta \right)x \nn \\
                & & -4(4\pi-\beta)x^{3/2} 
                 + \left( \frac{15293365}{508032}+\frac{27145}{504}\eta+\frac{3085}{72}\eta^2-10\sigma \right) x^2  \biggr]\,. \\
\Psi_{\times}(f) &= & \Psi_{+}(f) + \frac{\pi}{2}\,. 
%\label{Psi}
\ea
%

%%%%%%%%%%%%%%%%%%%%%%%%%%%%%%%%%%%%%%%%%%%%%%%%%%%%%%%%%%
\subsection{Gravitational Waves from Binaries with Slightly Eccentric Orbit}
\label{sec:ecc}

In this section, we consider gravitational waves from a binary that have a small eccentricity $e$.
We can express the eccentric orbit as
\be
r_\mrm{orb}=\frac{a(1-e^2)}{1+e\cos \Omega t}, \label{r-ecc}
\ee
where the orbital angular velocity $\Omega $ can be written as
\be
\Omega=\frac{\sqrt{M_t a(1-e^2)}}{r^2}. \label{omega-ecc}
\ee
By using the quadrupole formula, the radiated energy flux can be estimated as
\be
\frac{dE_{\mathrm{GW}}}{dt}=\frac{32}{5}\frac{\eta^2 M_t^5}{a^5(1-e^2)^{7/2}}\left( 1+\frac{73}{24}e^2+\frac{37}{96}e^4 \right)\,,
\label{Edot-ecc}
\ee
and similarly, the radiated angular momentum flux is calculated as
\ba
\frac{dJ_z^{\mathrm{GW}}}{dt}&=& \frac{2}{5}\varepsilon_{ijk}
         \left\langle\dddot{Q}_{xy}(\ddot{Q}_{yy}-\ddot{Q}_{xx})+\ddot{Q}_{xy}(\dddot{Q}_{xx}-\dddot{Q}_{yy})\right\rangle \nn \\
&=& \frac{32}{5}\frac{4\eta^2 M_t^{9/2}}{a^{7/2}(1-e^2)^{2}}\left( 1+\frac{7}{8}e^2 \right)\,.
\label{Jdot-ecc}
\ea

Next, we derive the evolution of the semi-major axis $a$ and the eccentricity $e$.
Since the orbital energy of the binary is given in Eq.~\eqref{energy} and the angular momentum is given as
\be
J_z=\mu r_\mrm{orb}^2 \Omega =\frac{\eta M_t^2}{\sqrt{M_t}}\sqrt{a(1-e^2)} \label{Jz}
\ee
at the Newtonian order,
by using Eqs.~\eqref{Edot-ecc} and \eqref{Jdot-ecc}, the evolutions of $a$ and $e$ can be calculated as
\ba
\frac{da}{dt} &=& \frac{da}{dE}\frac{dE}{dt} \nn \\
                 &=& -\frac{64}{5}\frac{\eta M_t^3}{a^3(1-e^2)^{7/2}}\left( 1+\frac{73}{24}e^2+\frac{37}{96}e^4 \right)\,, \\ 
\frac{de}{dt} &=& \frac{1-e^2}{2e} \left( \frac{1}{a} \frac{da}{dt} - 2\frac{1}{J_z} \frac{dJ_z}{dt} \right) \nn \\
&=& -\frac{304}{15}e\frac{\eta M_t^3}{a^4(1-e^2)^{5/2}}\left( 1+\frac{121}{304}e^2 \right)\,.
\label{dedt}
\ea
From these equations, we obtain
\be
\frac{da}{a}=\frac{19}{12}\frac{1+\frac{73}{24}e^2+\frac{37}{96}e^4}{e(1-e^2)( 1+\frac{121}{304}e^2 )} de.
\ee
Therefore, by integrating once, we obtain the relation between $a$ and $e$ as
\be
\frac{a}{a_i}=\frac{1-e_i^2}{1-e^2}\left( \frac{e}{e_i} \right)^{12/19}
                   \left[ \frac{ 1+\frac{121}{304}e^2 }{ 1+\frac{121}{304}e_i^2 } \right], 
\label{a(e)1}
\ee
where $a_i$ and $e_i$ denote the initial semi-major axis and eccentricity, respectively.

Finally, we derive the gravitational waveform (in the Fourier domain) from the compact object with the small eccentric orbit~{\cite{krolak,cutlerharms}}.
By substituting Eq.~\eqref{a(e)1} into the GW frequency $f=\Omega/\pi$,
%
%\be
%f=\frac{\Omega}{\pi}=\frac{M_t^{1/2}}{\pi a^{3/2}},
%\ee
%
we obtain
\ba
f(e)&=&f_i\frac{\zeta(e_i)}{\zeta(e)}, \\
\zeta(e)&\equiv&e^{18/19}\frac{(1+\frac{121}{304}e^2)^{\frac{1305}{2299}}}{(1-e^2)^{3/2}},
\ea
where $f_i$ is the frequency at $e=e_i$.
Especially, when $e\ll 1$, we have 
\be
e \approx e_i \left( \frac{f}{f_i} \right)^{-19/18}, \label{e(f)}
\ee
so that up to $\mathcal{O}(e)$, the \textit{asymptotic eccentricity invariant},
\be
I_e \equiv e^2f^{19/9}
\ee
is conserved.

The evolution the frequency $\dot{f}$ is given by
\be
\dot{f} = \frac{96}{5} \pi^{8/3} \mc^{5/3} f^{11/3}  \left(1+ \frac{157}{24}e^2\right) + \mathcal{O}(e^4)\,, 
\ee
for $e \ll 1$.
By substituting Eq.~\eqref{e(f)} into the above equation, we have
\be
\dot{f}=\frac{96}{5} \pi^{8/3} \mc^{5/3} f^{11/3}  \left(1+ \frac{157}{24}I_e f^{-19/9}\right).
\ee
From this equation, we can calculate $t(f)$ and $\phi(f)$, leading to the GW phases in the Fourier domain as
\ba
\Psi_{+}(f)&=&2\pi ft(f)-\phi(f)-\frac{\pi}{4} \nn \\
         &=&2\pi ft_0-\phi_0-\frac{\pi}{4}+\Psi_\mrm{0PN}(f) \left(1-\frac{2355}{1462}I_ef^{-19/9}\right)\,, \\
\Psi_{\times}(f)&=& \Psi_{+}(f) + \frac{\pi}{2}\,.
\ea
% 

%%%%%%%%%%%%%%%%%%%%%%%%%%%%%%%%%%%%%%%%%%%%%%%%%%%%%%%%%%
\section{Observed PN Inspiral Waveforms}
\label{sec:obsGW}

In this section, we explain the response of two-armed interferometers to GWs from compact binaries within the context of GR.
Expressions below can be applied to LISA and DECIGO/BBO.
(See Sec.~\ref{app:1arm} for DPF.)  
As explained in the previous section, we use the so-called \textit{restricted PN waveforms} where we keep the higher PN terms in the phase but only take the leading terms (leading PN and dominant harmonic contributions) for the amplitude~{\cite{cutlerflanagan}}. 
%We only consider the dominant harmonic contributions. 
In Sec.~\ref{spin-aligned}, we consider the case where the orbital angular momentum $\bm{L}$ and the spin vectors $\bm{S}_i \ (i=1,2)$ are all aligned (or anti-aligned).
Then, in Sec.~\ref{precession}, we include the effect of precession and see how this modifies the waveforms.
Finally in Sec.~\ref{hybrid}, we explain the fitted waveforms including all of inspiral, merger and ringdown phases for the spin-aligned binaries with circular orbits.

%%%%%%%%%%%%%%%%%%%%%%%%%%%%%%%%%%%%%%%%%%%%%%%%%%%
\subsection{Spin-Aligned Binaries}
\label{spin-aligned}

\begin{figure}[thbp]
% \begin{center}
  \centerline{\includegraphics[scale=.4,clip]{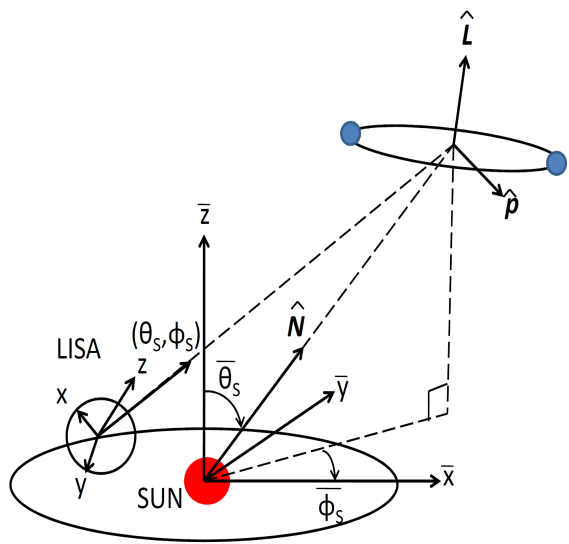}} 
%   \end{center}
\vspace*{8pt}
 \caption{\label{fig1} We use two types of coordinates: (i)a barred barycentric frame $(\bar{x},\bar{y},\bar{z})$ tied to the ecliptic and centered in the solar system barycenter, and (ii) an unbarred detector frame $(x,y,z)$, centered in the barycenter of the triangle and attached to the detector. (This figure is taken from Ref.~{\cite{yagiLISA}}.)}
\end{figure}

Following Ref.~{\cite{cutler1998}}, we introduce two Cartesian reference frames: (i) a barred barycentric frame $(\bar{x},\bar{y},\bar{z})$ tied to the ecliptic and centered in the solar system barycenter, with $\hat{\bar{\bm{z}}}$ (unit vector in ${\bar{\bm{z}}}$ direction) normal to the ecliptic and $\bar{x}\bar{y}$-plane aligned with the ecliptic and (ii) an unbarred detector frame $(x,y,z)$, centered in the barycenter of the triangle and attached to the detector, with $\hat{\bm{z}}$ normal to the detector plane (see Fig.~\ref{fig1}). 
The orbit of the detector barycenter in the barred frame can be written as
\begin{equation}
\bar{\theta}(t)=\pi/2, \qquad \bar{\phi}(t)=2\pi t/T,
\end{equation}
where $T=1$ yr and we have assumed $\bar{\phi}(t=0)=0$.

A triangular detector has effectively two individual interferometers.
Therefore, it is possible to measure both polarizations with a single detector.
We first focus on the interferometer I which consists of the arms 1 and 2.
This detector measures 
\begin{equation}
h_{\mathrm{I}}(t) \equiv \frac{\delta L_1(t)-\delta L_2(t)}{L} 
        = \frac{\sqrt{3}}{2} \left( \frac{1}{2}h_{xx}-\frac{1}{2}h_{yy} \right), \label{output}   
\end{equation}
where $\delta L_1(t)$ and $\delta L_2(t)$ are the differences in lengths in arms 1 and 2 due to GWs.
$L$ is the armlength where GWs are not present.
The factor $\sqrt{3}/2$ comes from the fact that the opening angle of adjacent arms is 60$^\circ$.
Next, we introduce two principal axes for the wave;
$\hat{\bm{p}}=\hat{\bm{N}}\times\hat{\bm{L}}/\lvert \hat{\bm{N}}\times\hat{\bm{L}} \rvert$ and 
$\hat{\bm{q}}=\hat{\bm{p}}\times\hat{\bm{N}}$, where $\hat{\bm{L}}$ is the unit vector parallel to the orbital angular momentum and $\hat{\bm{N}}$ is the unit vector pointing towards the center of mass of the binary.
The two polarizations are exactly $\pi/2$ out of phase and the waveform becomes
\begin{equation}
h_{ab}(t)=A_{+} H^+_{ab} \cos\phi(t) + A_{\times} H^{\times}_{ab} \sin\phi(t), 
\end{equation}
where $H^+_{ab}$ and $H^{\times}_{ab}$ are the polarization basis tensors, defined as
\begin{equation}
H^{+}_{ab}=p_a p_b-q_a q_b, \qquad H^{\times}_{ab}=p_a q_b+q_a p_b. \label{Hpol}
\end{equation}
From Eqs.~(\ref{output})--(\ref{Hpol}), the detector output $h_{\mathrm{I}}(t)$ becomes
%
%\begin{widetext}
\begin{equation}
h_{\mathrm{I}}(t)=\frac{\sqrt{3}}{2} A_{+} F_{\mathrm{I}}^+(\theta_{\mathrm{S}},\phi_{\mathrm{S}},\psi_{\mathrm{S}}) \cos\phi(t)
          +\frac{\sqrt{3}}{2} A_{\times} F_{\mathrm{I}}^{\times}(\theta_{\mathrm{S}},\phi_{\mathrm{S}},\psi_{\mathrm{S}}) \sin\phi(t). \label{output2}
\end{equation}
%\end{widetext}
Here, $F_{\mathrm{I}}^+(\theta_{\mathrm{S}},\phi_{\mathrm{S}},\psi_{\mathrm{S}})$ and $F_{\mathrm{I}}^{\times}(\theta_{\mathrm{S}},\phi_{\mathrm{S}},\psi_{\mathrm{S}})$ are the detector beam-pattern coefficients for the plus and cross polarization modes, respectively, and when the detector is an interferometer, they are given by
%
%\begin{widetext}
\begin{eqnarray}
\label{beam-pattern}
F_{\mathrm{I}}^{+}(\theta_{\mathrm{S}},\phi_{\mathrm{S}},\psi_{\mathrm{S}}) 
                &=&\frac{1}{2}(1+\cos^2 \theta_{\mathrm{S}}) \cos(2\phi_{\mathrm{S}}) \cos (2\psi_{\mathrm{S}})
                  -\cos(\theta_{\mathrm{S}}) \sin(2\phi_{\mathrm{S}}) \sin(2\psi_{\mathrm{S}}), \nn \\
\\
F_{\mathrm{I}}^{\times}(\theta_{\mathrm{S}},\phi_{\mathrm{S}},\psi_{\mathrm{S}})
                &=&\frac{1}{2}(1+\cos^2 \theta_{\mathrm{S}}) \cos(2\phi_{\mathrm{S}}) \sin (2\psi_{\mathrm{S}})
                  +\cos(\theta_{\mathrm{S}}) \sin(2\phi_{\mathrm{S}}) \cos(2\psi_{\mathrm{S}}). \nn \\
\end{eqnarray} 
%\end{widetext}
$(\theta_{\mathrm{S}},\phi_{\mathrm{S}})$ represents the direction of the source in the detector frame and $\psi_{\mathrm{S}}$ is the polarization angle defined as 
\begin{equation}
\tan\psi_{\mathrm{S}}=\frac{\hat{\bm{q}}\cdot\hat{\bm{z}}}{\hat{\bm{p}}\cdot\hat{\bm{z}}}
                                =\frac{\hat{\bm{L}}\cdot\hat{\bm{z}}-(\hat{\bm{L}}\cdot\hat{\bm{N}})(\hat{\bm{z}}\cdot\hat{\bm{N}})}
                                  {\hat{\bm{N}}\cdot(\hat{\bm{L}}\times\hat{\bm{z}})}. \label{tanpsi}
\end{equation}

Also, another interferometer can be constructed from arms 2 and 3. We call this interferometer II' and its signal can be written as $h_{\mathrm{II}'}=(\delta L_2(t)-\delta L_3(t))/L$.
However, since $h_{\mathrm{I}}$ and $h_{II'}$ have some correlations, they are not independent interferometers.
We combine interferometers I and II' to construct interferometer II which is uncorrelated with detector I.
The signal of interferometer II is 
\begin{equation}
h_{\mathrm{II}}(t) \equiv \frac{1}{\sqrt{3}}[h_{\mathrm{I}}(t)+2h_{II'}(t)]=\frac{\sqrt{3}}{2}\left[ \frac{1}{2}(h_{xy}+h_{yx}) \right].
\end{equation}
This interferometer II corresponds to the one that is rotated by 45$^\circ$ with respect to interferometer I.
Thus the beam-pattern coefficients for the interferometer II are
\begin{eqnarray}
F_{\mathrm{II}}^{+}(\theta_{\mathrm{S}},\phi_{\mathrm{S}},\psi_{\mathrm{S}})&=&F_{\mathrm{I}}^{+}(\theta_{\mathrm{S}},\phi_{\mathrm{S}}-\pi/4,\psi_{\mathrm{S}}), \\
F_{\mathrm{II}}^{\times}(\theta_{\mathrm{S}},\phi_{\mathrm{S}},\psi_{\mathrm{S}})&=&F_{\mathrm{I}}^{\times}(\theta_{\mathrm{S}},\phi_{\mathrm{S}}-\pi/4,\psi_{\mathrm{S}}).
\end{eqnarray}

When we perform parameter estimation, we include the direction $(\bar{\theta}_{\mathrm{S}},\bar{\phi}_{\mathrm{S}})$ and the orientation $(\bar{\theta}_{\mathrm{L}},\bar{\phi}_{\mathrm{L}})$ of the source measured in the solar barycentric frame in binary parameters.
Therefore we need to express the angles $\theta_\mrm{S}(t)$, $\phi_\mrm{S}(t)$  and $\psi_\mrm{S}(t)$ 
%the waveforms (especially $\hat{\bm{L}}\cdot\hat{\bm{N}}$ and the beam-pattern functions $F_{\alpha}^{+}$ and $F_{\alpha}^{\times}$ which appear in Eqs.~(\ref{Apol})-(\ref{phipol})) 
in terms of $\bar{\theta}_{\mathrm{S}},\bar{\phi}_{\mathrm{S}},\bar{\theta}_{\mathrm{L}}$ and $\bar{\phi}_{\mathrm{L}}$.
Explicit formulas are given in~\ref{app-spin-aligned}.

%The interferometer having three arms corresponds to having two individual detectors.
%We label each detector as detector $\mathrm{I}$ and $\mathrm{II}$, respectively.
%The waveforms measured by each detector are given as
Re-expressing the waveform measured by each interferometer in terms of an amplitude and phase, it becomes
\begin{equation}
%\begin{split}
h_{\alpha}(t)=\frac{\sqrt{3}}{2} \frac{2\eta M_t^2}{aD_L} 
                   A_{\mathrm{pol},\alpha}(t) \cos \left[ \phi(t) + \varphi_{\mathrm{pol},\alpha}(t)+\varphi_{D}(t) \right], 
                               \label{waveform1}
%\end{split}
\end{equation}
where $\alpha=\mathrm{I}, \mathrm{II}$ is the label to specify the interferometer.
The polarization amplitude $A_{\mathrm{pol},\alpha}(t)$, the polarization phases $\varphi_{\mathrm{pol},\alpha}(t)$ and the Doppler phase $\varphi_{D}(t)$ are defined as 
%
%%%\begin{widetext}
\begin{eqnarray}
A_{\mathrm{pol},\alpha}(t)&=&\sqrt{(1+(\hat{\bm{L}}\cdot\hat{\bm{N}})^2)^2F_{\alpha}^{+}(t)^2+4(\hat{\bm{L}}\cdot\hat{\bm{N}})^2F_{\alpha}^{\times}(t)^2}, \label{Apol} \\
 \cos(\varphi_{\mathrm{pol},\alpha}(t))&=&\frac{(1+(\hat{\bm{L}}\cdot\hat{\bm{N}})^2)F^{+}_{\alpha}(t)}{A_{\mathrm{pol},\alpha}(t)}, \label{phi-pol-cos} \\
 \sin(\varphi_{\mathrm{pol},\alpha}(t))&=&\frac{2(\hat{\bm{L}}\cdot\hat{\bm{N}}) F^{\times}_{\alpha}(t)}{A_{\mathrm{pol},\alpha}(t)}, \label{phipol} \\
\varphi_{D}(t)&=&2\pi f(t) R \sin \bar{\theta}_{\mathrm{S}} \cos[\bar{\phi}(t)-\bar{\phi}_{\mathrm{S}}], \label{doppler-phase}
\end{eqnarray}
%%%\end{widetext}
where $R=1$AU. 
The Doppler phase denotes the difference between the phase of the wavefront at the detector and the phase of the wavefront at the solar system barycenter.
It arises from the fact that the detector orbits the Sun.
%The factor $\sqrt{3}/2$ in Eq.~(\ref{waveform1}) comes from the 60$^\circ$ opening angle of adjacent detector arms of LISA.

Later, we estimate the measurement accuracies of the binary parameters using the matched filtering analysis, where we work in the Fourier domain.
Therefore we calculate the Fourier transform of the signal using the stationary phase approximation to yield~\cite{bertibuonanno}
\begin{equation}
\tilde{h}_\mrm{GR}(f)=\frac{\sqrt{3}}{2}\mathcal{A}f^{-7/6}e^{i\Psi (f)} \left[ \frac{5}{4}A_{\mathrm{pol},\alpha}(t(f)) \right] e^{-i \left( \varphi_{\mathrm{pol},\alpha}+\varphi_D \right)}, \label{waveform}
\end{equation}
where the amplitude $\mathcal{A}$ and the phase $\Psi(f)$ are given by
\begin{equation}
\mathcal{A}=\frac{1}{\sqrt{30}\pi^{2/3}}\frac{\mathcal{M}^{5/6}}{D_L},  \label{amp-noangle}
\end{equation}
%
%%\begin{widetext}
\begin{eqnarray}
% \begin{split}
\Psi(f) & = & 2\pi ft_0-\phi_0 -\frac{\pi}{4}+\frac{3}{128}(\pi \mathcal{M}f)^{-5/3} \biggl[1-\frac{2355}{1462}I_e x^{-19/6} \nn \\
                & &+ \left( \frac{3715}{756}+\frac{55}{9}\eta \right)x 
                  -4(4\pi-\beta)x^{3/2} \nn \\
         &&  + \left( \frac{15293365}{508032}+\frac{27145}{504}\eta+\frac{3085}{72}\eta^2-10\sigma \right) x^2  \biggr]. \nn \\ \label{Psi-noangle}
% \end{split}
\end{eqnarray}
%%\end{widetext}
Also, when we take the average of  the waveform (Eq.~\eqref{waveform}) over the angles, the sky-averaged (or pattern-averaged)  waveform becomes
\begin{equation}
\tilde{h}_\mrm{GR}(f)=\frac{\sqrt{3}}{2}\mathcal{A}f^{-7/6}e^{i\Psi (f)}. \label{wave-noangle}
\end{equation}

%%%%%%%%%%%%%%%%%%%%%%%%%%%%%%%%%%%%%%%%%%%%%%%%%%%%
\subsection{Precessing Binaries}
\label{precession}

In this section, we introduce an additional effect, the precession.
The spin-orbit interaction and the spin-spin interaction change the orientations of the orbital angular momentum vector $\bm{L}$ and the spin vectors $\bm{S}_i$.
These vectors precess over a time scale longer than the orbital period but shorter than the observation period.
This effect drastically changes the detected waveforms.

\subsubsection{\label{sec-simple}Simple Precession}

The precession equations are given in~\ref{app-prec}.
In this review, we assume that one of the spins of the binary constituents is negligible (i.e. $\bm{S}_1\sim 0$).
Then, there do not exist spin-spin interactions.
We also assume that the orbital angular momentum $\bm{L}$ is neither parallel nor anti-parallel to the total spin angular momentum $\bm{S}$ ($=\bm{S}_1+\bm{S}_2$).
Then, the precession equations are simplified and $\hat{\bm{L}}$ is obtained analytically up to some approximate orders.
This is the so-called \textit{simple precession approximation}~{\cite{apostolatos}}.
(This also holds when the masses of the binary constituents are equal ($m_1\sim m_2$) and spin-spin interactions are negligible, instead of $\bm{S}_1\sim 0$.)
Under this approximation, the precession equations simplify to 
%Eqs.~(\ref{40a})-(\ref{40d}).
%
\begin{eqnarray}
\dot{L}&=&-\frac{32}{5}\frac{\mu^2}{a}\left( \frac{M_t}{a} \right)^{5/2} \label{40a}, \\
\dot{S}&=&0, \\
\dot{\hat{\bm{L}}}&=&\left( 2+\frac{3}{2}\frac{m_2}{m_1}\right)\frac{\bm{J}}{a^3}\times \hat{\bm{L}}, \label{40c} \\
\dot{\hat{\bm{S}}}&=&\left( 2+\frac{3}{2}\frac{m_2}{m_1}\right)\frac{\bm{J}}{a^3}\times \hat{\bm{S}}, \label{40d} 
\end{eqnarray}
where $\bm{J}$ is the total angular momentum $\bm{J}\equiv \bm{L}+\bm{S}$.
Under this approximation, the following quantities become constant during the inspiral, $\bm{S}_1\cdot\bm{S}_2$, $\kappa\equiv\hat{\bm{L}}\cdot\hat{\bm{S}}$, and the magnitude of the total spin angular momentum $S\equiv\lvert \bm{S}_1+\bm{S}_2 \rvert$.
%In general, the precessing time scale $\Omega_p^{-1}$ is shorter than the inspiral time scale $L/\lvert \dot{L}\rvert$.
%Therefore $\bm{J}$ changes in magnitude but the direction is almost constant. 
%Then, the analytic form of $\hat{\bm{L}}$ can be obtained up to some approximate orders. %(see Appendix~\ref{app-prec}).

\begin{figure}[t]
 \begin{center}
  \centerline{\includegraphics[scale=.8,clip]{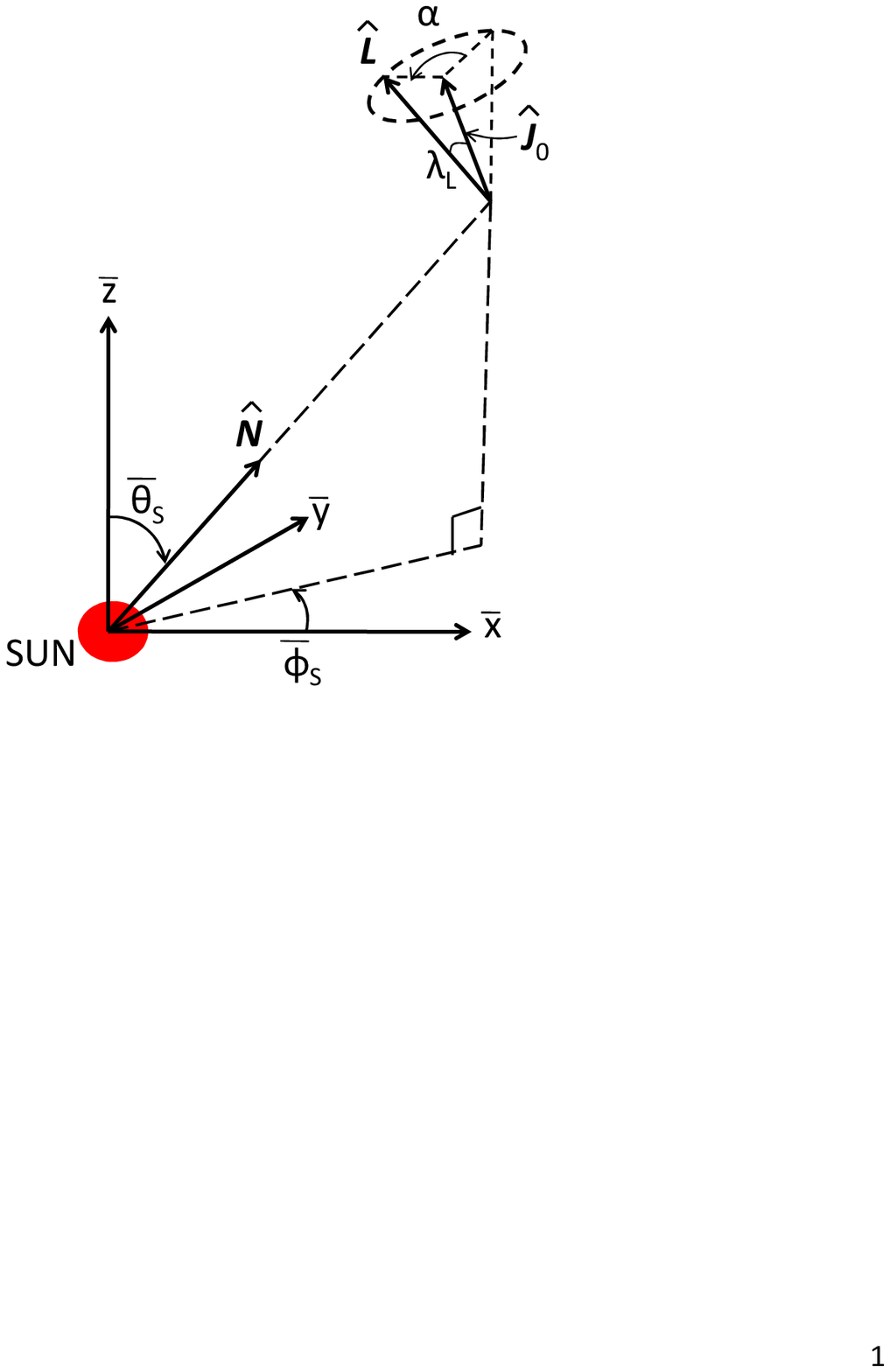}} 
   \end{center}
\vspace*{8pt}
 \caption{The unit orbital angular momentum vector $\hat{\bm{L}}$ precesses around a fixed vector $\hat{\bm{J}}_0$ \ {\cite{yagiLISA}}. The opening angle of the precession cone is given by $\lambda_{\mathrm{L}}$ and the precession angle is denoted as $\alpha$.}
\label{fig2}
\end{figure}

Next, we define a quantity $\gamma (t)$ as 
\begin{equation}
\gamma(t) \equiv \frac{S}{L(t)}.
\end{equation}
Then, the magnitude and the direction of the total angular momentum $\bm{J}$ can be expressed in terms of  $\kappa$, $\gamma(t)$, $L(t)$, $\hat{\bm{L}}$ and $\hat{\bm{S}}$ as
\begin{eqnarray}
J&=&L\sqrt{1+2\kappa \gamma+\gamma^2}, \label{48a} \\
\hat{\bm{J}}&=&\frac{\hat{\bm{L}}+\gamma \hat{\bm{S}}}{\sqrt{1+2\kappa \gamma+\gamma^2}}. \label{48b}
\end{eqnarray}
%From Eqs.~(\ref{40c}),~(\ref{40d}) and~(\ref{48b}), the precession equation of $\hat{\bm{J}}$ can be derived as
%
%\begin{equation}
%\dot{\hat{\bm{J}}}=\frac{\dot{\gamma}[\hat{\bm{S}}(1+\kappa\gamma)-\hat{\bm{L}}(\kappa+\gamma)]}
%                                     {(1+2\kappa \gamma+\gamma^2)^{3/2}}. \label{Jdothat}
%\end{equation}
%
From Eqs.~\eqref{40c} and~\eqref{40d}, it can be seen that both $\hat{\bm{L}}$ and $\hat{\bm{S}}$ precess around $\bm{J}$ with the precession angular velocity 
\begin{equation}
\Omega_p=\left( 2+\frac{3}{2}\frac{m_2}{m_1}\right)\frac{J}{a^3}. \label{omega-precess}
\end{equation}
%From Eqs.~(\ref{40c}) and~(\ref{40d}), it can be seen that the vectors $\hat{\bm{L}}$ and $\hat{\bm{S}}$ precess around $\hat{\bm{J}}$ with the angular velocity $\Omega_p$ given as Eq.~(\ref{omega-precess}).
In general, the precessing time scale $\Omega_p^{-1}$ is shorter than the radiation reaction time scale $L/\lvert \dot{L}\rvert$.
Therefore from $\dot{\bm{J}}=\dot{L}\hat{\bm{L}}$, $\bm{J}$ changes in magnitude but the direction is almost constant. Actually, if $J$ is much smaller than $L$ (as this can happen when $\bm{L}$ and $\bm{S}$ are anti-aligned with almost the same magnitudes), $\hat{\bm{J}}$ can change significantly in one precession period.
Therefore, we introduce the following small parameter, 
\begin{equation}
\varepsilon\equiv\frac{L}{J}\frac{\lvert \dot{L} \rvert/L}{\Omega_p}=\frac{\lvert \dot{\bm{J}} \rvert/J}{\Omega_p}.
\end{equation}
Then, $\bm{J}$ precesses around the fixed direction $\hat{\bm{J}}_0$ with
\begin{equation}
\hat{\bm{J}}_0=\hat{\bm{J}}-\varepsilon\hat{\bm{J}}\times\hat{\bm{L}} + \mathcal{O}(\varepsilon^2)\,.
\end{equation}
To the same order, the precession equation for $\hat{\bm{L}}$ becomes
\begin{equation}
\dot{\hat{\bm{L}}}=\Omega_p\hat{\bm{J}}_0\times\hat{\bm{L}}
                                 +\varepsilon\Omega_p(\hat{\bm{J}}_0\times\hat{\bm{L}})\times\hat{\bm{L}} + \mathcal{O}(\varepsilon^2). \label{57}
\end{equation}
The solution of this equation can be obtained geometrically~{\cite{apostolatos}}.
%We take the barycentric Cartesian frame $(\bar{x},\bar{y},\bar{z})$ which is tied to the ecliptic and centerd in the solar system barycenter.
Let us assume that $\hat{\bm{J}}_0$ points in the $(\bar{\theta}_{\mathrm{J}},\bar{\phi}_{\mathrm{J}})$ direction.
We denote $\lambda_{\mathrm{L}}$ as the opening angle of the cone on which $\hat{\bm{L}}$ precesses (i.e. the angle between $\hat{\bm{L}}$ and $\hat{\bm{J}}_0$; see Fig.~\ref{fig2}).
This can be regarded as the angle between $\hat{\bm{L}}$ and $\hat{\bm{J}}$ apart from the errors of order $\varepsilon^2$ and is given by
\begin{eqnarray}
\cos\lambda_{\mathrm{L}}&=&\hat{\bm{L}}\cdot\hat{\bm{J}}=\frac{1+\kappa \gamma}{\sqrt{1+2\kappa \gamma+\gamma^2}}, \\
\sin\lambda_{\mathrm{L}}&=&\frac{\lvert \dot{\hat{\bm{L}}} \rvert}{\Omega_p}
                                      =\frac{\gamma\sqrt{1-\kappa^2}}{\sqrt{1+2\kappa \gamma+\gamma^2}}.
\end{eqnarray}
Then, $\hat{\bm{L}}$ can be expressed as
%
%\begin{widetext}
\begin{equation}
\hat{\bm{L}}=\hat{\bm{J}}_0\cos\lambda_{\mathrm{L}}
                         +\frac{\hat{\bar{\bm{z}}}-\hat{\bm{J}}_0\cos\bar{\theta}_{\mathrm{J}}}{\sin\bar{\theta}_{\mathrm{J}}}
                           \sin\lambda_{\mathrm{L}}\cos\alpha
                         +(\hat{\bm{J}}_0\times\hat{\bar{\bm{z}}})\frac{\sin\lambda_{\mathrm{L}}\sin\alpha}{\sin\bar{\theta}_{\mathrm{J}}}, \label{L-prec}
\end{equation}
%\end{widetext}
where $\alpha$ is the \textit{precession angle} defined as the solution of
\begin{equation}
\frac{d\alpha}{dt}\equiv\Omega_p. \label{alpha-omega}
\end{equation} 
We assume that $\alpha=0$ is realized when $\hat{\bm{L}}\cdot\hat{\bar{\bm{z}}}$ is maximum (see Fig.~\ref{fig2}).
By solving the above equation, we obtain
%
%\begin{widetext}
\begin{equation}
\begin{split}
\alpha=&\alpha_c-\frac{5}{96}\frac{1}{\mu^3M_t^3}\left( 1+\frac{3}{4}\frac{m_1}{m_2} \right) \\
           &\times\left[ 2(\mathcal{G}L)^3-3\kappa S(L+\kappa S)\mathcal{G}L
             -3\kappa S^3(1-\kappa^2)\mathrm{arcsinh}\left( \frac{L+\kappa S}{S\sqrt{1-\kappa^2}}\right) \right],
\end{split}
\end{equation}
%\end{widetext}
where $\alpha_c$ is a quantity which characterizes $\alpha$ at $t=t_c$ and $\mathcal{G}$ is defined as
\begin{equation}
\mathcal{G}\equiv \sqrt{1+2\kappa\gamma+\gamma^2}.
\end{equation}

%

%In this review, we assume that one of the spins of the binary constituents is negligible (i.e. $\bm{S}_1\sim 0$).
%We also assume that the orbital angular momentum $\bm{L}$ is neither parallel nor antiparallel to the total spin angular momentum $\bm{S}$($=\bm{S}_1+\bm{S}_2$).
%Then, the precession equations become
%

\subsubsection{\label{sec-det-prec}Detector Response}

When precession is taken into account, the principal axis $\hat{\bm{p}}$ varies with time so that the GW phase $\phi(t)$ no longer equals to twice the orbital phase $\phi_{\mathrm{orb}}(t)\equiv\int\Omega(t) dt$.
We define this difference 2$\delta\phi(t)$ by
\begin{equation}
\phi(t) =2\phi_{\mathrm{orb}}(t)+2\delta\phi(t). \label{phi-prec}
\end{equation}
This $\delta\phi(t)$ is the so-called \textit{Thomas precession phase}.
We absorb the constant of integration in the definition of $\phi_\mrm{orb}(t)$ so that $\phi_{\mathrm{orb}}(t_c)=\phi_c$.
In general, $\delta\phi(t_c)\neq 0$ so that $\phi_c$ does not equal to $\phi(t_c)$ anymore in this case. 
From Eqs.~(\ref{output2}) and~(\ref{phi-prec}), the detector output $h_{\alpha}(t)$ becomes
%
%\begin{widetext}
\begin{eqnarray}
h_{\alpha}(t)&=&\frac{\sqrt{3}}{2} A_{+} F_{\alpha}^+(\theta_{\mathrm{S}},\phi_{\mathrm{S}},\psi_{\mathrm{S}}) 
                      \cos2 (\phi_{\mathrm{orb}}(t)+\delta\phi(t)) \nn \\
                    & & +\frac{\sqrt{3}}{2} A_{\times} F_{\alpha}^{\times}(\theta_{\mathrm{S}},\phi_{\mathrm{S}},\psi_{\mathrm{S}}) 
                      \sin2 (\phi_{\mathrm{orb}}(t)+\delta\phi(t)) \notag \\ 
                 &=&\frac{\sqrt{3}}{2} \frac{2\eta M_t^2}{a D}(F^{\cos}_{\alpha}\cos 2\phi_{\mathrm{orb}}+F^{\sin}_{\alpha}\sin 2\phi_{\mathrm{orb}}), \label{out-prec}
\end{eqnarray}
%%\end{widetext}
where 
%$\alpha=\mathrm{I}, \mathrm{II}$ labels the detector number and 
$F^{\cos}_{\alpha}$ and $F^{\sin}_{\alpha}$ are defined as
%
%%\begin{widetext}
\begin{eqnarray}
F^{\cos}_{\alpha}&\equiv &(1+(\hat{\bm{L}}\cdot\hat{\bm{N}})^2)F_{\alpha}^+\cos 2\delta\phi
                             -2(\hat{\bm{L}}\cdot\hat{\bm{N}})F_{\alpha}^{\times}\sin 2\delta\phi, \\
F^{\sin}_{\alpha}&\equiv &-(1+(\hat{\bm{L}}\cdot\hat{\bm{N}})^2)F_{\alpha}^+\sin 2\delta\phi
                             -2(\hat{\bm{L}}\cdot\hat{\bm{N}})F_{\alpha}^{\times}\cos 2\delta\phi.
\end{eqnarray} 
%\end{widetext}
Following Eq.~(\ref{waveform1}), we express this output~(\ref{out-prec}) in terms of an amplitude and phase form, and we also take the motions of the detectors into account to obtain
\begin{equation}
h_{\alpha}(t)=\frac{\sqrt{3}}{2} \frac{2\eta M_t^2}{a D}A_{\mathrm{pol},\alpha}^{\mathrm{prec}}(t)
                     \cos \left[ 2\phi_{\mathrm{orb}}(t)+\varphi_{\mathrm{pol},\alpha}^{\mathrm{prec}}(t)+\varphi_{D}(t) \right],
\end{equation}
where $A_{\mathrm{pol},\alpha}^{\mathrm{prec}}(t)$ and $\varphi_{\mathrm{pol},\alpha}^{\mathrm{prec}}(t)$ are given by
\begin{eqnarray}
A_{\mathrm{pol},\alpha}^{\mathrm{prec}}(t)&=&\sqrt{(F^{\cos}_{\alpha}(t))^2+(F^{\sin}_{\alpha}(t))^2}, \\
\cos(\varphi_{\mathrm{pol},\alpha}^{\mathrm{prec}}(t))&=&\frac{F^{\cos}_{\alpha}(t)}{A_{\mathrm{pol},\alpha}^{\mathrm{prec}}(t)}, \\
\sin(\varphi_{\mathrm{pol},\alpha}^{\mathrm{prec}}(t))&=&-\frac{F^{\sin}_{\alpha}(t)}{A_{\mathrm{pol},\alpha}^{\mathrm{prec}}(t)}. 
\end{eqnarray}
The Fourier transform of this waveform is given by
%
%\be
%
\begin{equation}
\tilde{h}_\mrm{GR}(f)=\frac{\sqrt{3}}{2}\mathcal{A}f^{-7/6}e^{i\Psi (f)} \left[ \frac{5}{4}A_{\mathrm{pol},\alpha}^{\mathrm{prec}}(t(f)) \right] e^{-i \left( \varphi_{\mathrm{pol},\alpha}^{\mathrm{prec}}+\varphi_D \right)}, \label{waveform-prec}
\end{equation}
%
%\ee
%
which looks similar to Eq.~\eqref{waveform}, except $A_{\mathrm{pol},\alpha}(t(f))$ and $\varphi_{\mathrm{pol},\alpha}(t(f))$ are replaced with $A_{\mathrm{pol},\alpha}^{\mathrm{prec}}(t(f))$ and $\varphi_{\mathrm{pol},\alpha}^{\mathrm{prec}}(t(f))$, respectively.
%The quantities $(\hat{\bm{L}}\cdot\hat{\bm{N}})$,$(\hat{\bm{L}}\cdot\hat{\bm{z}})$ and $[\hat{\bm{N}}\cdot(\hat{\bm{L}}\times\hat{\bm{z}})]$, which are needed to compute the polarization angle $\psi_{\mathrm{S}}(t)$ in the beam-pattern coefficients $F_{\alpha}^+$ and $F_{\alpha}^{\times}$, are expressed in Appendix~\ref{app-prec}.

Finally, we need to calculate the Thomas precession phase $\delta\phi(t)$.
Apostolatos \textit{et al.}~\cite{apostolatos} derived an explicit form as
\begin{equation}
\delta\phi(t)=-\int^{tc}_{t}dt \, \left( \frac{\hat{\bm{L}}\cdot\hat{\bm{N}}}{1-(\hat{\bm{L}}\cdot\hat{\bm{N}})^2} \right)
                            (\hat{\bm{L}}\times\hat{\bm{N}})\cdot\dot{\hat{\bm{L}}}. \label{delta-phi}
\end{equation}
%
%(See App.~\ref{app:thomas} for its derivation.)
This expression includes an integration which makes the computational time very long. 
Vecchio~\cite{vecchio} estimated the binary parameter accuracies by choosing a few random sources with and without including $\delta\phi(t)$ and found that this term did not affect the results, and concluded that it could be neglected.
This is true for the binaries for which $\hat{\bm{L}}\cdot\hat{\bm{N}}$ never becomes close to $\pm1$.
However, when $\hat{\bm{L}}\cdot\hat{\bm{N}}\approx \pm1$, the direction of the principal axis $\hat{\bm{p}}$ changes rapidly with time, and hence the polarization angle $\psi_{\mathrm{S}}(t)$ also changes rapidly.
Since it is the Thomas precession phase $\delta\phi(t)$ that cancels this rapid change, $\delta\phi(t)$ cannot be neglected in this case.

\begin{figure}[t]
 \begin{center}
  \centerline{\includegraphics[scale=.4,clip]{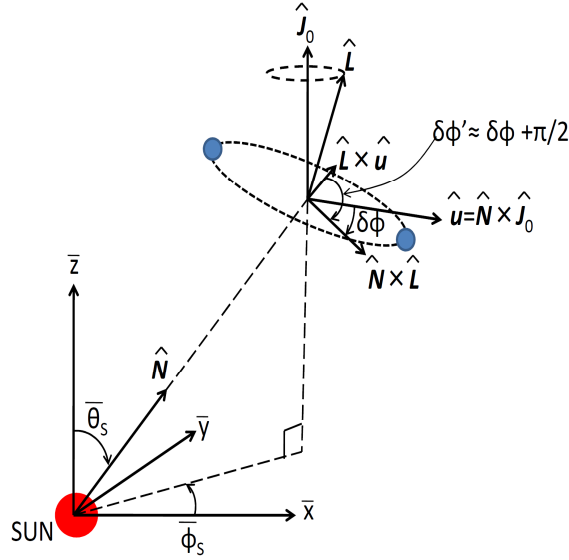}} 
   \end{center}
\vspace*{8pt}
 \caption{The Thomas precession phase $\delta\phi(t)$ is the angle from the vector $\hat{\bm{u}}$ to the principal axis $\hat{\bm{p}} \equiv \hat{\bm{N}}\times\hat{\bm{L}}$ \ {\cite{yagiLISA}}. We define $\delta\phi '$ as the angle from the vector $\hat{\bm{L}}\times\hat{\bm{u}}$ to the one $\hat{\bm{p}}$. Notice that these two vectors always lie in the orbital plane. $\delta\phi '$ equals to $\delta\phi + \pi /2$ up to $O(\lvert \delta\bm{L} \bm{} \rvert/L)$.}
\label{fig3}
\end{figure}

The direction which $\delta\phi$ is measured from is arbitrary.
It can be seen from Eq.~(\ref{delta-phi}) that Apostolatos \textit{et al.}~\cite{apostolatos} defined $\delta\phi$ as the angle measured from the principal axis at the time of coalescence.
In general, since $\hat{\bm{p}}|_{t=t_0}$ does not lie in the orbital plane at a time $t$, we need to follow the evolution of the principal axis to calculate the Thomas precession $\delta\phi(t)$.
This is the reason why the integration appears in Eq.~(\ref{delta-phi}). 

Here, we try to derive an approximate expression of $\delta\phi(t)$ that is not in the integral form.
To do so, we use a specific vector $\hat{\bm{L}}\times\hat{\bm{u}}$ that always lies in the orbital plane, where $\hat{\bm{u}}$ is a constant unit vector.
In the limit of a large separation (i.e. $t\rightarrow -\infty$ or $L\rightarrow\infty$), $\hat{\bm{L}}$ approaches to $\hat{\bm{J}}_0$.
Using this $\hat{\bm{J}}_0$, we take $\hat{\bm{u}}=\hat{\bm{N}}\times\hat{\bm{J}}_0$ (see Fig.~\ref{fig3}).
We define an approximate Thomas precession phase $\delta\phi(t)$ by the angle from this vector $\hat{\bm{u}}$ to the principal axis $\hat{\bm{p}}$. 
(We may take the vector $\hat{\bm{u}}$ to be the principal axis at the time of coalescence $\hat{\bm{p}}|_{t=t_0}$. Then, $\delta\phi(t_0)$ becomes 0 and $\phi(t_0)=\phi_0$. However, the computation is easier for choosing $\hat{\bm{u}}$ to be $\hat{\bm{N}}\times\hat{\bm{J}}_0$.)
We denote the angle from $\hat{\bm{L}}\times\hat{\bm{u}}$ to $\hat{\bm{p}}$ as $\delta\phi '$, and the difference between $\bm{J}_0$ and $\bm{L}$ as $\delta\bm{L}$; $\bm{L}=\bm{J}_0+\delta\bm{L}$.  
Then, $\delta\phi '$ can be related to $\delta\phi$ as $\delta\phi '=\delta\phi+\pi/2$ up to $O(\lvert \delta\bm{L} \bm{} \rvert/L)$.
As $\delta\phi '$ always lie in the orbital plane, by using this angle we can express $\delta\phi(t)$ without integration.
$\sin\delta\phi$ and $\cos\delta\phi$ can be written as 
\begin{eqnarray}
\sin\delta\phi &\approx &-\cos\delta\phi ' \notag \\
                      &=&-\frac{(\hat{\bm{L}}\times\hat{\bm{N}})\cdot(\hat{\bm{L}}\times\hat{\bm{u}})}
                       {\lvert \hat{\bm{L}}\times\hat{\bm{N}} \rvert \lvert \hat{\bm{L}}\times\hat{\bm{u}} \rvert}  \notag \\
                      &=&\frac{(\hat{\bm{L}}\cdot\hat{\bm{N}})(\hat{\bm{L}}\cdot\hat{\bm{u}})}
                       {\sqrt{(1-(\hat{\bm{L}}\cdot\hat{\bm{N}})^2)(1-(\hat{\bm{L}}\cdot\hat{\bm{u}})^2)}}, \\
\cos\delta\phi &\approx &\sin\delta\phi ' \notag \\
                       &=&-\frac{\hat{\bm{N}}\cdot(\hat{\bm{L}}\times\hat{\bm{u}})}
                        {\sqrt{(1-(\hat{\bm{L}}\cdot\hat{\bm{N}})^2)(1-(\hat{\bm{L}}\cdot\hat{\bm{u}})^2)}}.
\end{eqnarray}
Therefore the Thomas precession phase $\delta\phi(t)$ can be expressed as
\begin{equation}
\delta\phi\approx \arctan \left( -\frac{(\hat{\bm{L}}\cdot\hat{\bm{N}})(\hat{\bm{L}}\cdot\hat{\bm{u}})}
                  {\hat{\bm{N}}\cdot(\hat{\bm{L}}\times\hat{\bm{u}})} \right). \label{thomas-new}
\end{equation}
Here, the explicit forms of $\hat{\bm{L}}\cdot\hat{\bm{N}}$, $\hat{\bm{L}}\cdot\hat{\bm{u}}$ and $\hat{\bm{N}}\cdot(\hat{\bm{L}}\times\hat{\bm{u}})$ are given in~\ref{app-prec}.

%%%%%%%%%%%%%%%%%%%%%%%%%%%%%%%%%%%%%%%%%%%%%%%%%%%%%%%%%%%%%%%
\subsection{Inspiral-Merger-Ringdown Hybrid Waveform}
\label{hybrid}

In the previous 2 sections, we explained how inspiral waveforms can be modeled using PN theory.
One possible extension would be to include merger and ringdown.
The phenomenological inspiral-merger-ringdown hybrid waveforms have developed by Ajith \textit{et al.}~{\cite{ajith,ajithspin}}.
They first performed numerical simulations to generate merging binary BH waveforms from 8 cycles before merger including non-precessing BH spins.
Then they matched these waveforms with the Taylor T1 post-Newtonian (PN) inspiral waveforms~\cite{damour} at 3.5PN phase accuracy~{\cite{blanchet3.5}}.
They included the 3PN amplitude corrections~\cite{blanchet1996,arun2004,blanchet3} and the 2.5PN corrections coming from the spins~\cite{arunhigher} to the dominant quadrupole mode (see Ref.~\cite{ajith} for detailed procedure of the matching).
The Fourier component of the (observed) matched waveform 
\be
\tilde{h}(f)=A(f)e^{i\{\Psi(f)-\varphi_\mrm{pol}(f)-\varphi_\mrm{D}(f)\} }
\ee
is fitted by using phenomenological parameters as~\footnote{The polarization phase $\varphi_\mrm{pol}(f)$ and the Doppler phase $\varphi_\mrm{D}(f)$ are given in Eqs.~\eqref{phi-pol-cos}--\eqref{doppler-phase} without the index $\alpha$.}
\beqa
A(f)\equiv C(t(f)) f_1^{-7/6} \label{hybrid_amp}
\left\{ \begin{array}{ll}
f'^{-7/6} \lmk 1+\sum_{i=2}^3 \alpha_i x^{i/2}\rmk & (f \le f_1 ) \\
w_m f'^{-2/3} \lmk 1+\sum_{i=1}^2 \epsilon_i x^{i/2}\rmk & (f_1 \le f \le f_2 ) \\
w_r \mathcal{L} (f,f_2,\sigma) & (f_2 \le f \le f_3 ) \\
\end{array} \right.,
\eeqa
and
\beq
\Psi(f) \equiv 2\pi f t_0 - \phi_0 -\frac{\pi}{4} + \Psi_\mrm{0PN} \lmk 1+\sum_{k=2}^7 x^{k/2} \psi_k \rmk \label{hybrid_phase}
\eeq
with $f'\equiv f/f_1$.
$f_1$ and $f_2$ are the transition frequencies from inspiral to merger and merger to ringdown phases, respectively, with $w_m$ and $w_r$ corresponding to the normalization factors to make $A(f)$ continuous at these frequencies.
$f_3$ corresponds to the cutoff frequency.
$C$ is a numerical factor discussed in the next paragraph.
$\epsilon_i$ are given as
\beqa
\epsilon_1 &=& 1.4547\chi -1.8897, \\
\epsilon_2 &=& -1.8153\chi +1.6557.
\eeqa
Here, the spin parameter $\chi$ is defined as 
\beq
\chi\equiv \frac{1+\delta}{2}\chi_1+ \frac{1-\delta}{2}\chi_2,
\eeq
where $\delta\equiv (m_1-m_2)/M_t$ and $\chi_i\equiv S_i/m_i^2$ with $m_i$ and $S_i$ being the mass and the spin angular momentum of $i$-th BH, respectively. 
For the equal spin case, $\chi$ becomes $\chi=\chi_1=\chi_2$.
On the other hand, $\alpha_i$ are obtained from $(l,m)=(2,\pm2)$ PN waveform~\cite{arunhigher} as
\beqa
\alpha_2 &=& -\frac{323}{224} +\frac{451}{168}\eta, \\
\alpha_3 &=& \lmk \frac{27}{8}-\frac{11}{6} \eta\rmk \chi.
\eeqa
%where $\eta\equiv \mu/M$ represents the symmetric mass ratio with the reduced mass $\mu$.
$\mathcal{L} (f,f_2,\sigma)$ is the Lorentzian function centered around the frequency $f_2$ with width $\sigma$.
%$t_0$ represents the coalescence time and $\phi_0$ denotes the corresponding phase.
The phenomenological parameters $\psi_k$ and $\mu_k\equiv (f_1,f_2,\sigma,f_3)$ are fitted against physical parameters $(M_t,\eta,\chi)$ as
\beqa
\psi_k &=& \sum_{i=1}^3 \sum_{j=0}^N x_k^{(i,j)} \eta^i\chi^j+\psi_k^0, \\
\mu_k &=& \sum_{i=1}^3 \sum_{j=0}^N \frac{y_k^{(i,j)}\eta^i\chi^j+\mu_k^0}{\pi M_t},
\eeqa
where $N\equiv \min(3-i,2)$ while $x_k^{(i,j)}$, $y_k^{(i,j)}$, $\psi_k^0$ and $\mu_k^0$ are listed in Table I of Ref.~{\cite{ajithspin}}. 

The numerical factor $C(t(f))$ is given by~\cite{bertibuonanno} 
\beq
C(t(f))=\sqrt{\frac{5}{96\pi^{4/3}}}\frac{\mch^{5/6}}{D_L}A_\mrm{pol}(t(f)), \label{c}
\eeq
where $A_\mrm{pol}(t(f))$ is given in Eq.~\eqref{Apol} without the index $\alpha$.
%
%\beq
% A_\mrm{pol}(t(f)) \equiv \sqrt{(1+\cos^2 i)^2 F^{+}(\theta,\psi)^2+4(\cos i)^2 F^{\times}(\theta,\psi)^2}
%\eeq
%
%with $i$ representing the inclination angle of the binary.
%$F^{+}(\theta,\psi)$ and $F^{\times}(\theta,\psi)$ are the plus and cross mode polarization functions given in Eq.~(\ref{beam}) together with the direction of the source $(\theta(t(f)),\psi(t(f)))$ measured from the detector expressed in Eqs.~(\ref{theta}) and~(\ref{psi}).
For the time $t(f)$, we take up to 3.5PN order. 
The expression for $df/dt$ up to this order is given in Arun \et~{\cite{arun35}}.
We integrate this to yield
\beqa
t(f)&=&t_0  -  \frac{5}{256}\mch (\pi \mch f)^{-8/3} \biggl[1+\frac{4}{3}\left( \frac{743}{336}+\frac{11}{4}\eta \right) x  \nonumber  \\
      & &-\frac{8}{5}(4\pi-\beta) x^{3/2} 
       +2\left( \frac{3058673}{1016064}+\frac{5429}{1008}\eta+\frac{617}{144}\eta^2 -\sigma \right) x^2  \nonumber \\
     &  &+ \frac{8\pi}{3} \lmk -\frac{7729}{672}+\frac{13}{8}\eta \rmk x^{5/2}+ 15 \biggl( -\frac{10817850546611}{93884313600} \nonumber \\
      & &+\biggl( \frac{15335597827}{60963840}-\frac{616\lambda}{9}-\frac{451\pi^2}{48}+\frac{88\theta}{3} \biggr)\eta -\frac{15211}{6912}\eta^2 \nonumber \\
       & & +\frac{25565}{5184}\eta^3 +\frac{1712\gamma}{105}+\frac{32\pi^2}{3} +\frac{3424}{1575}\log{(32768x)} \biggr)x^3 \nonumber \\ 
       & & 8 \biggl( -\frac{15419335\pi}{1016064}-\frac{75703\pi}{6048}\eta +\frac{14809\pi}{3024}\eta^2 \biggr)x^{7/2} \biggr].
\eeqa
Here, $\gamma=0.577$ is the Euler's constant, $\lambda=-\frac{1987}{3080}$ and $\theta=-\frac{11831}{9240}$.
We take the contribution of spins into account up to 2PN order with the spin-orbit coupling $\beta$ and the spin-spin coupling $\sigma$.
%Here, $\hat{\bm{S}}_i$ is the unit spin angular momentum vector of the $i$-th BH.
%In this review, we assume that orbital angular momentum and spins are all aligned.
%$\hat{\bm{L}}\cdot \hat{\bm{S}}_i=\hat{\bm{S}}_1 \cdot \hat{\bm{S}}_2=1$.
The sky-averaged values of the beam-pattern functions are $\lla F^{+2} \rra =\lla F^{\times 2} \rra=4/15$ which yield the sky-averaged value of $C$ as $\lla C \rra =\sqrt{\frac{2}{45\pi^{4/3}}}\frac{\mch^{5/6}}{D_L}$. We use this hybrid waveform when we perform binary parameter estimation with DPF in Sec.~\ref{sec:results-DPF}.

%%%%%%%%%%%%%%%%%%%%%%%%%%%%%%%%%%%%%%%%%%%%%%%%%%%%%%%
\section{Brans-Dicke Theory}
\label{sec:BD}

Currently, solar system experiments put stringent constraints on gravitational theories in the weak field regime~{\cite{will-living}}, but general relativity (GR) is not the only one that passes this test.
Among the modified gravity theories that pass these tests, in this section, we consider one of the simplest extension of GR called the \textit{scalar-tensor} theory~{\cite{fujii}}.
As its name tells, in this theory, we add scalar degrees of freedom to gravity.
Among various types of scalar-tensor theories, we here focus on  \textit{Brans-Dicke} theory~{\cite{brans}}.
We note here that the scalar-tensor theory had already been proposed 20 years before the paper by Brans and Dicke~{\cite{brans-history,goenner}}. 

The reason we consider scalar-tensor theory is that it may give clues to solving unknown problems in GR and cosmology: 
(I) Current observations show the acceleration expansion of the Universe~{\cite{riess}}. 
However, the cause of this cosmic accelerating is unknown.
If we are to explain this phenomenon with a matter field, we need to introduce the \textit{dark energy} which has a negative pressure~{\cite{copeland,tsujikawa-book}}.
Another approach is to modify gravitational theory from GR.
In scalar-tensor theory, the additional scalar field called \textit{quintessence} can explain the current acceleration of the Universe~{\cite{perrotta}}.
(II) Inflation can be explained in the context of scalar-tensor theory.
This inflation model is called the \textit{hyper-extended} inflation~\cite{steinhardt} where the additional scalar field acts as the inflaton.
(III) Scalar-tensor theory also appears in the particle-physics.
Superstring theory~\cite{polchinski1,polchinski2} predicts that the dimension of the spacetime is 10 or 11, and in the low energy effective theory (supergravity theory), scalar fields called \textit{dilatons} appear.
In this theory, the gravitational theory is of the scalar-tensor type.
Also, in Kaluza-Klein theory where GR is extended to 5 dimension with the extra dimension being compactified, it reduces to the scalar-tensor type with $g_{55}$ component acting as a scalar field~{\cite{fujii}}.

As for GWs from compact binaries, there exists scalar dipole radiation which modifies the binary evolution from GR.
After introducing the action in Sec.~\ref{bd-action}, we derive the correction to GW phase in Sec.~\ref{bd-corr}.
Then, in the final section, we describe the current constraints from solar system experiment and binary pulsar observations.

%%%%%%%%%%%%%%%%%%%%%%%%%%%%%%%%%%%%%%%%%%%%%%%%%%%
\subsection{The Action}
\label{bd-action}

In general, the action for the scalar-tensor theory in the \emph{Jordan} frame~\footnote{Some literature uses a different conformal frame (the \emph{Einstein} frame)~{\cite{damour1992,damour-review}}.} is given as
\be
S=\frac{1}{16\pi}\int d^4x \sqrt{-g} \left( \phi R
                             -\omega(\phi) \frac{1}{\phi}g^{\mu\nu}\partial_{\mu}\phi \partial_{\nu}\phi +16\pi L_{\mathrm{matter}}(\Psi) \right).
\label{action-ST}
\ee
Here, $\phi$ is the scalar field that carries the scalar gravitational degrees of freedom and $\Psi$ is the matter field.
The first term is the Einstein-Hilbert term non-minimally coupled to the scalar field.
The second term is the kinetic term of the scalar field.
The coefficient $\omega(\phi)$ is the parameter that characterizes this theory.
 %In the self-gravitating system, $L_\mrm{matter}$ also depends on the scalar field $\phi$.
  Especially, when $\omega(\phi) = \omega_\mrm{BD}=\mrm{const.}$, this theory is called Brans-Dicke theory and $\omega_\mrm{BD}$ is called Brans-Dicke parameter.
(The inverse of) this parameter expresses the strength of the coupling between the scalar and the matter fields (c.f. Eq.~\eqref{phi-eq-bin}).
 %The scalar field can be divided into the background homogeneous and stationary field $\phi_0$ and the perturbation $\delta \phi$.
 If we take the limit $\omega_\mrm{BD} \rightarrow \infty$, the theory reduces to GR.
 
The third term in Eq.~\eqref{action-ST} is the Lagrangian for the matter field.
At the microscopic level, this term does not depend on $\phi$. However, if one describes the orbital dynamics of a system of strongly self-gravitating objects by means of an effective point-particle theory, in which the internal structure of the objects is integrated out, then $L_\mrm{matter}$ consists of a sum of point-particle Lagrangians which {\it do} depend on $\phi$, on account of strong equivalence principle violation. For example, see Eq.~\eqref{action-BD} below.

In this theory, the gravitational constant is not a constant but a function of $\phi$.
It is given as~\cite{fujii}~\footnote{This can be derived by perturbing the scalar and the metric fields around the flat background to obtain the gravitational potential. 
}
\be
G_\mrm{BD}(\phi) = \frac{4+2\omega_\mrm{BD}}{3+2\omega_\mrm{BD}}\frac{1}{\phi}\,. 
\ee
If we assume that the theory reduces to GR at spatial infinity, we have
\ba
G_\mrm{BD} (\bm{x} \rightarrow \infty) &=& \frac{4+2\omega_\mrm{BD}}{3+2\omega_\mrm{BD}}\frac{1}{\phi_0} \nn \\
&=& G=1\,,
\ea
where $\phi_0$ is the value of the scalar field at spatial infinity.
Therefore, we obtain
\be
\phi_0 = \frac{3+2\omega_\mrm{BD}}{4+2\omega_\mrm{BD}}\equiv \frac{1}{2} (1+\gamma)\,,
\label{phi0gamma}
\ee
where $\gamma$ is a \textit{parameterized post-Newtonian} (PPN) parameter~\cite{will-book} defined as
\be
\gamma \equiv \frac{1+\omega_\mrm{BD}}{2+\omega_\mrm{BD}}\,.
\label{gamma}
\ee 
In GR, $\gamma$ takes the value $\gamma=1$.
 
%The current strongest constraing on $\omega_\mrm{BD}$ is obtained by the solar system experiment as $\omega_\mrm{BD} > 4\times 10^4$~\cite{cassini}.

%%%%%%%%%%%%%%%%%%%%%%%%%%%%%%%%%%%%%%%%%%%%%%%%%%
\subsection{Correction to GW Phase of the Binary}
\label{bd-corr}

In this section, we derive the correction to the gravitational waves from compact binaries in Brans-Dicke theory, assuming that $\omega_\mrm{BD} \gg 1$.
First, the action for the binary system in this theory is given as~\cite{eardley,zaglauer}~\footnote{See also Refs.~\cite{goldberger1,goldberger2} where the authors put the point-particle action into the modern context of Effective Field Theory, and interpreted $m_{a}(\phi)$ as the leading-order term in the derivative expansion of the point-particle Lagrangian of body $a$.}
\be
S=\frac{1}{16\pi}\int d^4x\sqrt{-g}\left(\phi R-\frac{\omega_{\mathrm{BD}}}{\phi}\phi_{,\alpha}\phi^{,\alpha} \right)
        -\sum_{a=1,2}\int d\tau_a m_a(\phi),
\label{action-BD}
\ee
where $\tau_a$ is the proper time measured along the trajectory of $a$-th body.
%In this theory, the gravitational constant is given as
%
%\be
%G_{\mathrm{BD}}(\phi)=\frac{4+2\omega_{\mathrm{BD}}}{3+2\omega_{\mathrm{BD}}}\frac{1}{\phi}.
%\ee
%
%At spatial infinity, 
The field equations are given as
\ba
G_{\mu\nu}=8\pi\frac{1}{\phi}T_{\mu\nu}+\omega_{\mathrm{BD}}\frac{1}{\phi^2}
                  \left(\phi_{,\mu}\phi_{,\nu}-\frac{1}{2}g_{\mu\nu}\phi_{,\mu}\phi^{,\mu}\right)
                  +\frac{1}{\phi}(\phi_{;\mu\nu}-g_{\mu\nu}\Box \phi), \label{bd-ein2}
\ea
where $G_{\mu\nu} = R_{\mu\nu}-\frac{1}{2} R g_{\mu\nu} $ is the Einstein tensor and $T_{\mu\nu}$ is the matter energy-momentum tensor.
The scalar wave equation can be obtained as
\be
\Box \phi=\frac{8\pi}{3+2\omega_{\mathrm{BD}}}\left[ T-2\phi\frac{\partial T}{\partial \phi} \right]. \label{phi-eq-bin}
\ee
In the point-particle limit, $T_{\mu\nu}$ and $\partial T/\partial \phi$ can be expressed as
\ba
T^{\mu\nu}&=&\frac{1}{\sqrt{-g}}\sum_{a=1,2}m_a(\phi)\frac{u^{\mu}u^{\nu}}{u^0}\delta^3(\bm{x}-\bm{x_a}), \\
\frac{\partial T}{\partial \phi} &=& -\frac{1}{\sqrt{-g}}\sum_{a=1,2}\frac{\partial m_a(\phi)}{\partial \phi}\frac{1}{u^0}\delta^3(\bm{x}-\bm{x_a}).
\ea

Next, let us consider the perturbation of $g_{\mu\nu}$ and $\phi$ around spatial infinity~\cite{will1977}:
\be
g_{\mu\nu}=\eta_{\mu\nu}+h_{\mu\nu}, \qquad \phi=\phi_0+\tilde{\phi}.
\ee
We define a new tensor field
\be
\chi_{\mu\nu}=h_{\mu\nu}-\frac{1}{2}h\eta_{\mu\nu}-\frac{\tilde{\phi}}{\phi_0}\eta_{\mu\nu},
\ee
and impose the Lorentz gauge condition 
\be
\chi^{\mu\nu}{}_{,\nu}=0.
\ee
By perturbing the field equations (Eqs.~\eqref{bd-ein2} and \eqref{phi-eq-bin}), we get
\ba
\Box \chi^{\mu\nu} &=& -16\pi \frac{1}{\phi_0}T^{\mu\nu},  \label{wave-chi} \\
\Box \tilde{\phi} &=& -16\pi S,
\ea
where the box operator represents the d'Alembertian operator in Minkowski background and $S$ is defined as
\be
S \equiv -\frac{1}{6+4\omega_{\mathrm{BD}}}\left[ T-2\phi_0\left(\frac{\partial T}{\partial \phi}\right)_0 \right],
\ee
where the index 0 denotes the value at spatial infinity.

At a distance $r$ that is sufficiently larger than the size of the binary system, we can perform the multipolar expansion to yield~\cite{zaglauer}
\ba
\chi^{\mu\nu} &=& \frac{4}{r}\frac{1}{\phi_0}\sum_{m=0}^{\infty}\frac{1}{m!}\left(\frac{\partial}{\partial t}\right)^m
                       \int \, T^{\mu\nu}(t-r,x')(\bm{n}\cdot\bm{x'})^md^3x', \\
\tilde{\phi} &=& \frac{4}{r}\sum_{m=0}^{\infty}\frac{1}{m!}\left(\frac{\partial}{\partial t}\right)^m
                       \int \, S(t-r, x')(\bm{n}\cdot\bm{x'})^md^3x', \label{tilde-phi}
\ea
First, we consider the correction to the tensor part.
From the gauge condition and the field equations, we have $T^{\mu\nu}{}_{,\nu}=0$, and hence at the leading order,
\ba
\chi^{ij}&=&\frac{4}{r}\frac{1}{\phi_0}\int \, T^{ij}d^3x', \nn \\
         &=&\frac{1+\gamma}{r}\frac{d^2}{dt^2}\int \, T^{00}x'^ix'^jd^3x',
\ea
where we have used Eq.~\eqref{phi0gamma}.
%
%\be
%\gamma \equiv \frac{1+\omega_\mrm{BD}}{2+\omega_\mrm{BD}}.
%\ee
%
At the lowest order, we have $T^{00} \approx \rho$ so that
\be
\chi^{ij}=\frac{1+\gamma}{r}\frac{d^2}{dt^2}\sum_{a=1,2}m_ax^i_ax^j_a. \label{chi-ij}
\ee
The term proportional to $\gamma$ represents the correction compared to GR.
However, we now show that this 0PN correction is sub-leading compared to the one from the scalar field.

Next, let us consider the contribution from the scalar field.
First, we define the \textit{sensitivity} (which is related to the scalar charge) as
\be
s_a\equiv -\left[ \frac{\partial(\ln m_a)}{\partial(\ln G)} \right]_0.
\ee
Then, $S$ can be rewritten as
\be
S=\frac{1}{6+4\omega_{\mathrm{BD}}}\sum_{a=1,2}m_a\delta^3(\bm{x}-\bm{x_A})(1-2s_a). \label{bd-7-64}
\ee
We substitute this into the scalar wave equation and solve for $\tilde{\phi}$.
At spatial infinity, we can assume that $\phi_0$ is conserved.
This means that the mass is conserved, and hence the leading order scalar monopole radiation vanishes.
Therefore the leading radiation appears at dipole order.
When we move to the center of mass frame with
\be
x^i=x_2^i-x_1^i,\qquad v^i=v_2^i-v_1^i,   \\ 
\ee
the scalar perturbation $\tilde{\phi}$ can be expressed as
\be
\tilde{\phi}=-\frac{4}{3+2\omega_{\mathrm{BD}}}\frac{\mu}{r}\mathcal{S}\,\bm{n}\cdot\bm{v}
\label{phi-tilde}
\ee
with $\mathcal{S} \equiv s_2-s_1$.
This $\mathcal{S}$ in Eq.~\eqref{phi-tilde} comes from ``-2$s_a$'' part in Eq.~\eqref{bd-7-64}. 
The first term in Eq.~\eqref{bd-7-64} is sourced by the mass of each component, but because of the linear momentum conservation, there is no dipole radiation from this term.
%In contrast, the second term is source

Next, we calculate the correction to the radiated energy~{\cite{will1977}}.
In order to perform this, we need to expand perturbation up to the second order.
Then, Eq.~\eqref{wave-chi} becomes
\be
\Box \chi^{\mu\nu}=-16\pi \tau^{\mu\nu}, 
\ee
where
\be
\tau^{\mu\nu}=\frac{1}{\phi_0}T^{\mu\nu}+t^{\mu\nu}\,.
\ee
Here, $t^{\mu\nu}$ is a second order tensor composed of $\chi_{\mu\nu}$ and $\tilde{\phi}$
%, and it is given as
%
%
%
%
and corresponds to the energy-momentum tensor of GWs.
(The explicit form is given in~\ref{app-tmunu}.)
By taking the TT gauge and performing integration by parts, we obtain the expression for the GW energy flux as
\ba
\frac{dE_{\mathrm{GW}}}{dt}&=& -\left\langle\int dA \,t^{00}\right\rangle \nn \\
                                       &=& -\frac{r^2}{32\pi}\phi_0\left\langle\int[\dot{\chi}_{kl}^{\mathrm{TT}}\dot{\chi}_{kl}^{\mathrm{TT}}
                                           +(4\omega_{\mathrm{BD}}+6)\dot{\varphi}\dot{\varphi}]d\Omega\right\rangle
\ea
with $\varphi\equiv\tilde{\phi}/\phi_0$.
By substituting Eqs.~\eqref{chi-ij} and \eqref{phi-tilde} into the above equation and keeping only the leading PN order for the correction, we obtain
\be
\frac{dE_{\mathrm{GW}}}{dt}=-\frac{32}{5} \left\langle\frac{\mu^2M_t^2}{a^4}v^2
                                            \left[ 1+\frac{5}{48}\mathcal{S}^2\bar{\omega} v^{-2} \right]\right\rangle 
\label{power-rad-bd}
\ee
with $\bar{\omega} \equiv \omega_{\mathrm{BD}}^{-1}$.
The second term corresponds to the correction due to the scalar dipole radiation.
The monopolar radiation (not included in the above equation) appears at 0PN order relative to GR~\cite{damour1992}, which is subdominant compared to the dipolar radiation. 

Finally, we derive the correction to the gravitational waveform phase~{\cite{scharre}}.
By using Eq.~\eqref{power-rad-bd}, the time evolution of frequency can be calculated as
\ba
\dot{f}&=&\frac{\dot{\Omega}}{\pi}= \frac{1}{\pi} \frac{d\Omega}{da} \frac{da}{dE} \frac{dE}{dt} \nn \\
& =&-\frac{3}{2\pi}\frac{M_t^{1/2}}{a^{5/2}}\frac{2a^2}{\mu M_t^{1/2}}\dot{E}_\mrm{GW} \nn \\
         &=&\frac{96}{5}\eta\frac{1}{\pi M_t^2}\left( \frac{M_t}{a} \right)^{11/2} \left( 1+\frac{5}{48}\mathcal{S}^2\bar{\omega}x^{-1} \right)\,.
\ea
By performing the integration once, we derive the correction to $t(f)$ as
\be
t(f)= t_0-5\mc(8\pi \mc f)^{-8/3}
                                         \left(1 -\frac{1}{12}\mathcal{S}^2\bar{\omega} x^{-1} \right), \\
\ee
and the gravitational waveform phase in the Fourier domain is given as~\footnote{See Ref.~\cite{ohashi} for higher PN corrections at linear order in $\eta$ obtained by applying Teukolsky formalism. Also, very recently, Yunes \et~\cite{yunes-massiveBD} solved the Teukolsky equations numerically and obtained a fitted PN formula.} 
\be
\Psi(f)=2\pi ft_0-\phi_0-\pi/4+\Psi_\mrm{0PN}(f)
                  \left(1-\frac{5}{84}\mathcal{S}^2\bar{\omega} x^{-1} \right)\,.
 \label{bd-gw-phase}
\ee
Higher PN terms in GR that appear in Eq.~\eqref{Psi-noangle} can be added to this equation accordingly.

From the corrections above, we see that the effect of the scalar dipole radiation is greater for binaries with larger $\mathcal{S}$.
The sensitivity $s_a$ roughly corresponds to the self-gravitating energy of the body 
with $s_\mrm{WD} \sim 10^{-3}$, $s_\mrm{NS} \sim 0.2$ and $s_\mrm{BH} =0.5$ for WD, NS and BH, respectively~{\cite{eardley}}.
Therefore, binaries whose components consist of different types of compact objects give larger BD effect.
In this theory, the (rescaled) scalar charge is given as $q = 1-2s_a$.
Hence, BHs do not have scalar charges and this is a consequence of the no hair theorem~\cite{robinson,israel,israel2,hawking-uniqueness0,hawking-uniqueness,carter-uniqueness} that also holds in BD theory~\cite{hawking-no-hair}~\footnote{See a related work by Sotiriou and Faraoni~\cite{sotiriou} where they extend Hawking's result to more general class of scalar-tensor theories. }.
However, the proof of the no-hair theorem relies strongly on the assumption of stationarity, and there are astrophysically interesting situations where this assumption is violated and BHs grow scalar hair~{\cite{jacobson-hair,horbatsch-hair}}.

%%%%%%%%%%%%%%%%%%%%%%%%%%%%%%%%%%%%%%%%%%%%%%%%%%%
\subsection{Current Constraints}

\subsubsection{Solar System Experiment}

In this subsection, we explain the current constraint on BD theory obtained from the Shapiro time delay measurement by Cassini satellite~{\cite{cassini}}.
First, the metric for the solar system can be written in PN gauge as~\cite{will-book,will-living}
\be
ds^2 = -(1-2 \Phi )dt^2 + (1-2\gamma \Phi)\delta_{ij} dx^i dx^j + \mathcal{O}(\epsilon^{3/2})\,,
\ee
where the dimensionless parameter $\epsilon$ is defined as $\epsilon = M_{\odot}/r$ with $r$ denoting the typical length scale and $\Phi = -M_{\odot}/r + \mathcal{O}(\epsilon^2)$.
%$\gamma$ is a parameterized post-Newtonian (PPN)  parameter, which gives $\gamma=1$ in GR while $\gamma = (1+\omega_\mrm{BD})/(2+\omega_\mrm{BD}) $ in BD theory. 

The most stringent constraint on $\gamma$ has been obtained by Cassini spacecraft while on its way to Saturn.
The light from the satellite that passes close to the sun gets deflected and there arises a time delay called \textit{Shapiro time delay}.
(See e.g. Ref.~\cite{amendola} and references therein.)
We take this path of the light to be a straight line parameterized by $\bm{x} = (x,b,0)$ where $b$ is the impact parameter and $x=-x_e,x_{\oplus}$ for the satellite and the earth, respectively.
For a round trip, the additional time delay can be expressed as
\ba
\Delta t & = & -2\int^{x_{\oplus}}_{-x_e} \left( 1+\gamma \right) \Phi|_{r=\sqrt{x^2 + b^2}}dx\,, \nn \\
&=& (1+\gamma) r_g \ln \frac{a_{\oplus} r_e}{4b^2}\,,
\ea
 where $r_g = 2M_\odot$ is the Schwarzschild radius of the sun, $a_{\oplus}=1$AU and $r_e=8.43$AU is the distance from the satellite to the sun.
 Here, we have assumed that $x_{\oplus} \approx a_{\oplus} \gg b$ and similar condition for the satellite.
Cassini measured the frequency shift
\ba
y_\mrm{gr} &=&  \frac{d\Delta t}{dt} = \frac{d\Delta t}{db} \frac{db}{dt}\,, \nn \\
&=& -\frac{4r_g}{b}\frac{db}{dt}\,,
\ea
and obtained the result that $(1+\gamma)/2$ must be within 0.0012$\%$ of unity.
This gives the constraint~\footnote{See related works by Perivolaropoulos~{\cite{peri}} and  Alsing \et~\cite{alsing} for the current constraints on \emph{massive} BD theories.} 
\be
\omega_\mrm{BD} > 4\times10^4\,.
\ee 
This constraint is expected to be improved by 3 orders of magnitude using ASTROD~{\cite{ASTROD-I}}.
\subsubsection{Binary Pulsar Test}

%In the previous subsection, we explained the constraint obtained in the weak field regime.
There are also stringent constraints obtained by the binary pulsar tests.
% in the strong field regime.
%Among them, the orbital decay rate of the pulsar-WD system PSR  J1141-6545 puts the strongest constraint.
%The orbital decay rate of the pulsar-WD system PSR  J1141-6545 has been measured as $\dot{P}=(-4.01\pm 0.25) \times 10^{-13}$ which is $1.04 \pm 0.06$ times larger than the prediction in GR.
As we explained in Sec.~\ref{bd-corr}, there exists a scalar dipole radiation in BD theory, and hence we can place constraints on the theory from the observations of the orbital decay rate of the binary pulsar systems. 
All accurately-timed binary pulsars place a bound on $\omega_\mrm{BD}$ as shown in e.g. Fig.7 of Ref.~{\cite{freire}}. (See also Ref.~{\cite{damour-review}}.)
Among them, the binary pulsars that consist of a NS and a WD perform better in constraining the theory compared to the ones consist of 2 NSs, for the reason explained at the end of Sec.~{\ref{bd-corr}}. 
The WD-pulsar systems PSR J1141-6545 \ \cite{bhat} and PSR J1738+0333 \ \cite{freire} place the bounds as
%This additional radiation modifies the binary orbital evolution, and hence the observation of PSR J1141-6545 strongly constrains this dipole radiation, with the lower bound on $\omega_\mrm{BD}$ set as~\cite{bhat} 
%
\be
\omega_\mrm{BD} >
\begin{cases}
2\times 10^4 & \mrm{(J1141-6545)}\,, \\
2.5 \times 10^4 & \mrm{(J1738+0333)}\,,
\end{cases}
\ee
respectively.
These are slightly weaker than the one obtained from the solar system experiment, but still they are meaningful in the sense that they constrain the dipole radiation of the binary in BD theory.
%Recently, more stringent constraint has been obtained with PSR J1738+0333~{\cite{freire}}, but this is still slightly weaker than the Cassini bound~{\cite{cassini}}.

%%%%%%%%%%%%%%%%%%%%%%%%%%%%%%%%%%%%%%%%%%%%%%%%%%%%%%%
\section{Massive Gravity Theories}
\label{sec:MG}

Massive gravity theories are simple extension of GR where we add finite mass to graviton.
There are many different types of massive gravity theories (for recent reviews, see e.g. Refs.~\cite{rubakov,hinterbichler}).
Originally, Fierz and Pauli~\cite{fierz} proposed a Lorentz-invariant massive gravity by simply adding a quadratic graviton mass term to the Einstein-Hilbert action. 
However, it was found that the linearized Fierz-Pauli theory does not reduce to the linearized GR in the massless limit (van Dam-Veltman-Zakharov (vDVZ) discontinuity~\cite{vdv,z}).
%This is due to the fact that the helicity-0 component of the graviton does not decouple from matter.
If this discontinuity is valid in the solar system, Fierz-Pauli theory is ruled out by the solar system experiments.
However, Vainshtein proposed that the effect of nonlinearity cannot be neglected~\cite{vainshtein} and unlike GR, linear approximation breaks down already at a distance much larger than the Schwarzschild radius (Vainshtein radius).
Later, Nicolis and Rattazzi~\cite{nicolis} showed that this Vainshtein mechanism indeed works for the DGP braneworld model~{\cite{dvali}}.
Rubakov~\cite{rubakov2} and Dubovsky~\cite{dubovsky} proposed Lorentz-violating massive gravity theories which evade pathologies related to the vDVZ discontinuity.
Chamseddine and Mukhanov came up with a massive gravity model that is analogous to the Higgs mechanism~{\cite{chamseddine}}.
They introduced 4 scalars with global Lorentz symmetry.
When it is broken, graviton absorbs scalar degrees of freedom and acquires a finite mass.
Although Vainshtein mechanism seems to work in this theory~{\cite{alberte}}, it is realized that the appearance of ghosts cannot be evaded~{\cite{chamseddine2}}.

Recently, de Rahm \et~\cite{derham1,derham2} proposed a novel massive gravity (the so-called \textit{non-linear massive gravity}) under a flat reference metric, which is a non-linear generalization of Fierz-Pauli theory.
In Ref.~{\cite{derham2}}, it has been shown in the unitary gauge that the Hamiltonian constraint which is required to kill BD ghost exists up to fourth order in non-linearities for a certain range of free parameters.
Later, this theory is generalized to generic reference metric by Hassan and Rosen~{\cite{hassan-generic}}. 
Then, the existence of the Hamiltonian constraint has been shown to exist to all orders in the unitary gauge with a reference metric taken as (i) flat~{\cite{hassan-flat}}, (ii) generic but non-dynamical~\cite{{hassan-curv-nondyn}}, and (iii) generic and dynamical (non-linear bimetric gravity)~{\cite{hassan-bi}}.
Moreover, the secondary constraint is also shown to exist~\cite{hassan-secondary} for both non-linear massive and bimetric gravities, concluding that BD ghost is evaded up to all orders in these theories.
See Refs.~\cite{derham-stuckel,derham-helicity} for the proofs of existence of the secondary constraint in generic gauge.
%Therefore it is also important to put constraint on the graviton mass (or the graviton Compton wavelength $\lambda_g\equiv h/m_g c$).

Current constraint from solar system experiment puts a model-independent constraint as $\lambda_g > 2.8 \times 10^{17}$cm \ {\cite{talmadge}}.
As for GWs, the propagation speed is modified from the speed of light when graviton acquires a finite mass.
In the next subsection, we derive the correction to the GW phase due to this deviation in the GW propagation speed from $c$.
Then, in Sec.~\ref{mg-current}, we explain the current constraints on the mass of graviton both from the solar system experiment and binary pulsar tests.

\begin{figure}[t]
 \begin{center}
  \centerline{\includegraphics[scale=.6,clip]{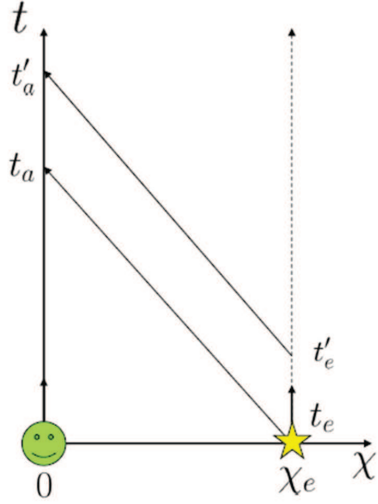}} 
   \end{center}
\vspace*{8pt}
 \caption{Observer detects a GW signal at $t=t_a$ that is emitted from $\chi=\chi_e$ at $t_e$. Then, he detects another signal at $t=t'_a$ that is emitted at $t=t'_e$.}
\label{fig-massive}
\end{figure}

%%%%%%%%%%%%%%%%%%%%%%%%%%%%%%%%%%%%%%%%%%%%%%%%
\subsection{Correction to GW Phase of the Binary}

When the mass of graviton is non-vanishing, the propagation speed of GW becomes less than the speed of light, which shifts GW phase from GR.
In this section, we derive this correction to GW phase based on Will~{\cite{will1998}}.

Let us first assume that the graviton propagates in a homogeneous and isotropic Friedmann-Lema\^itre-Robertson-Walker spacetime.
Its metric is given as
\be
ds^2 = -dt^2 + a^2(t) \left[ d\chi^2 + \Sigma^2 (\chi) (d\theta^2 + \sin^2 \theta d\phi^2) \right]\,,
\ee
where $a(t)$ is the scale factor of the background universe and $\Sigma (\chi)$ equals $\sin\chi$, $\chi$ or $\sinh \chi$ for spatially closed, flat or open universe, respectively.
When a graviton is radially propagating from $\chi = \chi_e$ to the detector $\chi = 0$, we can show from the geodesic equation that $p_{\chi} = \mrm{const.}$, where $p^{\mu}$ is a 4-momentum vector~{\cite{mukhanov}}. 
The norm of $p^{\mu}$ is given as
\be
p^{\mu} p^{\nu} g_{\mu\nu} = -E^2 +\frac{p_{\chi}^2}{a^2} = -m_g^2\,,
\ee
where $E \equiv p^0$.
From this equation, we obtain
\be
\frac{d\chi}{dt} = \frac{p^{\chi}}{E} = -\frac{1}{a} \left( 1+\frac{m_g^2 a^2}{p_{\chi}^2} \right)^{-1/2}
\label{dchidt}
\ee
with 
\be
p_{\chi}^2 = a^2(t_e) (E_e^2 - m_g^2)\,,
\ee
where $E_e$ is the emitted energy of the graviton.
Let us assume that $E_e \gg m_g$ and expanding Eq.~\eqref{dchidt} to first order in $(m_g/E_e)^2$. 
After taking integration, we obtain
\be
\chi_e = \int^{t_a}_{t_e} \frac{dt}{a(a)} - \frac{1}{2} \frac{m_g^2}{a^2(t_e) E_e^2} \int^{t_a}_{t_e} a(t) dt\,,
\ee
where $t_a$ is the arrival time of graviton at $\chi=0$.
Now, we consider two gravitons emitted at $t_e$ and $t'_e$ with energies $E_e$ and $E'_e$, and arrive at $t_a$ and $t'_a$, respectively.
(See Fig.~\ref{fig-massive}.)
Assuming that $\Delta t_e \equiv t_e - t'_e \ll a/\dot{a}$, we obtain
\be
\Delta t_a = (1+z) \left[ \Delta t_e + \frac{D}{2 \lambda_g^2} \left( \frac{1}{f_e{}^2} - \frac{1}{f'_e{}^2} \right) \right]\,,
\label{deltata}
\ee
where $\Delta t_a \equiv t_a - t'_a$ and $f_e$ is the emitted frequency which can be expressed as $f_e = E_e/(m_g \lambda_g)$.
The new distance parameter $D$ is defined as
\ba
D & \equiv & \frac{(1+z)}{a_0} \int^{t_a}_{t_e} a(t) dt \nn \\
& = & (1+z) \int_0^{z} \frac{dz'}{(1+z'^2) H(z')}
\ea
with $a_0 = a(t_a)$ representing the present scale factor.

Under the stationary phase approximation, the phase of GW in Fourier domain is written as
\be
\Psi(f) = 2\pi \int^f_{f_0} (t-t_0) df + 2\pi f t_0 - \phi_0 - \frac{\pi}{4}\,,
\ee
where $f$ is the observed frequency and $f_0$ is the one at coalescence.
By substituting Eq.~\eqref{deltata} into the above equation and using $f_e = (1+z) f$, we have
\be
\Psi(f) = 2\pi \int^{f_e}_{f_{e0}} (t_e-t_{e0}) df_e -\frac{\pi D}{f_e \lambda_g^2}  + 2\pi f \bar{t}_0 - \bar{\phi}_0 - \frac{\pi}{4}\,,
\label{psi-grav}
\ee
where $t_{e0}$ and $f_{e0}$ are the coalescing time and frequency measured in the source frame and 
\be
\bar{t}_0 \equiv t_0 -\frac{D}{2(1+z) \lambda_g^2 f_0^2}, \qquad \bar{\phi}_0 \equiv \phi_0 - \frac{2\pi D}{(1+z) \lambda_g^2 f_0}\,. 
\ee
By performing integration and re-expressing Eq.~\eqref{psi-grav} in $f$, we obtain
%
%\be
%\Psi(f) = \Psi_\mrm{GR} (f) -\beta_g \eta^{-3/5} x^{-3/2}\,,
%\ee
%
\be
\Psi(f) = 2 \pi f \bar{t}_0 - \bar{\phi}_0 - \frac{\pi}{4} + \Psi_\mrm{0PN}(f) \left( 1-\frac{128}{3}\beta_g \eta^{2/5} x \right)\,,
\label{phase-massive}
\ee
where 
%$\Psi_\mrm{GR}(f)$ is given in Eq.~\eqref{Psi-noangle} and 
%
\be
\beta_g \equiv \frac{\pi^2 D \mc}{ \lambda_g^2 (1+z)}\,.
\ee
(When we perform Fisher analysis later, we replace $\bar{t}_0$ and $\bar{\phi}_0$ with $t_0$ and $\phi_0$, respectively.)

%%%%%%%%%%%%%%%%%%%%%%%%%%%%%%%%%%%%%%%%%%%%%%%%%%%
\subsection{Current Constraints}
\label{mg-current}

In this section, following Ref.~{\cite{bertisesana}}, we explain the current \textit{static} and \textit{dynamical} constraints on the mass of graviton. We first explain the current static constraints mainly from solar system experiment~{\cite{talmadge}}.
When the mass of graviton $m_g$ is finite, the gravitational potential is given as a Yukawa type:
\be
V(r) = -\frac{M}{r} \exp \left(-\frac{r}{\lambda_g} \right)\,.
\ee
Under the assumption $r \gg \lambda_g$, the gravitational acceleration $g^i$ is estimated by taking the derivative of the above equation with respect to $r$ as 
\be
g^i = -\frac{n^i M_t}{r^2} \mathcal{G}(r)\,,
\ee
where $n^i$ is the unit radial vector and 
\ba
\mathcal{G}(r) &\equiv& -\left( 1+\frac{r}{\lambda_g} \right) \exp\left( -\frac{r}{\lambda_g} \right) \nn \\
&=& 1-\frac{1}{2} \left( \frac{r}{\lambda_g} \right)^2 + \mathcal{O}\left( \left( \frac{r}{\lambda_g} \right)^3 \right)\,. 
\ea
%
%The constraint on $\lambda_g$ can be put as 
%
%\be
%\lambda_g > \sqrt{ \frac{1-a^2}{6|\eta_p|} }\,,
%\ee
%
%where we have defined the parameter $\eta_p$ as
%
%\be
%1+\eta_p \equiv \left( \frac{\mu(a)}{\mu(a_{\oplus})}  \right)^{1/3}\,.
%\ee
%
%Here, $\lambda_g$ and $a$ are both measured in the astronomical unit (1.5$\times 10^{13}$cm).
From the observation of the Kepler's third law ($a(2\pi/P)^{2/3} = \mathcal{G}(a)^{1/3}$ with $P$ denoting the orbital period) for Mars, the bound on $\lambda_g$ has been obtained as
\be
\lambda_g > 2.8 \times 10^{17} \mathrm{cm}\,.
\ee

Other static constraints include the evidences for the bound clusters and tidal interactions between galaxies~{\cite{goldhaber}}, and the bound from weak lensing~{\cite{choudhury}}.
Although these bounds are considerably stronger than the solar system bound, there remain large ambiguities in these bounds due to the uncertainties in the amount and dynamics of dark matter. 
Recently, Sj$\ddot{\mrm{o}}$rs and M$\ddot{\mrm{o}}$rtsell~\cite{sjors} calculated the constraint on $\lambda_g$ in non-linear massive gravity~\cite{derham1,derham2} using galactic lensing and velocity dispersion data.
Under the assumption of static, spherically symmetric spacetime, the bound is given by $\lambda_g > 10^{26} \mrm{cm}$.
We stress here that this is a model-dependent constraint while the one from the solar system experiment does not rely on the specific choice of the massive gravity theories.

Other than the static constraints, there is a dynamical one obtained from a binary pulsar test~{\cite{finnsutton}}.
For a specific type (Fierz-Pauli type) of the massive gravity theory, they estimated how radiated GW luminosity is modified in a binary pulsar, and obtained a constraint from PSR B1913+16 and PSR B1534+12 as
\be
\lambda_g > 1.6 \times 10^{15} \mrm{cm}\,,
\ee
which is 2 orders of magnitude weaker than the solar system experiment.
%(See App.~\ref{massive-pulsar-app} for more details.)
These current constraints are summarized in Table~\ref{table-mg-current}.

\begin{table}
\tbl{ 
The current constraints on $\lambda_g$.
}
%\begin{center}
%%\begin{ruledtabular}
{\begin{tabular}{c||c}  \hline\hline
%%%%%%%%%%%%%%%%%%%%%%%%%%%%%%%%%%%%%%%%%%%%
Current Bound & $\lambda_g (\mrm{cm})$ \\ \hline
%%%%%%%%%%%%%%%%%%%%%%%%%%%%%%%%%%%%%%%%%%%%
Solar System~\cite{talmadge} (model-independent) & $2.8\times 10^{17}$  \\ 
%\multicolumn{9}{c||} \\
Clusters~\cite{goldhaber} (model-independent) & $6.2\times 10^{24} h_0$    \\ 
Weak Lensing~\cite{choudhury} (model-independent) & $1.8\times 10^{27}$    \\ 
Galaxies~\cite{sjors} (Nonlinear Massive Gravity) & $10^{26}$ \\
Binary Pulsars~\cite{finnsutton} (Fierz-Pauli type) & $1.6\times 10^{15}$    \\ \hline\hline
\end{tabular} \label{table-mg-current}}
%%\end{ruledtabular}
%\end{center}
\end{table}

%%%%%%%%%%%%%%%%%%%%%%%%%%%%%%%%%%%%%%%%%%%%%%%%%%%%%%%%%%%%%
\section{Observing GWs from IMBHs with DPF}
\label{sec:results-DPF}

\begin{figure}[t]
  \centerline{\includegraphics[scale=1.5,clip]{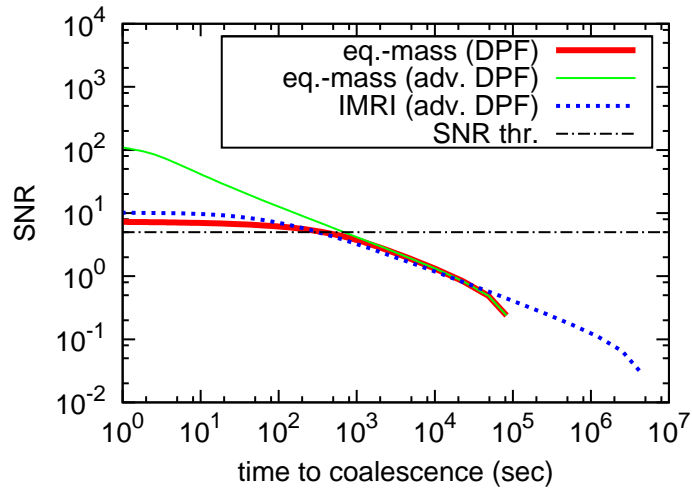} }
 \vspace*{8pt}
\caption{\label{fig:t}
Accumulated sky-averaged SNRs with DPF (red thick solid) and with adv.~DPF (green thin solid) against the time to coalesce for a $10^3\so$ equal-mass IMBH binary with $\chi=0.2$ in NGC 6752.
We also show the one for a $(10+2\times 10^3)\so$ binary in NGC 6752 with adv.~DPF (blue dotted). 
%The vertical dotted-dashed line corresponds to $f=1$Hz and the horizontal one corresponds to the threshold of  SNR=5.
The horizontal dotted-dashed line corresponds to the detection threshold of  $\rho_\mathrm{thr}=5$. 
(This figure is taken from~{\cite{yagiDPF}}.)
}
\end{figure}

Although the event rate is not so promising, in this section, we consider how DPF might help ground-based detectors as a complementary observation and how accurately one can measure binary parameters if the signal is detected. In this section, we do not consider deviations away from GR.

%%%%%%%%%%%%%%%%%%%%%%%%%%%%%%%%%%%%%%%%%%%%%%%%%%%%%%%%%%
\subsection{Possible Joint Search with Ground-Based Detectors}
\label{sec:joint}

From Fig.~\ref{noise-dpf}, one sees that the GW signals from an equal-mass IMBH binary in NGC 6752 can be detected by both DPF and ground-based interferometers, and hence the joint searches of these signals with these detectors may be possible. 
To clarify this possibility, we show the accumulated SNRs of DPF and adv.~DPF against time to coalesce for an equal-mass IMBH binary and an IMRI in Fig.~\ref{fig:t}. 
Let us focus on the former case, where the accumulated SNR reaches $\rho = 5$ at $t \approx 4 \times 10^2$s.
(The definition of SNR is given in Eq.~\eqref{SN}.)
This means, in principle, DPF will be able to give an alert to ground-based detectors about 7 mins before coalescence to make sure that they are operating at the time of merger.
By combining the DPF and the ground-based data, we would be able to observe the whole history of the late inspiral, merger and ringdown.
(See Ref.~\cite{amarojoint} for a related work on possible joint observations of IMBH binaries with LISA, ET and adv.~LIGO.)
DPF data may help in confirming GW signals for the ground-based detectors when only the ringdown signals have been detected.
Notice that the times the accumulated SNR reaches $\rho=5$ are almost identical for DPF and adv.~DPF.
This shows that it is more important to improve the sensitivity at lower frequency to claim earlier detection.
For example, if we improve DPF sensitivity by a factor of 2, the accumulated SNR would reach $\rho =5$ about 1 hour before coalescence.

\subsection{Fisher Analysis}

We use the matched filtering analysis to estimate the determination errors of the binary parameters $\bm{\theta}$ \ {\cite{finn,cutlerflanagan}}. 
We assume that the detector noise is stationary and Gaussian.
%It can be seen that the noise r.m.s. is proportional to $\frac{1}{2}S_n(f)$.
The noise takes the Gaussian probability distribution given by 
\be
p(n_0) \propto \exp \left[-\frac{1}{2}(n_0|n_0)\right]\,, \label{gauss}
\ee%
where we have defined the inner product as 
\be
(A|B)=4 \mathrm{Re}\int ^{\infty}_{0}df \, \frac{\tilde{A}^{*}(f)\tilde{B}(f)}{S_n(f)}. \label{scalar-prod}
\ee
The signal to noise ratio (SNR) for a given GW signal $h$ is  
\begin{equation}
\rho[h]\equiv \sqrt{(h|h)}.
\label{SN}
\end{equation}

\begin{figure}[t]
  \centerline{\includegraphics[angle=-90, scale=.3,clip]{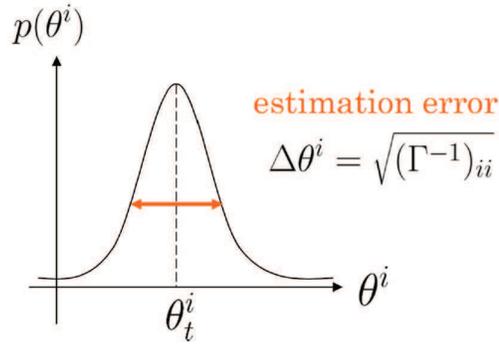}}
\vspace*{8pt}
\caption{Gaussian probability distribution of binary parameters $\theta^i$.
}
\label{fig:fisher}
\end{figure}

The detected signal $s(t)$ is the sum of the gravitational wave signal $h(t;\bm{\theta}_t)$ and the noise $n(t)$, 
 with $\bm{\theta}_t$ representing the true binary parameters.
Then, Eq.~(\ref{gauss}) can be rewritten as
\begin{equation}
p(\bm{\theta}_t|s)\propto p^{(0)}(\bm{\theta}_t)\exp\left[ (h_t|s)-\frac{1}{2}(h_t|h_t) \right], \label{prob}
\end{equation}
where $p^{(0)}(\bm{\theta}_t)$ denotes the distribution of prior information and $h_t\equiv h(\bm{\theta}_t) $.
We determine the binary parameters as $\hat{\bm{\theta}}$ that maximizes the probability distribution $p(\bm{\theta}_t|s)$. 
$\hat{\bm{\theta}}$ satisfies the following equation,
\begin{equation}
(\partial_ih_t|s)-(\partial_ih_t|h_t)=0
\end{equation}
with $\partial_i \equiv \frac{\partial}{\partial\theta^i_t}$.
We can express $\theta^i$ as $\theta^i=\hat{\theta}^i+\Delta\theta^i$ with $\Delta\theta^i$ representing the error in the determination of $\theta^i$.
Next, we expand Eq.~(\ref{prob}) in powers of $\Delta\theta^i$ up to quadratic order and find
\begin{equation}
p(\bm{\theta}|s)\propto p^{(0)}(\bm{\theta})\exp\left[ -\frac{1}{2}\Gamma_{ij}\Delta\theta^i\Delta\theta^j \right],
\end{equation}
where $\Gamma_{ij}=(\partial_i\partial_jh|h-s)+(\partial_ih|\partial_jh)$ is called the {\emph{Fisher matrix}}
(see Fig.~\ref{fig:fisher}).
Since $h-s=-n$, we can neglect the first term of $\Gamma_{ij}$ in the limit of large SNR and 
\begin{equation}
\Gamma_{ij}=(\partial_ih|\partial_jh).
\label{gammaij}
\end{equation} 
When we perform sky-averaged analysis under the assumption that $N_\mrm{int}$ effective interferometer are placed on one site, the Fisher matrix can be written as
\begin{equation}
\Gamma_{ij}=N_\mrm{int} (\partial_ih|\partial_jh).
\label{gammaij-Nint}
\end{equation} 

We set the prior information that the dimensionless spin has to be less than 1.
This can be achieved by setting~\cite{bertibuonanno} 
\begin{equation}
p^{(0)}(\bm{\theta})\propto \exp\left[ -\frac{1}{2}\left( \frac{\beta}{9.4} \right)^2- \frac{1}{2}\left( \frac{\sigma}{2.5} \right)^2\right]. \label{prior}
\end{equation}
The root-mean-square of $\Delta\theta^i$ can be calculated by taking the square root of the diagonal elements of the covariance matrix $\Sigma^{ij}$, which is the inverse of the Fisher matrix:
\begin{equation}
\Sigma^{ij}\equiv \left\langle\Delta\theta^i\Delta\theta^j\right\rangle =(\tilde{\Gamma}^{-1})^{ij}
\end{equation}
with $\tilde{\Gamma}_{ij}$ defined by
\begin{equation}
p^{(0)}(\bm{\theta})\exp\left[ -\frac{1}{2}\Gamma_{ij}\Delta\theta^i\Delta\theta^j \right] \equiv \exp\left[ -\frac{1}{2}\tilde{\Gamma}_{ij}\Delta\theta^i\Delta\theta^j \right]\,.
\end{equation}

We take the integration range of the inner product in Eq.~\eqref{scalar-prod} as $(f_\mrm{in},f_\mrm{fin})$ where 
\beq
f_{\mathrm{in}}=\max \bigl\{ f_{\mathrm{low}},  f_{\mrm{tobs}} \bigr\}, \qquad
f_{\mathrm{fin}}=\min \bigl\{ f_{\mathrm{high}},  f_{\mathrm{ISCO}} \bigr\}.
\eeq
$f_{\mrm{tobs}}$ is the frequency at $t_\mrm{obs}$ before the binary reaches ISCO and it is given by
\ba
f_{t_\mrm{obs}}&=& \left( \frac{5}{256} \right)^{3/8} \pi^{-1} \mc^{-5/8} t_\mrm{obs}^{-3/8} \nn \\
&=&5.5 \times 10^{-2} \left[ \left(\frac{\mathcal{M}}{10
M_{\odot}}\right)^{-5/8} \left( {t_\mrm{obs}\over 1 \mrm{yr}} \right)^{-3/8} \right] \mrm{Hz}. \label{fT}
\ea
Here, we only take the leading contribution from GR into account.
The frequency at ISCO is given by
\beq
f_{\mathrm{ISCO}}=\frac{1}{6^{3/2}\pi M_t}  =4.3\times 10^{2} \lmk \frac{10 \so}{M_{t}} \rmk \mrm{Hz}\,.
\eeq
We recall that $f_\mrm{low}$ and $f_\mrm{high}$ correspond to the low and high cutoff frequencies of the detector, respectively.

%%%%%%%%%%%%%%%%%%%%%%%%%%%%%%%%%%%%%%%%%%%%%%%%%%
\subsection{Binary Parameter Estimation}

In this subsection, we explain how accurately DPF can determine binary parameters by using the inspiral-merger-ringdown hybrid waveform.

\begin{figure}
  \begin{center}
    \begin{tabular}{cc}
      \resizebox{60mm}{!}{\includegraphics{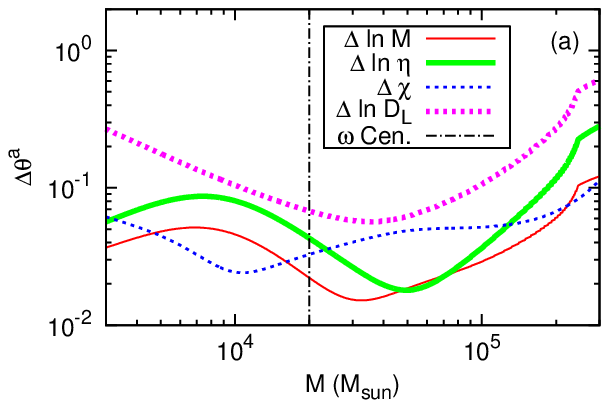}} &
      \resizebox{60mm}{!}{\includegraphics{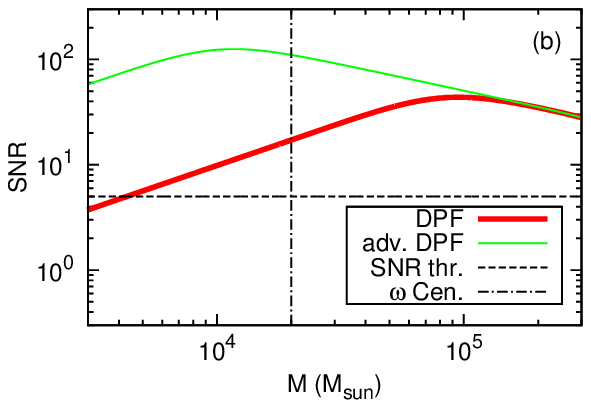}} 
    \end{tabular}
\vspace*{8pt}
    \caption{(a) DPF Parameter determination accuracies of $\ln M$ (red thin solid), $\ln \eta$ (green thick solid), $\chi$ (blue thin dotted) and $\ln D_L$ (magenta thick dotted) against the total mass $M_t$ for equal-mass binaries of $\chi=0.2$ at $D_L=5$kpc, obtained from pattern-averaged analysis.
The vertical dotted-dashed line represents a possible IMBH binary in $\omega$ Centauri. 
(b) The SNRs for equal-mass binaries of $\chi=0.2$ at $D_L=5$kpc against the total mass $M_t$ with DPF (red thick solid) and with adv.~DPF (green thin solid).
The horizontal dashed line corresponds to the detection threshold $\rho_{\mrm{thr}}=5$.
(This figure is taken from Ref.~{\cite{yagiDPF}}.) }
    \label{par}
  \end{center}
\end{figure}

\subsubsection{Pattern-Averaged Analysis}

First, we show the results for the Pattern-averaged analysis, where we take the sky-average of the waveform.
There are 6 parameters in total:
\be
\theta^i = (\ln M_t, \ln \eta, \chi, t_0,\phi_0,D_L)\,.
\ee
We take the fiducial values of $m_1/m_2=1$, $\chi=0.2$, $t_0 = \phi_0 =0$, and $D_L=5$kpc.
The panel (a) of Fig.~\ref{par} shows the DPF measurement errors of $\ln M_t$ (red thin solid), $\ln \eta$ (green thick solid), $\chi$ (blue thin dotted) and $\ln D_L$ (magenta thick dotted) against the total mass $M_t$.
The dotted-dashed vertical line corresponds to a possible equal-mass IMBH binary in $\omega$ Centauri.
One can see that DPF has an ability to measure binary parameters within several $\%$ accuracies. 
The corresponding SNRs for DPF and adv.~DPF are shown in the panel (b).
The parameter estimation for adv.~DPF can be roughly estimated by linearly scaling the DPF results with the SNR ratio.
For example, the ones for an equal-mass IMBH binary in $\omega$ Centauri improves roughly by a factor of 7.
Since the SNRs are not so high, the Fisher analysis can only give rough estimates~{\cite{cutlervallisneri}}.

\begin{figure}[t]
  \centerline{\includegraphics[scale=.7,clip]{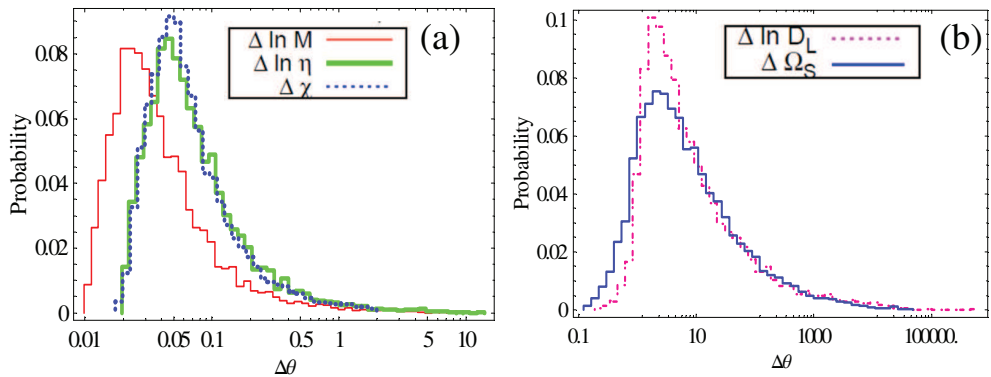} }
\vspace*{8pt}
 \caption{\label{par_mc}
(a) Histograms showing the parameter determination accuracies of $\ln M_t$ (red thin solid), $\ln \eta$ (green thick solid) and $\chi$ (blue dotted) using DPF obtained from Monte Carlo simulations.
We assumed $10^4\so$ equal-mass IMBH binaries in $\omega$ Centauri with $\chi =0.2$ and random orientations for the orbital angular momenta.
(b) Histograms showing the parameter determination accuracies of $\ln D_L$ (magenta dotted-dashed) and $\Omega_\mrm{S}$ (blue solid) using DPF. 
(This figure is taken from Ref.~{\cite{yagiDPF}}.) }
\end{figure}

\subsubsection{Monte-Carlo Simulations}

Next, we perform Monte-Carlo simulations, where we randomly distribute the directions and orientations of the sources.
This time, we have 10 parameters in total:
\be
\theta^i = (\ln M_t, \ln \eta, \chi, t_0, \phi_0, D_L, \bar{\theta}_\mrm{S}, \bar{\phi}_\mrm{S}, \bar{\theta}_\mrm{L}, \bar{\phi}_\mrm{L})\,,
\ee
We set $(m_1,m_2)=(10^4,10^4)M_\odot$, $\chi=0.2$ and $D_L=5$kpc.
We use the galactic latitude and the longitude of $\omega$ Centauri, which can be converted into $(\bar{\theta}_\mrm{S}, \bar{\phi}_\mrm{S})$ using the formulas given in Appendix A of Ref.~{\cite{yagiDPF}}.
We randomly generate $10^4$ sets of $\cos\bar{\theta}_\mrm{L}$ in the range [-1,1] and $\bar{\phi}_L$ in the range [0, 2$\pi$].
We also distribute $\varphi_{\mrm{E}0}$ and $\varphi_{\mrm{D}0}$ (the detector position at $t=0$) in the range [0,2$\pi$].
For each set, we calculate the Fisher matrix.
$10^4$ sets of results allow us to construct the probability distribution of each binary parameter. 

In the panel (a) of Fig.~\ref{par_mc}, we show the probability distributions of $\ln M_t$, $\ln \eta$ and $\chi$.
One sees that these parameters can be determined within several $\%$ accuracies.
In the panel (b), we show the results for $\ln D_L$ and $\Omega_\mrm{S}$ where the latter is defined as
\beq
\Delta\Omega_S\equiv 2\pi | \sin\bar{\theta}_{\mathrm{S}} | \sqrt{\Sigma_{\bar{\theta}_{\mathrm{S}}\bar{\theta}_{\mathrm{S}}}
                              \Sigma_{\bar{\phi}_{\mathrm{S}}\bar{\phi}_{\mathrm{S}}}-\Sigma^2_{\bar{\theta}_{\mathrm{S}}\bar{\phi}_{\mathrm{S}}}}.
\eeq
Unfortunately, these parameters cannot be determined with high accuracy.
The angular resolution can be roughly estimated as
\beq
\Delta\Omega_\mrm{s}\sim \lmk \frac{1}{\rho}\frac{\lambda}{D} \rmk^2 \sim 2.4\lmk \frac{30}{\rho} \rmk^2 \lmk \frac{\lambda}{3\times 10^{10}\mrm{cm}} \rmk^2 \lmk \frac{6.4\times 10^8 \mrm{cm}}{D} \rmk^2\,.
\eeq
Here, $\lambda$ is the wavelength of GWs and $D$ is the effective size of the detector which corresponds the diameter of Earth for DPF observation.
This estimate agrees with the Monte-Carlo results.
$D_L$ is determined from the amplitudes of GWs, but since the angular resolution is of $\mathcal{O}(1)$, $\ln D_L$ is only determined to the same accuracy.
This shows that distance and angular parameters are poorly determined by DPF.
However, because the correlation between the amplitude and the phase parameters are very weak, DPF would be able to determine the mass and spin parameters with good accuracy.
Since these signals cannot be detected by ground-based detectors, DPF would perform unique observations.

%%%%%%%%%%%%%%%%%%%%%%%%%%%%%%%%%%%%%%%%%%%%%%%%%%%%%%
\section{Testing Alternative Theories of Gravity with Gravitational Waves}
\label{sec:testing-GR}

One of the most interesting scientific goals of GW observations is to test alternative theories of gravity in the strong-field regime. In this section, we review the proposed constraints on Brans-Dicke and massive gravity theories by observing gravitational waves from compact binaries.

\subsection{Brans-Dicke}

%%%%%%%%%%%%%%%%%%%%%%%%%%%%%%%%%%%%%%%%%%%%%%%%%%%
\subsubsection{Proposed Constraints with ground-based detectors and LISA}

Constraining Brans-Dicke theory using gravitational waves from compact binaries was first discussed by Eardley~{\cite{eardley}}.
In this theory, the non-vanishing scalar dipole radiation exists~{\cite{will1977,zaglauer}},
which modifies the binary orbital evolution from the one in GR.
%Eardley mentioned that it is possible to constrain $\omega_{\mathrm{BD}}$ by measuring the rate of the secular decrease in the orbital period.
The change in the orbital evolution due to this dipole radiation modifies the phasing of the gravitational waveform.
Will~\cite{will1994} carried out the matched filtering analysis and calculated how accurately one can measure the binary parameters including $\omega_{\mathrm{BD}}$ with adv.~LIGO.
Damour and Esposito-Far\`ese~\cite{damour-GW-ST} extended Will's work to generic scalar-tensor theories.
Scharre and Will~\cite{scharre} and Will and Yunes~\cite{willyunes} followed Will's analysis using LISA.
Will and Yunes improved the previous works by investigating how the constraints depend on LISA position noise, acceleration noise and arm lengths.
These works consider non-spinning binaries under pattern-averaged (sky-averaged) analysis.
%They also used slightly improved noise curve than the one used by Scharre and Will. 
However, these calculations do not include binary spins and also they used the pattern-averaged waveforms.
%(having been averaged over the source direction $(\theta_{\mathrm{S}},\phi_{\mathrm{S}})$ and the direction of the orbital angular momentum $(\theta_{\mathrm{L}},\phi_{\mathrm{L}})$).
Berti \textit{et al.}~\cite{bertibuonanno} calculated the determination accuracies using LISA, taking the effect of the spin-orbit coupling into account for spin-aligned binaries.
They performed both the pattern-averaged and the Monte Carlo simulations.
In the latter simulations, they randomly distribute the directions and the orientations of $10^4$ sources over the sky, calculate the constraint on $\omega_{\mathrm{BD}}$ from each binary and take the average at the end.
%According to their analysis, for a (1.4+1000)M$_{\odot}$ BH/NS binary of SNR=$\sqrt{200}$, 1 yr observation by LISA can put the bound $\omega_{\mathrm{BD}}>10799$ on average.
These results are summarized in Tables~\ref{table-bd-previous1} and~\ref{table-bd-previous2}.
There are also some works on constraining massive BD theory~{\cite{bertialsing,yunes-massiveBD}} and generic theories that produce dipolar gravitational radiation~{\cite{arun-dipole}} with GW interferometers.
Very recently, merger simulations of BH~\cite{healy-merger-ST} and NS~\cite{barausse-merger-ST} binaries have been performed.

\begin{table}
\tbl{ 
Summary of the works on probing BD theory using GWs from compact binaries.
The second column denotes up to which PN order they take. 
The third, fourth and fifth columns represent whether they take the spin-orbit coupling $\beta$, precession or eccentricity into account.
The sixth column describes whether they assume multiple-source detections. 
The seventh column shows what kind of analyses they have performed (either the pattern-averaged (PA) or Monte Carlo (MC).)
Finally, the last column denotes whether generic scalar-tensor theories are considered.
}
%\begin{center}
%%\begin{ruledtabular}
{\begin{tabular}{c||cccccccc}  \hline\hline
 Reference & PN & $\beta$ &  prec. & ecc. & multi. & analy. & generic ST \\ \hline\hline 
%\multicolumn{9}{c||} \\
Will (1994)~\cite{will1994} &1.5  & $\times$ & $\times$ & $\times$ & $\times$ & PA & $\times$   \\ 
Damour \& Esposito-Far\`ese (1998)~\cite{damour-GW-ST} & 1.5  & $\times$ & $\times$ & $\times$ & $\times$ & PA & $\bigcirc$   \\ 
Scharre \& Will (2002)~\cite{scharre} & 1.5   & $\times$ & $\times$ & $\times$ & $\times$ & PA & $\times$ \\
Will \& Yunes (2004)~\cite{willyunes} & 1.5  & $\times$ & $\times$ & $\times$ & $\times$ & PA & $\times$ \\
Berti \et~(2005)~\cite{bertibuonanno} & 2 & $\bigcirc$  & $\times$ & $\times$ & $\times$ & MC & $\times$  \\ 
Yagi \& Tanaka (2010)~\cite{yagiLISA} & 2 & $\bigcirc$ & $\bigcirc$ & $\bigcirc$  & $\times$ & MC & $\times$  \\
Yagi \& Tanaka (2011)~\cite{yagiDECIGO} & 2 & $\bigcirc$ & $\bigcirc$ & $\bigcirc$ & $\bigcirc$ & MC & $\times$  \\ \hline
\end{tabular} \label{table-bd-previous1}}
%%\end{ruledtabular}
%\end{center}
\end{table}

%\begin{landscape}

\begin{table}
%\rotatebox{90}{
\tbl{
Summary of the proposed constraints on $\omega_\mrm{BD}$ using GWs.
The numbers in the brackets denote the total masses of the binaries in the unit of $M_\odot$. }
%\begin{center}
%%\begin{ruledtabular}
{\begin{tabular}{c||ccc}  \hline\hline
 Reference & adv.~LIGO &  LISA  & DECIGO/BBO  \\ \hline\hline 
%\multicolumn{9}{c||} \\
Will (1994)~\cite{will1994} & $2\times 10^3$ (3.7)  &  &    \\ 
Damour \& Esposito-Far\`ese (1998)~\cite{damour-GW-ST} & $2.2\times 10^2$ (11.4)  &  &    \\ 
Scharre \& Will (2002)~\cite{scharre} &  &   $2\times 10^5$ ($10^3$)   &  \\
Will \& Yunes (2004)~\cite{willyunes} &  &   $2\times 10^5$ ($10^3$)   &  \\
Berti \et~(2005)~\cite{bertibuonanno} &  &   $10^4$ ($10^3$)   &  \\
Yagi \& Tanaka (2010)~\cite{yagiLISA} &  &   $7\times 10^3$ ($10^3$)   &  \\
Yagi \& Tanaka (2011)~\cite{yagiDECIGO} &  &  & $4\times 10^8$ (11.4)     \\ \hline
\end{tabular} \label{table-bd-previous2}}
%%\end{ruledtabular}
%\end{center}
%}
\end{table}

%\end{landscape}

In Ref.~{\cite{yagiLISA}}, we further improved these analyses by (i) taking the spin-spin coupling effect into account, (ii) considering slightly eccentric binaries, and (iii) looking at precessing binaries~{\cite{yagiLISA}}.
We first performed the pattern-averaged analysis where we have 9 parameters in total:
\be
\theta^i = (\ln \mc, \ln \eta, \beta, \sigma, t_0, \phi_0, D_L, I_e, \bar{\omega})\,.
\ee
We set the fiducial values of $\beta = \sigma = t_0 = \phi= \bar{\omega}=0$ and set $\rho=10$.
For the initial eccentricity, we assume $e_i = 0.01$ at 1 yr before coalescence.
%(See App.~\ref{app1} for the justification of this choice.)
We consider BH/NS binaries and set $\mathcal{S}=0.3$.
We fix $m_{\mathrm{NS}}=1.4$M$_{\odot}$ and consider 4 different BH masses, $m_{\mathrm{BH}}=400$, 1000, 5000 and 10$^4$ M$_{\odot}$. 
For each binary, we show the results for circular and eccentric binaries in Table~\ref{table-bd-noangle-lisa}.
($\sigma$ is included in both cases.)
For comparison, we also show the results in Ref.~\cite{bertibuonanno} 
where the authors did not include $\sigma$ nor $I_e$ into binary parameters.

 From this table, one can see that the determination accuracies deteriorate as we increase the number of parameters.
This is because the parameters are strongly correlated and adding parameters dilutes the binary information in the detected GWs.
Inclusion of both $\sigma$ and $I_e$ into parameters increases the determination errors by roughly 1 order of magnitude.
In particular, the inclusion of $I_e$ has larger effect on determining $\omega_{\mathrm{BD}}$ than the inclusion of $\sigma$.
The reason can be understood as follows.
In the phase $\Psi(f)$, the term containing $\omega_{\mathrm{BD}}$ has a relative frequency dependence of $f^{-2/3}$ compared to the leading Newtonian term, while the ones containing $\sigma$ and $I_e$ have the relative frequency dependences of $f^{4/3}$ and $f^{-19/9}$, respectively.
The first (BD correction) and the third (eccentricity) terms both have negative power-law indices which makes the correlation between these 2 parameter stronger than the one between $\omega_{\mathrm{BD}}$ and $\sigma$.

The table also shows that the constraint becomes more stringent as the BH mass decreases.
This is because the orbital velocity of the binaries become smaller, which makes relative ``-1PN'' dipole correction larger.   
In Ref.~{\cite{yagiLISA}}, we also show that the measurement accuracy of $\omega_{\mathrm{BD}}$ would be similar to the one without including $I_e$ if we impose the prior information of $I_e > 0$.

\begin{table}
\tbl{The results of error estimation in Brans-Dicke theory using LISA~{\cite{yagiLISA}}.
These are calculated with pattern-averaged analysis for the spin-aligned BH/NS binaries with various BH masses under $\rho=10$. 
We fix NS masses to $1.4 M_{\odot}$.
For each binary, the first line represents the results obtained by Berti \et~\cite{bertibuonanno} which does not include $\sigma$ nor $I_e$ into parameters. 
The second line represents the results including $\sigma$, while the third line shows the ones including both $\sigma$ and $I_e$.
%Here $I_e$ is the value of $e^2$ at $f=0.3$Hz.
We used only one interferometer for the analyses.
(This table is taken from Ref.~{\cite{yagiLISA}}.)}
%\begin{center}
%%\begin{ruledtabular}
{\begin{tabular}{ccc||cccccccc}  \hline\hline
 & $\sigma$& $I_e$ & $\omega_{\mathrm{BD}}$ & $\Delta \ln\mathcal{M}$(\%) & 
                    $\Delta \ln\eta $ & $\Delta \beta $ &
                    $\Delta t_0(s)$ & $\Delta \phi_0$ & $\Delta \sigma$ & $\Delta I_e (10^{-12})$ \\ \hline\hline 
\multicolumn{3}{c||}{400$M_{\odot}$} & \multicolumn{7}{l}{ } \\
& $\times$ & $\times$ & 39190 & 0.00657 & 0.0250 & 0.0508 & 7.95 & 76.7 & - & - \\ 
& $\bigcirc$ & $\times$ & 24886 & 0.0130 & 0.0819 & 0.202 & 13.8 & 552 & 2.39 & -\\
& $\bigcirc$ & $\bigcirc$ & 4583 & 0.0396 & 0.142 & 0.280 & 16.7 & 552 & 2.49 & 1.09 \\ \hline
\multicolumn{3}{c||}{1000$M_{\odot}$} & \multicolumn{7}{l}{ } \\
& $\times$ & $\times$ & 21257 & 0.00764 & 0.0186 & 0.0557 & 7.99 & 58.4 & - & - \\
& $\bigcirc$ & $\times$ & 8210 & 0.0265 & 0.110 & 0.0692 & 23.5 & 919 & 1.96 & - \\
& $\bigcirc$ & $\bigcirc$ & 1881 & 0.0692 & 0.193 & 0.261 & 23.6 & 1059 & 2.41& 6.34 \\ \hline
\multicolumn{3}{c||}{5000$M_{\odot}$} & \multicolumn{7}{l}{ } \\
& $\times$ & $\times$ &  6486 & 0.0114 & 0.0133 & 0.0550 & 8.79 & 23.4 & - & - \\
& $\bigcirc$ & $\times$ & 1933 & 0.0503 & 0.0936 & 0.221 & 37.9 & 1108 & 0.595 & - \\
& $\bigcirc$ & $\bigcirc$ & 281 & 0.224 & 0.302 & 0.916 & 62.9 & 2438 & 1.30 & 173 \\ \hline
\multicolumn{3}{c||}{10000$M_{\odot}$} & \multicolumn{7}{l}{ } \\
& $\times$ & $\times$ &  3076 & 0.0178 & 0.0161 & 0.0706 & 13.6 & 15.5 & - & - \\
& $\bigcirc$ & $\times$ & 862 & 0.0827 & 0.114 & 0.350 & 82.9 & 1763 & 0.474 & - \\
& $\bigcirc$ & $\bigcirc$ & 113 & 0.412 & 0.418 & 1.51 & 160 & 4454 & 1.11& 797 \\ \hline\hline
\end{tabular} \label{table-bd-noangle-lisa}}
%%\end{ruledtabular}
%\end{center}
\end{table}

\begin{figure}[t]
  \centerline{\includegraphics[scale=.4,clip]{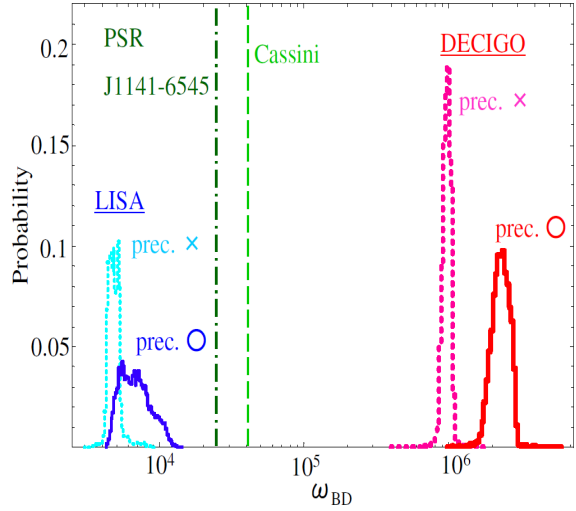}}
\vspace*{8pt} 
\caption{\label{bd} The histograms showing the probability distributions
 of the lower bounds of $\omega_{\mathrm{BD}}$ obtained from our Monte
 Carlo simulations of $10^4$ BH/NS binaries in Brans-Dicke theory~{\cite{yagiLISA,yagiDECIGO}}. 
We take the masses of the binaries as $(1.4+10^3)M_{\odot}$ for LISA and $(1.4+10)M_{\odot}$ for DECIGO/BBO, both with signal-to-noise ratio (SNR) of $\sqrt{200}$ and we assumed that the binaries are circular. 
The thin solid and dotted lines denote the bounds obtained using LISA with and without precessions, while the thick dotted and solid ones represent the estimate using DECIGO/BBO with and without precessions, respectively. 
The dashed and the dotted-dashed lines represent the constraints using Cassini satellite~\cite{cassini} and PSR J1141-6545 \ {\cite{bhat}}, respectively.
(This figure is taken from Ref.~{\cite{yagiDECIGO}}.)}
\end{figure}

Next, we show the results for the Monte Carlo simulations.
We have 13 parameters in total:
\be
\theta^i = (\ln \mc, \ln \eta, \beta, \sigma, t_0, \phi_0, D_L, I_e, \bar{\theta}_{\mathrm{S}},\bar{\phi}_{\mathrm{S}},\bar{\theta}_{\mathrm{L}},\bar{\phi}_{\mathrm{L}},\bar{\omega})\,.
\ee
We fix the masses of the binary constituents as $m_{\mathrm{NS}}=1.4 M_{\odot}$ and $m_{\mathrm{BH}}=10^4 M_{\odot}$, and consider a binary of $\rho=\sqrt{200}$. (This roughly corresponds to $\rho=10$ for a single interferometer).
We randomly distribute $\cos\bar{\theta}_{\mathrm{S}}$ and $\cos\bar{\theta}_{\mathrm{L}}$ in the range [-1,1], and $\bar{\phi}_{\mathrm{S}}$ and $\bar{\phi}_{\mathrm{L}}$ in the range [0,2$\pi$], generating $10^4$ sets of spin-aligned binaries.
We calculate the measurement accuracies of $\bar{\omega}$ for each binary using Fisher analysis and find the probability distribution of the future lower bound on $\omega_\mathrm{BD}$ shown as the (light blue) thin dotted histogram in Fig.~\ref{bd}.
For comparison, we also show the solar system~\cite{cassini} and the binary pulsar~\cite{bhat} constraints as the dashed and the dotted-dashed lines, respectively.
This histogram shows that LISA can only place weaker bounds than the current ones.
Notice that the histogram has ``tails'' on both weaker and stronger bounds.
The former corresponds to binaries with $\hat{\bm{L}} \cdot \hat{\bm{N}} = 0$. 
The latter is due to the Doppler phase
\be
\varphi_D(t) = 2\pi \frac{f(t)}{f_c} \sin \bar{\theta}_\mrm{S} \cos[\bar{\phi}(t)-\bar{\phi}_\mrm{S}]\,,
\ee
where 
\be
f_c \equiv \frac{c}{R} = 2.00 \mrm{mHz}\,,
\ee
is the critical frequency with $R=1$AU.
This Doppler phase makes the degeneracies between some parameters stronger, but for some specific directions, the effect of the Doppler phase vanish and the constraint becomes stronger.
For example, the derivative of $\varphi_D$ with respect to $\bar{\theta}_\mrm{S}$ is proportional to $\cos\bar{\theta}_\mrm{S}$, and hence the correlation between $\omega_\mrm{BD}$ and $\bar{\theta}_\mrm{S}$ due to the Doppler phase is disentangled at $\bar{\theta}_\mrm{S}=\pi/2$. 

Next, we show the results for the precessing binaries.
There are 15 parameters in total: 13 parameters for the spin-aligned case plus $\kappa$ and $\alpha_c$.
%We perform the Monte Carlo simulations as in the previous subsection.
%As in the previous subsection, we fix the masses of the binary constituents to $m_{\mathrm{NS}}=1.4 M_{\odot}$ and $m_{\mathrm{BH}}=10^4 M_{\odot}$ with $\sqrt{200}$.
For Monte Carlo simulations, we set the dimensionless spin parameter of a NS and a BH to 0 and 0.5, respectively.  
%(For a BH/NS binary, the lighter body corresponds to the NS $\chi_\mrm{NS} \sim 0$ is supported from observations~\cite{blanchet2pn}. )
We randomly distribute the fiducial values of $\kappa=\hat{\bm{L}}\cdot\hat{\bm{S}}$ in the range [-1,1] and $\alpha_c$ in the range $[0,2\pi]$.
%We use $L=\mu\sqrt{M_t a}$ for the calculation of orbital angular momentum, where the separation is given as $a=M_t^{1/3}/(\pi f)^{2/3}$.
%(This is derived from $v^2=\sqrt{M_t/a}=(\pi M_t f)^{2/3}$.)

\begin{table}
\tbl{The results of error estimation in Brans-Dicke theory for a  $(1.4+1000)M_{\odot}$ BH/NS binary obtained from the Monte Carlo simulations using LISA~{\cite{yagiLISA}}. We used two interferometers for the analyses and we fixed $\rho=\sqrt{200}$ ($\rho=10$ for each interferometer). 
%We performed the following Monte Carlo simulations. We distribute $10^4$ binaries, calculate the error of each parameter for each binary and take the average. 
The first half of the table shows the results for the spin-aligned case, and the second half represents the precessing case. The first line of each part shows the results without taking $I_e$ into parameters, while in the second line this is taken into account. $\sigma$ is included in all the cases. 
(This table is taken from Ref.~{\cite{yagiLISA}}.)}
%\begin{ruledtabular}
{\begin{tabular}{c||ccccccc}  \hline\hline
 Cases & $\omega_{\mathrm{BD}}$ &  $\Delta \ln\mathcal{M}(\%)$  & 
                    $\Delta\ln \eta$ & $\Delta \beta $ &
                    $\Delta \ln D_L$ & $\Delta \Omega_S(10^{-3}\mathrm{str})$ & $\Delta \sigma$  \\ \hline\hline
% &  & ($\%$) & & & & ($10^{-3}$str) & \\ \hline\hline 
%\multicolumn{6}{l}{(1.4+400)$M_{\odot}$} \\
\textbf{spin-aligned} & \multicolumn{6}{l}{ } \\
Excluding $I_e$ & 4844 &  0.0396 & 0.143 & 0.108 & 2.53 & 0.406 & 2.18   \\ 
Including $I_e$ & 1058 & 0.106 & 0.241 & 0.481 & 1.34 &1.05 & 2.45  \\ \hline
\textbf{precessing} & \multicolumn{6}{l}{ } \\
Excluding $I_e$ & 6944 & 0.0291 & 0.107 & 0.144 & 0.0809 & 0.341 & 1.66  \\
Including $I_e$ & 3523 & 0.0432 & 0.130 & 0.161 & 0.0851 & 0.589 & 1.86  \\ \hline\hline
\end{tabular} \label{table-bd-noprec-lisa}}
%\end{ruledtabular}
\end{table}

The results are shown in the lower half of Table~\ref{table-bd-noprec-lisa}.
The first and the second row in this table represents the results excluding and including eccentricity $I_e$ into parameters, respectively.
Figure~\ref{bd} shows that the constraints on $\omega_{\mathrm{BD}}$ become stronger by 20$\%$ for the precessing binaries.
As we stated before, imposing the prior distribution $\Delta I_e >0$ would increase the lower bound of $\omega_\mrm{BD}$ to the case without including $I_e$ into parameters.
In this case, the constraint on $\omega_{\mathrm{BD}}$ for the precessing binaries including $\sigma$ becomes $\omega_{\mathrm{BD}}>6944$, which is just 1.6 times lower than the one obtained in~{\cite{bertibuonanno}}.

%%%%%%%%%%%%%%%%%%%%%%%%%%%%%%%%%%%%%%%%%%%%%%%%%%5
\subsubsection{Proposed Constraints with DECIGO/BBO}
\label{num-BD-DECIGO}

%\subsection{Spin-Aligned Case}

%\subsection{Setup}

Next, we show the results obtained with DECIGO/BBO~{\cite{yagiDECIGO}}.
Here, we only consider circular binaries since as we explained in the previous section, the constraint on $\omega_\mrm{BD}$ is almost unaffected for a binary having an eccentric orbit provided that we impose a prior condition on eccentricity.
For the instrumental noise spectrum, we have used the one of DECIGO shown in Eq.~\eqref{inst-DECIGO}.
It is expected that BBO may give stronger constraints than the ones shown in this section by a factor of a few .

%\subsubsection{Pattern-Averaged Analysis}
%\label{num-BD-DECIGO-pattern}

\begin{table}
\tbl{The results of error estimation with DECIGO/BBO in
 Brans-Dicke theory for various mass spin-aligned BH/NS binaries. We performed 
pattern-averaged estimates, using only one
  interferometer with $\rho = 10$.
(This table is taken from Ref.~{\cite{yagiDECIGO}}.)}
%%\begin{ruledtabular}
%\begin{center}
{\begin{tabular}{c||ccccccc}  \hline\hline
 masses & $f_{\mathrm{in}}$ & $f_{\mathrm{fin}}$ & $\omega_{\mathrm{BD}}$ &  $\Delta \ln\mathcal{M}$ & $\Delta\ln \eta$ & $\Delta \beta $ & $\Delta \sigma$  \\ 
 &  (Hz) &  (Hz) & $(10^6)$ & $(10^{-5})$ & $(\%)$ & &  \\ \hline
% &  & ($\%$) & & & & ($10^{-3}$str) & \\ \hline\hline 
%\multicolumn{6}{l}{(1.4+400)$M_{\odot}$} \\
$(1.4+10)M_{\odot}$ & 0.118 & 100 & 1.342 & 0.978 & 2.78 & 0.190  & 2.18 \\
$(1.4+50)M_{\odot}$ & 0.0776 & 85.6 & 0.2662 &  2.34 & 2.64 & 0.106 & 1.09 \\ 
$(1.4+100)M_{\odot}$ & 0.0651 & 43.36 & 0.1899 & 2.34 & 1.87 & 0.0485 & 0.563  \\
$(1.4+400)M_{\odot}$ & 0.0460 & 10.95 & 0.04244 & 4.96 & 1.85 & 0.0133 & 0.250  \\ \hline\hline
\end{tabular} \label{table-bd-noangle}}
%\end{center}
%%\end{ruledtabular}
\end{table}

\begin{table}[t]
\tbl{Comparison of the constraints on $\omega_{\mathrm{BD}}$ ($\omega_\mrm{BD}^\mrm{uncor}$ corresponds to the bound assuming that $\omega_\mrm{BD}$ is not correlated with other parameters) and other physical quantities for a $(1.4, 10)M_{\odot}$ binary with DECIGO and a $(1.4, 1000)M_{\odot}$ binary with LISA. We performed pattern-averaged analyses and SNRs are fixed to 10 for both cases. $N_{\mathrm{GW}}$ is the number of GW cycles, $v_\mrm{1yr}$ denotes the velocity at 1yr before coalescence, and $f_\mrm{in}$ and $f_\mrm{fin}$ are the initial and final frequencies of observation, respectively.  This table is taken from Ref.~{\cite{yagiDECIGO}}.}
%\begin{ruledtabular}
%\begin{center}
{\begin{tabular}{c||cccccc}  %\hline
 masses and detector & $\omega_{\mathrm{BD}}$ &  $\omega_{\mathrm{BD}}^{\mathrm{uncor}}$ & $N_{\mathrm{GW}}$ & $v_{\mathrm{1yr}}$ & $f_{\mathrm{in}}$ & $f_{\mathrm{fin}}$ \\ 
 &  ($10^6$) & ($10^6$) & $(10^6)$ &  & (Hz) & (Hz)  \\ \hline
$(1.4, 10)M_{\odot}$, DECIGO & 1.34 & 332 & 5.9 & 0.027 & 0.118 & 100   \\
$(1.4, 1000)M_{\odot}$, LISA & 0.00821 & 21.6 & 1.8 &  0.083 & 0.0366 & 1.00 \\ 
\end{tabular} \label{table-bd-compare} }
%\end{center}
%\end{ruledtabular}
\end{table}

In Table~\ref{table-bd-noangle}, we show the pattern-averaged results of binary parameter estimation errors in Brans-Dicke theory with $(1.4+10)M_{\odot}$, $(1.4+50)M_{\odot}$, $(1.4+100)M_{\odot}$ and $(1.4+400)M_{\odot}$ spin-aligned BH/NS binaries of $\rho=10$.
%(It seems that SNRs of $O(10)$ are too small in performing the Fisher analysis.~\cite{vallisneri}
%However, constraints for higher SNR binaries are obtained by just scaling in proportional to SNR. )
Just like the results obtained using LISA, smaller mass binaries give stronger constraints on $\omega_{\mathrm{BD}}$.
%This can be understood as follows.
%The velocities of binaries at 1 yr before coalescences are slower for smaller mass binaries.
%Since Brans-Dicke theory gives dipole correction to binary gravitational waves, this correction is -1PN order.
%Therefore, when we fix SNRs, this contribution is larger for slower binaries, which makes the constraints stronger. 
Comparing these results with the ones using LISA, one sees that DECIGO/BBO has better ability in constraining $\omega_{\mathrm{BD}}$ compared to LISA.
This is because (i) the number of GW cycles are larger, (ii) the velocity at 1yr before coalescence is smaller and (iii) the effective frequency range is larger, as shown in Table~\ref{table-bd-compare}.

Here, we give a rough estimate of how to derive the constraint using DECIGO/BBO.
Matched filtering allows us to detect the effect due to BD correction if the correction term in the phase times SNR is of $\mathcal{O}(1)$. From Eq.~\eqref{bd-gw-phase} and Table~\ref{table-bd-compare}, we have
\be
\rho \times \Psi_\mrm{0PN} \frac{5}{84} \mathcal{S}^2 \bar{\omega} v^{-2} = 1.0 \left( \frac{\rho}{10} \right) \left( \frac{N_\mrm{GW}}{5.9 \times 10^6} \right) \left( \frac{\mathcal{S}}{0.3} \right)^2 \left( \frac{0.027}{v_\mrm{1yr}} \right)^2 \left( \frac{4.3 \times 10^8}{\omega_\mrm{BD}} \right)\,,
\ee
where we have used the fact that $\Psi_\mrm{0PN}$ roughly corresponds to the number of GW cycles $N_\mrm{GW}$ shown in Table~{\ref{table-bd-compare}}.
Therefore, if $\omega_\mrm{BD}$ is not correlated with other parameters, one can constrain the theory with DECIGO/BBO as $\omega_\mrm{BD}^\mrm{uncor} > 4.3 \times 10^8$. This roughly agrees with the value shown in Table~{\ref{table-bd-compare}}. However, since there are correlations, the bound would be reduced by 100 times. This reduction is not as large as the one for LISA because the effective frequency range of observation is broader for DECIGO/BBO (as shown in Table~\ref{table-bd-compare}). As we will discuss later, the expected event rate of BH/NS binaries with DECIGO/BBO would be $\mathcal{O}(10^4)/\mrm{yr}$. This allows us to improve the bound by roughly 100 times. Therefore, in total, the bound would be $\omega_\mrm{BD} > \mathcal{O}(10^8)$, which is 4 orders of magnitude stronger than the solar system bound~{\cite{cassini}}.

%\subsubsection{Monte Carlo Simulations}
%\label{num-BD-DECIGO-mc}

%\subsection{Precessing Case}

 \begin{table}
\tbl{The results of error estimation in
 Brans-Dicke theory for a $(1.4+10)M_{\odot}$ BH/NS binary using DECIGO/BBO. The first
 line shows the ones of pattern-averaged estimate. We used only one
 interferometer with $\rho=10$. The second and the third lines
 show the results of Monte Carlo simulations. We used two interferometers for
 the analyses and we set $\rho=\sqrt{200}$ (corresponding to SNR=10 for
 each interferometer). 
%We distribute $10^4$ binaries, calculate the
 %determination error of each parameter for each binary and take the
 %average. 
The second line shows the ones for the spin-aligned case, while the third line represents the ones including
 precession. $\sigma$ is included in the binary parameters for all the
 cases.
(This table is taken from Ref.~{\cite{yagiDECIGO}}.)}
%%\begin{ruledtabular}
%\begin{center}
{\begin{tabular}{c||ccccccc}  \hline\hline
 cases & $\omega_{\mathrm{BD}}$ &  $\Delta \ln\mathcal{M}$  & 
                    $\Delta\ln \eta$ & $\Delta \beta $ &
                    $\Delta \ln D_L$ & $\Delta \Omega_S$ & $\Delta \sigma$  \\ 
 & $(10^6)$ & $(10^{-5})$ & $(\%)$ & & & $(10^{-5}\mathrm{str})$ & \\ \hline
% &  & ($\%$) & & & & ($10^{-3}$str) & \\ \hline\hline 
%\multicolumn{6}{l}{(1.4+400)$M_{\odot}$} \\
pattern-averaged & 1.342 & 0.978 & 2.78 & 0.190 & 0.100 & - & 2.18 \\
no precession & 0.9774 &  1.22 & 3.06 & 0.186 & 1.24 & 3.27 & 2.15   \\ 
including precession & 2.317 & 0.350 & 0.295 & 0.0551 & 0.183 & 2.52 & 0.627  \\ \hline\hline
\end{tabular} \label{table-bd-noprec}}
%\end{center}
%%\end{ruledtabular}
\end{table}

In Table~\ref{table-bd-noprec}, we show the results of error estimation
in Brans-Dicke theory for a $(1.4+10)M_{\odot}$ BH/NS binary of $\rho=10$, for both the pattern-averaged analysis and of $\rho=\sqrt{200}$ for Monte Carlo simulations.
As expected, we see that inclusion of precession improves the constraint on $\omega_{\mathrm{BD}}$ by a factor of two. 
We show the probability distribution of the constraint on $\omega_{\mrm{BD}}$ using DECIGO/BBO in Fig.~\ref{bd}.
The thick dotted histogram represents the result for the spin-aligned binaries and the thick solid one shows the one for precessing binaries.
This shows that DECIGO/BBO can place 300 times stronger constraint than LISA.
The reasons are the same as the pattern-averaged analysis.

%%%%%%%%%%%%%%%%%%%%%%%%%%%%%%%%%%%%%%%%%%
%\subsubsection{Statistical Analysis}
%\label{stat-BD}

Unlike the case of LISA, BH/NS binaries are expected to be the definite GW targets for DECIGO/BBO.
The event rate of BH/NS is expected to be about one order of magnitude smaller than the NS/NS one~{\cite{shibata}}, which is given by $10^5$ yr$^{-1}$ \ \cite{cutlerharms} (see also Sec.~\ref{DECIGO-event}).
Therefore it is possible to place even stronger constraint by performing a
statistical analysis, where we take the benefit of the large event rate.
The total variance $\sigma_{\bar{\omega}}$ of $\bar{\omega}$ is given by~\cite{yagiDECIGO} 
\begin{equation}
\sigma_{\bar{\omega}}^{-2}%=\sum_i \sigma^{-2}_i
=\Delta T \int^{\infty}_{0} 4\pi [a_0 r(z) ]^2
\dot{n}(z)\frac{d\tau}{dz}\sigma (z)^{-2} dz. 
\label{statistics}
\end{equation}
Here, $\Delta T =1$ yr denotes the observation time, $a_0$ is
the current scale factor, $r(z)$ represents the comoving distance to the source,
$\dot{n}(z)$ shows the BH/NS merger rate at redshift $z$ and $\tau$ denotes
the proper look back time of the source.
$\sigma (z)$ represents the variance of $\bar{\omega}$ from a BH/NS binary at redshift $z$.
$a_0r(z)$ and $d\tau/dz$ are given in Eq.~\eqref{a0rz}.
%
%\be
%a_0 r(z)  = \int ^z_0 \frac{dz'}{H(z)}\,, \qquad 
%\frac{d\tau}{dz} = \frac{1}{(1+z) H(z) }\,, 
%\label{a0rz}
%\ee
%
%where $H(z)$ is the Hubble parameter at redshift $z$, given in Eq.~\eqref{Hz}.
$\dot{n}(z)$ is given by~\cite{cutlerharms} $\dot{n}(z)=\dot{n}_0 R(z)$, 
where $\dot{n}_0=10^{-8}$ Mpc$^{-3}$ yr$^{-1}$ is the estimated BH/NS
merger rate today and $R(z)$
encodes the time-evolution of this rate given in Eq.~\eqref{Rz}~\footnote{$R(z)$ is taken from the redshift evolution of NS/NS merger rate obtained in Ref.~{\cite{schneider}}. Although BH/NS merger rate has not been obtained in this reference, the formation rate of BH/NS follows the same evolution as that of NS/NS (see Fig.~6 of Ref.~\cite{schneider}). Therefore we adopt this evolution for BH/NS merger rate as well.}.
%This model gives the detection rate of $10^4$ yr$^{-1}$.
We assume that all BH/NS binaries have the same typical masses of $(1.4+10)M_{\odot}$ for simplicity. 
We first calculate the variance $\sigma (z)$ for each $z$ 
using pattern-averaged estimate and obtain the total variance 
$\sigma_{\bar{\omega}}$ using Eq.~(\ref{statistics}).
In order to take the effects of redshift into account, we use the redshifted masses: 
$m_{\mathrm{NS}}\rightarrow (1+z)m_{\mathrm{NS}}$
and $m_{\mathrm{BH}}\rightarrow (1+z)m_{\mathrm{BH}}$. 

From the pattern-averaged analysis, we find that 
$10^4$ BH/NS binaries enable us to place a new constraint 
$\omega_{\mathrm{BD}}>2.18\times 10^8$~\footnote{ET is expected to detect BH/NS signals out to $z \approx 1$ for the SNR threshold value of 10 and the expected event rate is more than $\mathcal{O}(10^3)$/yr. Following the analysis of DECIGO/BBO, we found that ET can place $\omega_{\mathrm{BD}}>7.25\times 10^4$ using a single interferometer for the pattern-averaged analysis. (See Ref.~\cite{arun-ASTROD} for the proposed constraint on $\omega_\mrm{BD}$ with ET without including spins into parameters.) If the event rate is about 10 times larger (which corresponds to the averaged value found in e.g. Ref.~\cite{abadie}) and if we assume that we use 2 interferometers, the above constraint further improves as $\omega_{\mathrm{BD}}>3.24\times 10^5$. This is roughly 1 order of magnitude stronger than the current solar system bound. For precessing binaries, this bound should improve by a factor of a few.}, which is 94 times stronger than the one from a single binary of SNR=10 (or $D_L=17$Gpc).
This implies that if we perform Monte Carlo simulations, the constraint would be
%We calibrate the result of this analysis by using the results of Monte Carlo simulation 
%to yield 
%
\be
\omega_{\mathrm{BD}}>3.77\times 10^8\,.
\ee
This is 4 orders of magnitude stronger than the current strongest bound from the solar system experiment~{\cite{cassini}}.  

Furthermore, we have assumed that we use only two interferometers, but since there are eight effective interferometers in total for DECIGO/BBO, the constraint would improve by a factor of $\sqrt{8/2}=2$.
Moreover, we note again that if we use BBO which is roughly three times more sensitive than the noise curve that we used in this section, the constraint would further be three times stronger.
In total, BBO should be able place nearly 5 orders of magnitude stronger constraint than the solar system bound!
This is still 2 orders of magnitude stronger than the expected bound from the future solar system mission ASTROD~{\cite{ASTROD-I}}.
%This leads to the conclusion that 

\subsubsection{Prospects for Other Space-Borne GW Interferometers}

 \begin{table}
\tbl{The proposed bounds on $\omega_\mrm{BD}$ with LISA, eLISA and ASTROD-GW. We assume spin-aligned BH/NS binaries of $\rho=10$ with a single interferometer and perform pattern-averaged Fisher analysis. We set NS masses to 1.4$M_\odot$ and include both $\beta$ and $\sigma$ into parameters, while we consider binaries with circular orbits and do not include $I_e$ into parameters. The LISA bounds are same as the ones shown in Table~\ref{table-bd-noangle-lisa}.}
%%\begin{ruledtabular}
%\begin{center}
{\begin{tabular}{c||ccc}  \hline\hline
 BH mass & LISA & eLISA  & ASTROD-GW \\ \hline
400 $M_\odot$ & 24886 & 25644 & 24809  \\
1000 $M_\odot$ & 8210 &  8759 & 8166   \\ 
5000 $M_\odot$ & 1933 & 2520 & 1861  \\ 
10000 $M_\odot$ & 862 & 1148 & 820  \\ \hline\hline
\end{tabular} \label{table-bd-elisa}}
%\end{center}
%%\end{ruledtabular}
\end{table}

In this subsection, we explain the proposed bounds on $\omega_\mrm{BD}$ using eLISA and ASTROD-GW.
The non-sky-averaged instrumental noise spectral density for eLISA is given by~\cite{elisa-science2}
\be
S_n^\mrm{inst,eLISA} (f) = \frac{4 S_\mrm{acc}(f)/(2\pi f)^4 + S_\mrm{sn}(f) + S_\mrm{omn}(f)}{L_e^2} \left[ 1+\left( \frac{f}{0.41 c/(2 L_e)} \right)^2 \right]\,,
\label{noise-eLISA}
\ee
where $L_e = 1 \ \mrm{Mkm}$ the acceleration noise, shot noise and all other measurement noises (for the position noises) are given by
\ba
S_\mrm{acc} (f) &=& 2.13 \times 10^{-29} \left( 1+\frac{10^{-4} \mrm{Hz}}{f} \right) \ \mrm{m^2 \ s^{-4} \ Hz^{-1}}\,, \\
S_\mrm{sn} (f) &=& 5.25 \times 10^{-23} \ \mrm{m^2 \ Hz^{-1}}\,, \\
S_\mrm{omn} (f) &=& 6.28 \times 10^{-23} \ \mrm{m^2 \ Hz^{-1}}\,,
\ea 
respectively.
For ASTROD-GW, the non-sky-averaged instrumental noise spectral density is given by~\cite{nipriv}
\be
S_n^\mrm{inst,ASTROD} (f) = \frac{[1+(f/f_a)^2/2] S_p + 4 S_a / (2 \pi f)^4}{L_a^2}\,,
\label{noise-ASTROD}
\ee
%
%1/La^2*((1+0.5*(f./fa).^2).*Spa + (4*Sa./(2*pi*f).^4) );
where $f_a = c/(2 \pi L_a)$ with $L_a = 52 \ \mrm{Mkm}$ and the acceleration and the position noises are given by
\ba
S_a &=& 1.0816 \times 10^{-18} \ \mrm{m^2 \ Hz^{-1}}\,, \\
S_p &=& 9.0 \times 10^{-30} \ \mrm{m^2 \ Hz^{-1}}\,,
\ea
respectively.
The total noise sensitivities including the WD/WD confusion noises can be obtained by substituting Eqs.~\eqref{noise-eLISA} or~\eqref{noise-ASTROD} into Eq.~\eqref{noise-BBO}, and they are shown in Fig.~\ref{noise}.

The bounds using these detectors, together with the ones with LISA, are shown in Table~\ref{table-bd-elisa}.
One can see that the proposed constraints are almost the same among these detectors.
(Ref.~{\cite{arun-ASTROD}} discusses that the bound from eLISA is worse by a factor of a few compared to the LISA one. This difference comes from the difference in the observational frequency range that we use for eLISA.)
Since we are setting SNR to be $\rho=10$, it is the shape of the sensitivity curves that matters and not the overall amplitude of the curves. 
For the mass range of BHs shown in Table~\ref{table-bd-elisa}, the frequency at 1yr before coalescence is higher than $10^{-2}$Hz. 
For this frequency range, the shape of the sensitivity curves of the 3 detectors are similar ($\sqrt{S_n} \propto f^{1}$), and hence the bounds turn out to be almost identical.  
The bounds would be similar even if we include the effect of precessions, as can be understood by comparing Tables~\ref{table-bd-noangle-lisa} and~\ref{table-bd-noprec-lisa}.

%%%%%%%%%%%%%%%%%%%%%%%%%%%%%%%%%%%%%%%
\subsection{Massive Gravity}

%%%%%%%%%%%%%%%%%%%%%%%%%%%%%%%%%%%%%%%%%%%%%%%%%%%
\subsubsection{Proposed Constraints with ground-based detectors, LISA, Pulsar Timings and CMB}

In massive gravity theories, the phase velocity of GW is given as~\cite{will1998}
\begin{equation}
v_{\mathrm{ph}}^2=\left( 1-\frac{1}{f^2\lambda_g^2}\right)^{-1}.
\label{vphase}
\end{equation}  
We see that it depends on its frequency, which modifies the time of arrival from general relativity.
This modifies the phasing of the gravitational waveforms.
Jones~\cite{jones} estimated how accurately LISA will be able to distinguish the difference in the arrival times of different harmonic signals from a binary with an eccentric orbit.
Larson and Hiscock~\cite{larson} proposed to place bounds on $\lambda_g$ by using both GW and EM signals from a WD binary.
This work was improved by Cutler \et~\cite{cutler-mg} and Cooray and Seto~\cite{cooray} where the latter considered $\sim$400 close WD binaries and obtained the bound $\lambda_g > 2.1\times 10^{19}\mrm{cm}$ which is two orders of magnitude stronger than the solar system experiment. 
Kocsis \et~\cite{kocsis-mg} proposed to use the correlation between GW and EM signals of SMBH binary coalescence.
Adopting the timing uncertainty as the inverse of the GW frequency at ISCO, they estimated the possible future bound as $\lambda_g > 2.8 \times 10^{20}$cm, which is three orders of magnitude stronger than solar system bound.
However, uncertainty remains on the systematic delay in the emission of EM bursts.
They claim that if the variability of EM signals before coalescence is identified and can be related to the orbital period, similar analyses as the ones using WD binaries mentioned above can be applied so that the systematic errors are reduced and one can obtain stronger constraint. 
These results are summarized in Table~\ref{table-mg-previous3}.

\begin{table}
\tbl{ 
The proposed bounds on $\lambda_g$ using GWs.
}
%\begin{center}
%%\begin{ruledtabular}
{\begin{tabular}{c||c}  \hline\hline
%%%%%%%%%%%%%%%%%%%%%%%%%%%%%%%%%%%%%%%%%%%%
Current Bound & $\lambda_g (\mrm{cm})$ \\ \hline
%%%%%%%%%%%%%%%%%%%%%%%%%%%%%%%%%%%%%%%%%%%%%%
Eccentric Binary & $3 \times 10^{21}$ \ \cite{jones} \\
WD &  $1.4\times 10^{19}$ \ \cite{cutler-mg} \\
 & $2.1 \times 10^{19}$ \ \cite{cooray} \\
EM counterparts & $2.8\times 10^{20}$ \ \cite{kocsis-mg} \\ 
Pulsar Timing &  $1.5 \times 10^{19}$ \ \cite{baskaran,polnarev}  \\
 & $4.1\times 10^{18}$ \ \cite{lee} \\ \hline\hline
\end{tabular} \label{table-mg-previous3}}
%%\end{ruledtabular}
%\end{center}
\end{table}

Will~\cite{will1998} included $\lambda_g$ into the binary parameters and performed the matched filtering analysis, estimating how accurately one can determine $\lambda_g$ using adv.~LIGO or LISA.
Will and Yunes~\cite{willyunes} followed similar analysis using the improved noise curve for LISA.  
As in the BD case, they did not include the spins and also they only performed the pattern-averaged analysis.
Berti \textit{et al.}~\cite{bertibuonanno} estimated the constraint on $\lambda_g$ by performing the Monte Carlo simulations, taking the spin-orbit coupling into account.
% including the additional parameters of the spin-orbit coupling $\beta$ and the angles $\bar{\theta}_{\mathrm{S}},\bar{\phi}_{\mathrm{S}},\bar{\theta}_{\mathrm{L}}$, and $\bar{\phi}_{\mathrm{L}}$.
They find that for a ($10^6+10^6$)M$_{\odot}$ BH/BH binary at 3Gpc, 1 yr observation with LISA can place the bound $\lambda_g>1.33\times 10^{21}$cm on average.
%~\footnote{Subsequent results estimated just at the same time or after Yagi and Tanaka~\cite{yagiLISA, yagiDECIGO} are summarized in App.~\ref{app:subsequentMG}.}.
Arun and Will~\cite{arunwill} included the effect of higher harmonics and found that for higher mass sources, the constraints become stronger compared to the analysis using the restricted waveforms.
The above results are summarized in the upper halves of Tables~\ref{table-mg-previous1} and~{\ref{table-mg-previous2}}.

\begin{table}
\tbl{ 
Summary of the works on probing MG theories using GWs from compact binaries with the matched filtering analysis.
The second column denotes up to which PN order they take. 
(Berti \et~(2011) is left blank because they do not state this information in their paper.)
The third, fourth, fifth and sixth columns represent whether they take the spin-orbit coupling $\beta$, higher harmonics (HH), precession or eccentricity into account.
The seventh column shows whether they consider all of the inspiral, merger and ringdown (IMR) phases or just the inspiral phase. 
The eighth column describes whether they assume multiple-source detections. 
The ninth column represent whether they perform model-independent (MI) calculations.
Finally, the last column shows what kind of analyses they have performed (either the pattern-averaged (PA), Monte Carlo (MC) or Bayesian (B).)
}
%\begin{center}
%%\begin{ruledtabular}
{\begin{tabular}{c||ccccccccc}  \hline\hline
 Reference & PN & $\beta$ & HH & prec. & ecc. & IMR & multi. & MI & analy. \\ \hline\hline 
%\multicolumn{9}{c||} \\
Will (1998)~\cite{will1998} &1.5 & $\times$ & $\times$ & $\times$ & $\times$ & $\times$ & $\times$ & $\times$ & PA   \\ 
Will \& Yunes (2004)~\cite{willyunes} & 1.5 & $\times$ & $\times$ & $\times$ & $\times$ & $\times$ & $\times$ & $\times$ & PA  \\
Berti \et~(2005)~\cite{bertibuonanno} & 2 & $\bigcirc$ & $\times$ & $\times$ & $\times$ & $\times$ & $\times$ & $\times$ & MC   \\ 
Arun \& Will (2009)~\cite{arunwill} & 3.5 & $\bigcirc$ & $\bigcirc$ & $\times$ & $\times$ & $\times$ & $\times$ & $\times$ & PA   \\ %\hline\hline
Stavridis \& Will (2009)~\cite{stavridis} & 2 & $\bigcirc$ & $\times$ & $\bigcirc$ & $\times$ & $\times$ & $\times$ & $\times$ & MC   \\
Yagi \& Tanaka (2010)~\cite{yagiLISA} & 2 & $\bigcirc$ & $\times$ & $\bigcirc$ & $\bigcirc$ & $\times$ & $\times$ & $\times$ & MC   \\
Yagi \& Tanaka (2011)~\cite{yagiDECIGO} & 2 & $\bigcirc$ & $\times$ & $\bigcirc$ & $\bigcirc$ & $\times$ & $\times$ & $\times$ & MC   \\
Keppel \& Ajith (2010)~\cite{keppel} & 3.5 & $\times$ & $\times$ & $\times$ & $\times$ & $\bigcirc$ & $\times$ & $\times$ & PA   \\
Del Pozzo \et~(2011)~\cite{delpozzo} & 2 & $\times$ & $\times$ & $\times$ & $\times$ & $\times$ & $\bigcirc$ & $\times$ & B   \\
Cornish \et~(2011)~\cite{cornish-PPE} & 3.5 & $\times$ & $\times$ & $\times$ & $\times$ & $\times$ & $\times$ & $\bigcirc$ & B   \\
Berti \et~(2011)~\cite{bertisesana} &  & $\times$ & $\times$ & $\times$ & $\times$ & $\times$ & $\bigcirc$ & $\times$ & MC   \\
Huwyler \et~(2011)~\cite{huwyler} & 2 & $\bigcirc$ & $\bigcirc$ & $\bigcirc$ & $\times$ & $\times$ & $\times$ & $\bigcirc$ & MC   \\ \hline
\end{tabular} \label{table-mg-previous1}}
%%\end{ruledtabular}
%\end{center}
\end{table}

%\if0%%%%%%%%%%%%%%%%%%%%%%%%%%

%\begin{landscape}

\begin{table}
%\rotatebox{90}{
\tbl{ 
Summary of the constraints on $\lambda_g$ using GWs.
The units in the table are $10^{18}$cm for adv.~LIGO and ET, while $10^{21}$cm for LISA, eLISA and DECIGO/BBO.
The numbers in brackets denote the total masses of the binaries in the unit of $M_\odot$.
}
%\begin{center}
%%\begin{ruledtabular}
{\begin{tabular}{c||ccccc}  \hline\hline
 Reference & adv.~LIGO & ET & LISA & eLISA & DECIGO/BBO  \\ \hline\hline 
%\multicolumn{9}{c||} \\
Will (1998)~\cite{will1998} &0.6 (20)  &  & 7 ($2\times 10^7$) &  &    \\ 
Will \& Yunes (2004)~\cite{willyunes} &   &  & 5 ($10^7$) &  &  \\
Berti \et~(2005)~\cite{bertibuonanno} &  &   & 1 ($2\times 10^6$) &  &    \\ 
Arun \& Will (2009)~\cite{arunwill} & 0.7 (60) & 10 (400) & 5 ($2\times 10^6$)  &  &    \\ %\hline\hline
Stavridis \& Will (2009)~\cite{stavridis} &   &  & 7 ($2\times 10^7$) &  &    \\
Yagi \& Tanaka (2010)~\cite{yagiLISA} &   &  & 3 ($1.1\times 10^7$) & &    \\
Yagi \& Tanaka (2011)~\cite{yagiDECIGO} &  &  &  &  & 0.3 ($1.1\times 10^6$)   \\
Keppel \& Ajith (2010)~\cite{keppel} & 8 (360) & 70 (3000) & 60 ($4.8\times 10^7$) &  &   \\
Del Pozzo \et~(2011)~\cite{delpozzo} & 0.5--2.5  &  &  &  &   \\
Cornish \et~(2011)~\cite{cornish-PPE} & 0.9 (18--24)   &  & 4 ($4-5\times 10^6$) &  &    \\
Berti \et~(2011)~\cite{bertisesana} &  &  & 6.5--7.5 & 3--5 &    \\
Huwyler \et~(2011)~\cite{huwyler} &   &  & 7 ($1.3\times 10^7$) &  &    \\ \hline
\end{tabular} \label{table-mg-previous2}}
%%\end{ruledtabular}
%\end{center}
%}
\end{table}

%\end{landscape}

%\fi%%%%%%%%%%%%%%%%%%%%%%%%%%%%%%%%%%%%%%%%%%%%

\begin{table}
\tbl{The results of error estimation in massive gravity theories using LISA~{\cite{yagiLISA}}. These are calculated with pattern-averaged analysis for spin-aligned SMBH binaries with masses $(10^7+10^7)M_{\odot}$, $(10^7+10^6)M_{\odot}$, $(10^6+10^6)M_{\odot}$ and $(10^6+10^5)M_{\odot}$ at 3Gpc. We used on only one interferometer. As in the Brans-Dicke case, the first line of each binary represents the estimation without taking $\sigma$ nor $I_e$ into parameters.
These results do not exactly match with the ones shown in~\cite{bertibuonanno} because we also included prior information. The meaning of second and third lines are the same as in Table~\ref{table-bd-noangle-lisa}.
(This table is taken from Ref.~{\cite{yagiLISA}}.)}
%\begin{ruledtabular}
{\begin{tabular}{ccc||cccccccc}  \hline\hline
 & $\sigma$& $I_e$ & $\lambda_g$ & $\Delta \ln\mathcal{M}$ & 
                    $\Delta \ln\eta $ & $\Delta \beta $ &
                    $\Delta t_c$ & $\Delta \phi_c$ & $\Delta \sigma$ & $\Delta I_e$ \\
 & & & ($10^{20}$cm) & (\%) & & & (s) & & &  $(10^{-10})$ \\ \hline\hline 
\multicolumn{3}{c||}{$(10^7+10^7)M_{\odot}$} & \multicolumn{7}{l}{ } \\
& $\times$ & $\times$ & 22.77 & 0.0669 & 0.467 & 2.93 & 75.7 & 1.06 & - & - \\ 
& $\bigcirc$ & $\times$ & 11.33 & 0.0687 & 0.960 & 7.10 & 77.8 & 1.09 & 1.73 & -\\
& $\bigcirc$ & $\bigcirc$ & 11.29 & 0.246 & 1.10 & 7.56 & 133 & 2.23 & 1.89 & 13.0 \\ \hline
\multicolumn{3}{c||}{$(10^7+10^6)M_{\odot}$} & \multicolumn{7}{l}{ } \\
& $\times$ & $\times$ & 9.629 & 0.0493 & 0.253 & 1.49 & 82.3 & 2.15 & - & - \\
& $\bigcirc$ & $\times$ & 4.061 & 0.0495 & 0.839 & 6.26 & 82.8 & 2.16 & 1.89 & - \\
& $\bigcirc$ & $\bigcirc$ & 4.052 & 0.202 & 0.927 & 6.54 & 159 & 5.01 & 1.94& 14.3 \\ \hline
\multicolumn{3}{c||}{$(10^6+10^6)M_{\odot}$} & \multicolumn{7}{l}{ } \\
& $\times$ & $\times$ &  12.41 & 0.00869 & 0.122 & 0.787 &  3.01& 0.316 & - & - \\
& $\bigcirc$ & $\times$ & 3.582 & 0.00871 & 0.926 & 6.97 & 3.01 & 0.316 & 1.68 & - \\
& $\bigcirc$ & $\bigcirc$ & 3.582 & 0.0288 & 0.933 & 7.00 & 4.75 & 0.589 & 1.69 & 1.14 \\ \hline
\multicolumn{3}{c||}{$(10^6+10^5)M_{\odot}$} & \multicolumn{7}{l}{ } \\
& $\times$ & $\times$ &  6.019 & 0.00586 & 0.0551 & 0.337 & 2.45 & 0.521 & - & - \\
& $\bigcirc$ & $\times$ & 1.286 & 0.00586 & 0.823 & 6.20 & 2.45 & 0.521 & 1.88 & - \\
& $\bigcirc$ & $\bigcirc$ & 1.285 & 0.0200 & 0.826 & 6.21 & 3.96 & 0.989 & 1.88 & 1.50 \\ \hline\hline %\hline
\end{tabular} \label{table-massive-noangle-lisa}}
%\end{ruledtabular}
\end{table}

Following the BD case, we improved the analysis of Ref.~\cite{bertibuonanno} by (i) including the spin-spin coupling, (ii) considering binaries with slight eccentricity and (iii) looking at precessing binaries.
%
%
%\subsection{Spin-Aligned Binaries}
%\subsubsection{Pattern-Averaged Analysis}
%
In Table~\ref{table-massive-noangle-lisa} we show the pattern-averaged results for various spin-aligned BH/BH binaries. 
%We also consider four BH/BH binaries in this case: $(10^7+10^7)M_{\odot}, (10^7+10^6)M_{\odot}, (10^6+10^6)M_{\odot}$ and $(10^6+10^5)M_{\odot}$.
%As in Brans-Dicke case, we perform three types of analyses whose results are shown in Table~\ref{table-massive-noangle-lisa}.
%Berti \textit{et al.}~\cite{bertibuonanno} reported that the effect of prior information on the maximum spin is negligible.
%We confirmed that inclusion of prior information changes the results just about a few percent.
%They have also claimed that when they include both $\sigma$ and $\bar{\omega}$ or $\beta_g$ into parameters, they cannot take the inverse of the Fisher matrix properly.
%This is true but, if we include the prior information, the matrix inverse can be performed successfully.
%Hence, we decided to include the prior information.
%This is the reason why the first column of each binary in Table~\ref{table-massive-noangle-lisa} is slightly different from the results shown in Ref.~\cite{bertibuonanno}.  
%
One sees that including both $\sigma$ and $I_e$ into parameters increases the measurement error of $\lambda_g$ only slightly, implying that $\beta_g$ is not so strongly correlated with other parameters.
In the massive gravity case, $\lambda_g$ is more degenerate with $\sigma$ than with $I_e$.
%The reason is the same as in the case of Brans-Dicke.
Unlike in the Brans-Dicke theory, both terms containing $\lambda_g$ and $\sigma$ in the phase $\Psi(f)$ have frequency dependences with positive power-law indices.
Therefore, their correlation is stronger than the one between $\lambda_g$ and $I_e$.
Notice that binaries with larger masses place stronger constraint on $\lambda_g$. 
This is because the correction to $v_{\mathrm{ph}}$ is larger when the frequency is lower.

%When the graviton has a finite mass, its phase velocity is given in Eq.~\eqref{vphase}. 
%expressed as~\cite{will1998}
%
%Therefore, at a lower frequency the difference between $v_{\mathrm{ph}}$ and the speed of light ($c=1$) is larger.
%This is why the heavier binaries put more stringent constraint on $\lambda_g$.

%\subsubsection{Monte Carlo Simulations}

\begin{table}
\tbl{The results of the Monte Carlo simulations in massive gravity theories for $(10^7+10^6)M_{\odot}$ BH/BH binaries at 3Gpc without pattern averaging using LISA~{\cite{yagiLISA}}. We used two interferometers for the analyses. As in Brans-Dicke case, we distribute $10^4$ binaries, calculate the error of each parameter for each binary and take the average. The first half of the table shows the results for the spin-aligned cases, and the second half represents the ones for the precessing binaries. The first line of each part shows the results without taking $I_e$ into parameters, while in the second line this is taken into account. $\sigma$ is included in the parameters for all the cases.
(This table is taken from Ref.~{\cite{yagiLISA}}.)}
%\begin{ruledtabular}
{\begin{tabular}{c||cccccccc}  \hline\hline
 Cases & $\lambda_g$ & SNR & $\Delta \ln\mathcal{M}$  & 
                    $\Delta\ln \eta$ & $\Delta \beta $ &
                    $\Delta \ln D_L$ & $\Delta \Omega_S$ & $\Delta \sigma$  \\ %\hline\hline
 & ($10^{21}$cm) & & ($\%$) & & & ($\%$) & ($10^{-4}$str) & \\ \hline\hline 
\textbf{Spin-Aligned} & \multicolumn{7}{l}{ } \\
Excluding $I_e$ & 0.40598 & 1540 & 0.0507 & 0.841 & 6.27 & 2.30 & 0.957 &  1.89 \\ 
Including $I_e$ & 0.40507 & 1540 & 0.191 & 0.927 & 6.54 & 2.33 & 0.972 &  1.94 \\ \hline
\textbf{Precessing} & \multicolumn{7}{l}{ } \\
Excluding $I_e$ & 4.8540 & 1596 & 0.00838 & 0.00675 & 0.0117 & 0.189 &  0.366 & 0.0508 \\
Including $I_e$ & 3.0570 & 1586 & 0.0269 & 0.00708 & 0.0120 & 0.192 & 0.364 &  0.0825 \\ \hline\hline
\end{tabular} \label{table-corr}}
%\end{ruledtabular}
\end{table}

\begin{figure}[htbp]
  \centerline{\includegraphics[scale=.4,clip]{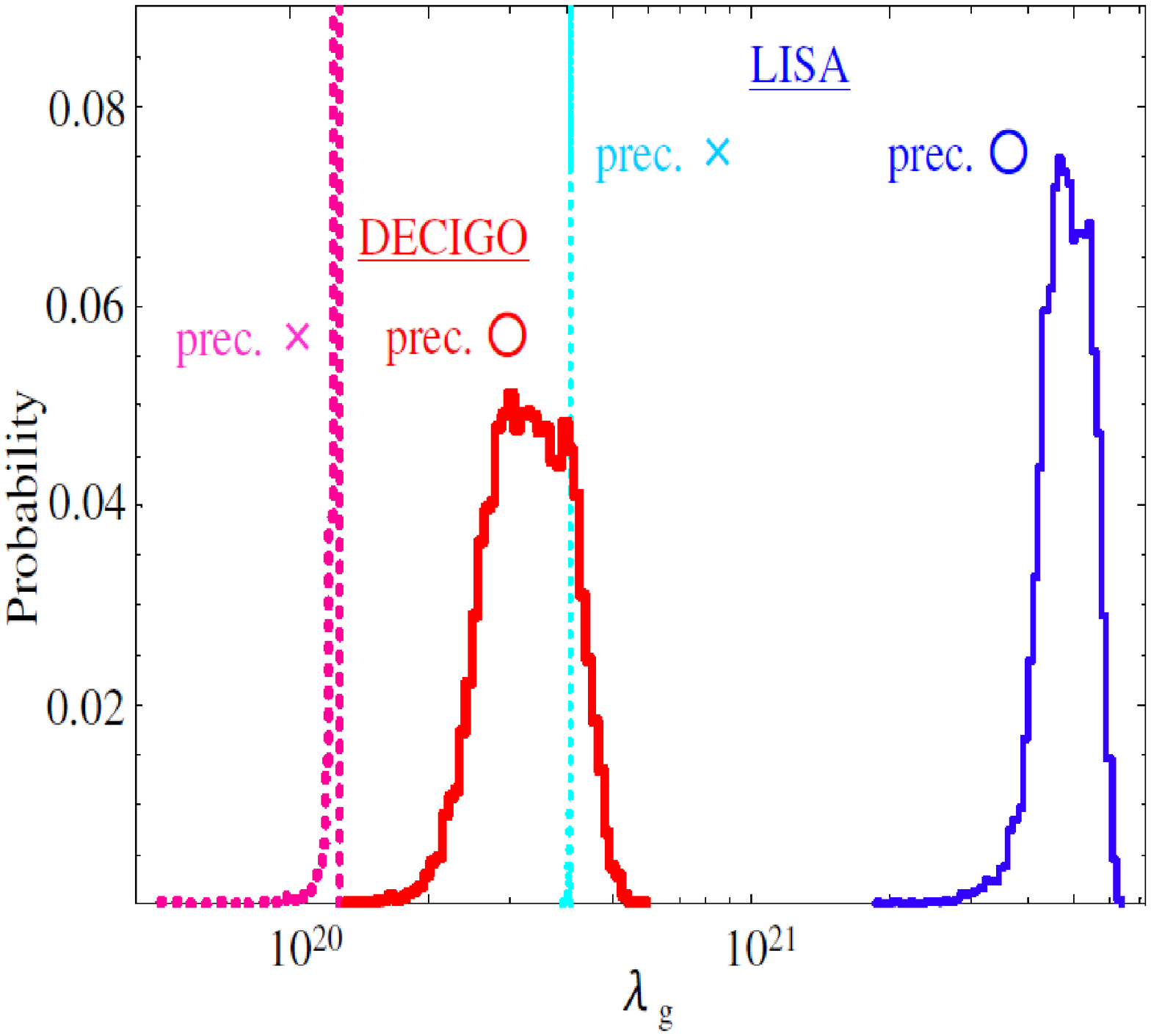} }
 \caption{\label{massive} The histograms showing the probability
 distributions of the lower bounds of $\lambda_g$ obtained from our Monte
 Carlo simulations of $10^4$ BH/BH binaries in massive gravity
 theories~{\cite{yagiLISA,yagiDECIGO}}. We take the masses of the binaries as $(10^7+10^6)M_{\odot}$ for LISA and $(10^6+10^5)M_{\odot}$ for DECIGO/BBO, both placed at the luminosity distance of 3Gpc.
The meaning of each histogram is the same as in Fig.~\ref{bd}.
(This figure is taken from Ref.~{\cite{yagiDECIGO}}.)}
\end{figure}

Next, we show in Table~\ref{table-corr} the results of Monte Carlo simulations.
The upper half shows the ones for the spin-aligned binaries.
We fix the binary mass to $(10^7 +10^6)M_{\odot}$, and the distance to $D_L=3$Gpc.
%As in Brans-Dicke case, the first row represents the error estimation without including $I_e$ into binary parameters and the second row shows the one including $I_e$.
The corresponding histogram (only for circular binaries) for the probability distribution of $\lambda_g$ with LISA is shown as the (light blue) thin dotted one in Fig.~\ref{massive}.
When including higher PN terms, the Fourier transform of the GW waveform for a circular binary can be expressed as
\ba
\Psi (f) &\equiv & 2\pi ft_0-\phi_0 -\frac{\pi}{4}+\Psi_\mrm{0PN}(f) \biggl[1 -\frac{128}{3} \beta_g \eta^{2/5} x + \left( \frac{3715}{756}+\frac{55}{9}\eta \right)x  \nn \\
               &  &
                  -4(4\pi-\beta)x^{3/2} 
           + \left( \frac{15293365}{508032}+\frac{27145}{504}\eta+\frac{3085}{72}\eta^2-10\sigma \right) x^2  \biggr]\,,
           \label{Psi-PPE}
\ea
with $x\equiv (\pi M_t f)^{2/3}$.
Although there are 5 parameters ($\mathcal{M},\eta,\beta,\sigma$, and $\beta_g$) inside the square brackets, there are only 4 PN terms in total (the leading quadrupole term, 1PN, 1.5PN, and 2PN), meaning that these parameters degenerate for the spin-aligned binaries.
Since the level of uncertainty of $\sigma$ is determined completely from the prior distribution, the uncertainty in $\beta_g$ is also determined by the one in $\sigma$, and hence there is a sharp peak in the histogram in Fig.~\ref{massive}.

%\subsection{Precessing Binaries}

On the other hand, the lower bounds on $\lambda_g$ for precessing binaries are shown in the second half of Table~\ref{table-corr}. 
We fix the masses of the binary component to $(10^7+10^6) M_{\odot}$ and assume the BH spins to be 0.5 and 0.
%As in the case of Brans-Dicke theory, we take the spins of heavier black holes to be $\chi=0.5$. 
Since one of the BH spins is 0, the situation reduces to the simple precession.
%Under these assumptions, we can apply the simple precession approximation. 
%The third row in this table represents the results without taking eccentricity $I_e$ into parameters and the fourth row shows the ones including $I_e$.
The corresponding histogram is shown as the (blue) thin solid one in Figs.~\ref{massive}. 
%The (purple) dotted thick  line represents the estimation without taking $I_e$ into parameters and the (red) solid thick one is the estimation with $I_e$.
One sees that the lower bounds on $\lambda_g$ increase by one order of magnitude for the precessing binaries.
%, showing the importance of including the effect of precession. % compared to the spin-aligned ones.
The lower tail in the histogram corresponds to the binaries with $\kappa \sim 1$, i.e. the binaries are almost spin-aligned.

Notice the impact that the precession has on massive gravity theories compared to Brans-Dicke theory.
As explained above, up to 2PN order, there are not enough number of PN terms in GW phase to determine all of binary parameters.
However, the spin precession brings additional information that helps in determining the spins, which solves the degeneracies between spins and other parameters.
Therefore, inclusion of spin precession is crucial when constraining massive gravity theories. 
On the other hand, there are enough number of PN terms contained in Brans-Dicke theory, so that the effect of the precession is relatively weak.

%%%%%%%%%%%%%%%%%%%%%%%%%%%%%%%%%%%%%%%%%%%%%%%%%%%%%%%%%%%%%
%\subsubsection{Subsequent Works}
%\label{app:subsequentMG}

Just before we completed our paper~{\cite{yagiLISA}}, similar paper by Stavridis and Will~\cite{stavridis} appeared.
Although we assumed the simple precession which can be applied only to binaries with equal-mass components or one of the spins are zero, they numerically solved precession equations so that their estimate is not limited to such binaries.
For precessing binaries of $(10^6+10^7)M_\odot$ at 3Gpc, LISA can place the bound $\lambda_g > 5\times 10^{21}$cm which is slightly larger than our estimate.
They found an interesting result that this constraint is almost the same as the one without spins, suggesting that precession makes the spin and other parameters degenerate.
Very recently, calculations including both precessions and higher harmonics has been performed by Huwyler \et~{\cite{huwyler}}.
Their constraint on $\lambda_g$ is comparable to the one by Stavridis and Will.
It seems that the precessions already solves the degeneracies between $\lambda_g$ and other binary parameters so that the effect of higher harmonics is not so strong. 

Keppel and Ajith~\cite{keppel} calculated the bound including merger and ringdown information for the first time. 
They used the phenomenological inspiral-merger-ringdown waveform~\cite{ajith,ajithspin} (see Sec.~\ref{hybrid}).
They performed the pattern-averaged analysis and obtained rather stringent bound $\lambda_g > 5.9 \times 10^{22}$cm for the total mass of $M_t=4.8 \times 10^7M_\odot$ equal-mass binary at 3Gpc with LISA, and $\lambda_g > 7.8 \times 10^{18}$cm for the total mass of $M_t=360 M_\odot$ equal-mass binary at 1Gpc with adv.~LIGO.
The latter result shows that the second-generation ground-based interferometers can put the bound a few times stronger than the current solar system one.
Although their analysis does not include spins, calculations by Stavridis and Will implies that the results should not change much when one includes spins and precession effects.

Instead of using Fisher analyses, Del Pozzo \et~\cite{delpozzo} performed Bayesian analyses and found that the second-generation ground-based interferometers can put the bound $\lambda_g > 2.6 \times 10^{18}$cm using 50 inspiral signals.
Similar Bayesian analysis has been performed by Cornish \et~\cite{cornish-PPE} for model-independent test of modified gravities applying PPE formalism. 
Their constraint is consistent with the one found by Del Pozzo \et~{\cite{delpozzo}}.

Most of the works mentioned above assumes a binary of masses $M_t\sim 10^6M_\odot$ at distances $\sim 3$Gpc for the constraint with LISA.
However,  BH merger tree simulations show that the typical binaries that are observable with LISA have smaller masses with larger distances~{\cite{volonteri,begelman,sesana-merger}}.
Recently, Berti \et~\cite{bertisesana} performed 1000 realizations of merger simulations for two BH formation scenarios, and found that for each realization, there are about 40 BH binary mergers that can be observed with LISA.
By combining signals from these detectable binaries, they found the constraint $\lambda_g > 6.5$--$7.5 \times 10^{21}$cm with LISA.
They also estimated the bound using newly-proposed LISA (eLISA) and found that the bounds are reduced roughly by a factor of two. 
These results are summarized in the lower halves of Tables~\ref{table-mg-previous1} and~\ref{table-mg-previous2}.

Apart from GW observations using interferometers, there are works that propose bounds on $\lambda_g$ using pulsar timings.
Baskaran \et~\cite{baskaran} considered the so-called \textit{surfing effect}~\cite{polnarev} which is the resonance effect arising when the speeds of light and GWs are unequal.
From current pulsar timing observation of PRS B1937+21 with the observation period $T_\mrm{obs}=4$yr and the timing residual $R_\mrm{rms} = 0.1 \mu \mrm{sec}$, they obtained the bound $\lambda_g > 1.5 \times 10^{19}$cm, assuming that the energy density of the SMBH stochastic GW background is~\cite{sesana} $\Omega_\mrm{GW}(T_\mrm{obs}^{-1}) = 4.2 \times 10^{-10}$ and the spectral tilt~\cite{wyithe} $n_T=2/3$.  
Due to this assumption, this bound is less robust compared to the solar system experiment.
By assuming the future 10-yr pulsar timing arrays of 300 pulsars with 100ns timing accuracy, Lee \et~\cite{lee} estimated the effect of massive graviton on the correlation between pulsar signals and obtained $\lambda_g > 4.1 \times 10^{18}$cm.
Also, the graviton mass can be investigated by CMB observations. 
For certain ranges of $\lambda_g$, the graviton mass has a remarkable effect on the large scale CMB power spectrum (for both temperature fluctuations and polarizations)~{\cite{bessada,dubovsky-cmb}}. 
The results mentioned in this paragraph are also included in Table~\ref{table-mg-previous3}.

%%%%%%%%%%%%%%%%%%%%%%%%%%%%%%%%%%%%%%%%%%%%%%%%%%%%%%%%%%
\subsubsection{Proposed Constraints with DECIGO/BBO}
\label{num-MG-DECIGO}
%\subsection{Spin-Aligned Case}

%\subsection{Setup}

In this section, we show the results using DECIGO/BBO.
Following Sec.~\ref{num-BD-DECIGO}, we have adopted DECIGO instrumental noise rather than the BBO one.
In Table~\ref{table-massive-noangle}, we show the pattern-averaged results for various spin-aligned BH/BH binaries.
As discussed in the LISA section, since the constraints would be stronger for larger mass binaries, LISA performs better than DECIGO in constraining massive gravity theories.
%In this case, stronger constraints on $\lambda_g$ are realized for larger mass binaries, with the same reason discussed in the LISA case.
%This is explained from the correction to the phase velocity $v_{ph}$ which is given as~\cite{yagiLISA}
%
%\begin{equation}
%v_{\mathrm{ph}}^2=\left( 1-\frac{1}{f^2\lambda_g^2}\right)^{-1}.
%\end{equation}  
%Since more massive binaries generate lower frequency gravitational waves, these binaries give larger corrections and make the constraints stronger.
(The reason a $(10^6+10^5)M_{\odot}$ binary gives stronger constraint than a $(10^6+10^6)M_{\odot}$ binary is because the GW signal shifts out of the observation band as we increase the mass.)

\begin{table}
\tbl{The results of error estimation in
 massive gravity theories for spin-aligned BH/BH binaries at 3 Gpc with various masses using DECIGO/BBO. We performed 
pattern-averaged analyses using only one
 interferometer.
(This table is taken from Ref.~{\cite{yagiDECIGO}}.)}
%%\begin{ruledtabular}
%\begin{center}
{\begin{tabular}{c||cccccccc}  \hline\hline
 masses & $f_{\mathrm{in}}$ & $f_{\mathrm{fin}}$ & SNR & $\lambda_g$ &  $\Delta \ln\mathcal{M}$ & $\Delta\ln \eta$ & $\Delta \beta $ & $\Delta \sigma$  \\ 
 &  (mHz) &  (mHz) & & $(10^{20}\mathrm{cm})$ & $(\%)$ &  & & \\ \hline
% &  & ($\%$) & & & & ($10^{-3}$str) & \\ \hline\hline 
%\multicolumn{6}{l}{(1.4+400)$M_{\odot}$} \\
$(10^6+10^6)M_{\odot}$ & 1.0 & 2.20 & 1338 & 1.014 & 14.0 & 2.46 & 9.40 & 2.50 \\
$(10^6+10^5)M_{\odot}$ & 1.0 & 4.00 & 2044 & 1.270 & 1.19 & 1.69 & 9.10 & 2.44 \\ 
$(10^5+10^5)M_{\odot}$ & 1.0 & 22.0 & 4909 & 1.133 & 0.0286 & 0.930 & 7.00 & 1.69  \\
$(10^5+10^4)M_{\odot}$ & 1.0 & 40.0 & 3021 & 0.4066 & 4.51$\times 10^{-3}$ & 0.823 & 6.20 & 1.88  \\
$(10^4+10^4)M_{\odot}$ & 1.0 & 220.0 & 29569 & 0.3852 & 3.54 $\times 10^{-4}$ & 0.924 & 6.96 & 1.68  \\ \hline\hline
\end{tabular} \label{table-massive-noangle}}
%\end{center}
%%\end{ruledtabular}
\end{table}

%\subsubsection{Monte Carlo Simulations}

%\subsection{Precessing Case}

\begin{table}
\tbl{The results of error estimation in
 massive gravity theories for $(10^6+10^5)M_{\odot}$ BH/BH binaries at
 3Gpc using DECIGO/BBO. The meaning of each line is the same as in
 Table~\ref{table-bd-noprec}.
(This table is taken from Ref.~{\cite{yagiDECIGO}}.)}
%%\begin{ruledtabular}
%\begin{center}
{\begin{tabular}{c||ccccccc}  \hline\hline
 cases & SNR & $\lambda_g$ & $\Delta \ln\mathcal{M}$  & 
                    $\Delta\ln \eta$ & $\Delta \beta $ & $\Delta \sigma$ &
                     $\Delta \Omega_S$   \\ 
 & & $(10^{20}\mathrm{cm})$  & $(\%)$ & &  & & (str) \\ \hline
% & ($10^{21}$cm) & & ($\%$) & & & & ($10^{-4}$str) & \\ \hline\hline 
pattern-averaged & 2044 & 1.270  & 1.19 & 1.69 & 9.10 & 2.44 & - \\ 
no precession & 2601  & 1.266 & 1.16 & 1.64 & 8.92 &  2.40 & 1.16  \\ 
including precession & 2666 & 3.349 & 0.314 & 0.0388 & 0.0612 & 0.529 &  0.0248 \\ \hline\hline
\end{tabular} \label{table-massive-noprec}}
%\end{center}
%%\end{ruledtabular}
\end{table}

In Table~\ref{table-massive-noprec}, we show the results of binary parameter estimation in massive gravity theories for both pattern-averaged analysis and Monte Carlo simulations with and without including spin precession.
We assume $(10^6+10^5)M_{\odot}$ BH/BH binaries at $D_L=3$Gpc.
%We have chosen an optimized mass parameter (see Table~\ref{table-massive-noangle}).
The meaning of each row is the same as in Table~\ref{table-bd-noprec}.
One sees that the constraint on $\lambda_g$ becomes twice stronger when we include precession.
The corresponding histograms are shown in Fig.~\ref{massive}.
%Figure~\ref{massive} represents the histograms showing the number fraction of binaries that give the constraint of each $\lambda_g$. 
%For comparison, we also show the results for $(10^7+10^6)M_{\odot}$ BH/BH binaries at 3 Gpc using LISA.
%SNRs for the GWs from these binaries correspond to 1600.
%The meaning of each histogram is the same as in Fig.~\ref{bd}.
Notice that the effect of precession is larger for LISA.
We expect that this is due to the wider effective frequency range.
% is wider for $(10^7, 10^6)M_{\odot}$ BH/BH binaries with LISA. 
The proposed lower bound on $\lambda_g$ with DECIGO/BBO is $\lambda_g > 3.35\times 10^{20}$cm.
Although this is one order of magnitude weaker than the LISA one, it is still three orders of magnitude stronger than the current solar system bound~{\cite{talmadge}}.

%%%%%%%%%%%%%%%%%%%%%%%%%%%%%%%%%%%%%%%%%%%%%%%%%%%%%%%%%%
\subsubsection{Proposed Constraints with DPF}
\label{num-MG-DPF}

%We include $\beta_g$ to the binary parameters and perform Fisher analyses to estimate how accurately we can determine binary parameters (especially $\beta_g$) with DPF observations.
%Then, we convert the upper bound on $\beta_g$ to the lower bound on $\lambda_g$.
In this section, we show the results obtained using DPF.
First, we give a rough estimate on how strong we can constrain $\lambda_g$ with DPF. 
It is not possible to detect the effect of massive gravity if the correction term in Eq.~(\ref{phase-massive}) is smaller than $\rho^{-1}$.
This leads to the constraint
\beq
\lambda_g \geq 6.6 \times 10^{17} \mrm{cm} \lmk \frac{D}{5\mrm{kpc}} \rmk^{1/2} \lmk \frac{f}{0.1\mrm{Hz}} \rmk^{-1/2} \lmk \frac{\mrm{SNR}}{30} \rmk^{1/2}. \label{rough_g}
\eeq

\begin{figure}[t]
  \centerline{\includegraphics[scale=1.3,clip]{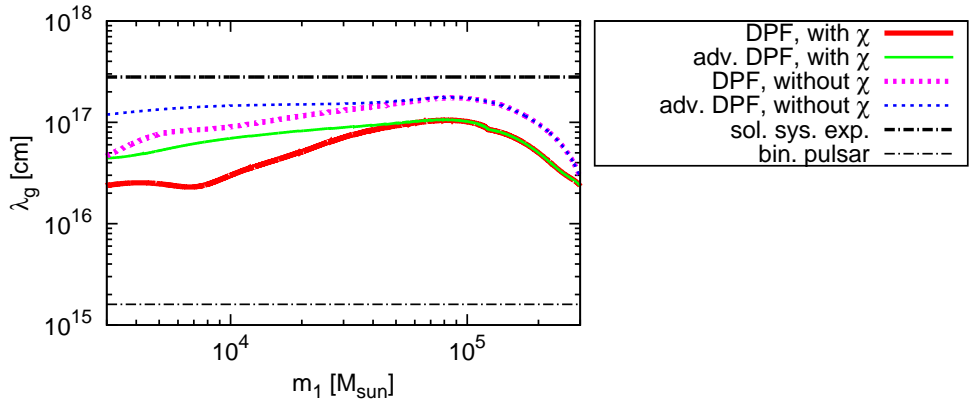} }
\vspace*{8pt}
 \caption{\label{lambda}
The lower bounds on $\lambda_g$ using DPF with (red thick solid) and without (magenta thick dotted) taking $\chi$ into parameters,  and using adv.~DPF with (green thin solid) and without (blue thin dotted) taking $\chi$ into parameters.
We assumed spin-aligned, equal-mass binaries with $\chi=0.2$ at $D_L=5$kpc.
We also assumed that the orbits are circular.
The (black) thin dotted horizontal line at $\lambda_g=2.8\times 10^{17}$cm represents the bound obtained from solar system experiment~\cite{talmadge} while the (black) thick dotted-dashed horizontal line at $\lambda_g=1.6\times 10^{15}$cm shows the one obtained from binary pulsar test~{\cite{finnsutton}}.
(This figure is taken from Ref.~{\cite{yagiDPF}}.)
}
\end{figure}

Next, we estimate the constraint numerically using Fisher analysis for a spin-aligned binary using inspiral-merger-ringdown hybrid waveform.
In Fig.~\ref{lambda}, we show the sky-averaged results for the lower bound on $\lambda_g$ against various BH masses, assuming spin-aligned, equal-mass binaries with circular orbits, with $\chi=0.2$ at $D_L=5$kpc. 
%The results using DPF are shown in the (red) thick solid curve.
For a $(10^4+10^4)\so$ binary, we find the constraint of $\lambda_g \geq 6\times 10^{16}$cm using DPF, which is weaker than our rough estimate in Eq.~(\ref{rough_g}).
This is due to the degeneracies between $\beta_g$ and other binary parameters. 
In Fig.~\ref{lambda}, we also show the constraints without including $\chi$ into variable parameters (i.e. assuming that $\chi$ is know \textit{a priori}).
This roughly gives how well one can constrain the theory when the binary is precessing~{\cite{stavridis}}.
This corresponds to the case where the degeneracies between $\lambda_g$ and $\chi$ are completely disentangled.
The dotted-dashed horizontal line at $\lambda_g =2.8\times10^{17}$cm represents the (static) lower bound obtained from the solar system experiment~{\cite{talmadge}}.
Although DPF constraint is slightly weaker, they are still meaningful since DPF measures the deviation in the propagation speed of GWs that appear in gravitational waveform phase, while the solar system experiment measures the deviation in the effective gravitational constant (or in the Kepler's third law).  % for some binary parameters.  
We find that DPF can place about 2 orders of magnitude stronger (dynamical) constraint than the one from the binary pulsar~\cite{finnsutton}  in the \emph{weak}-field regime,  shown as the dotted-dashed horizontal line at $\lambda_g =1.6\times10^{15}$cm.
Furthermore, Finn and Sutton~\cite{finnsutton} assumed Fierz-Pauli-type theory while the constraint obtained here is independent of the specific type of massive gravity theories.
Since the effect of the finite mass of graviton is larger for larger mass binaries, it would be important to reduce the acceleration noises, just like in the case of joint search with DPF and the network of ground-based detectors.

\begin{figure}[t]
  \centerline{\includegraphics[scale=.45,clip]{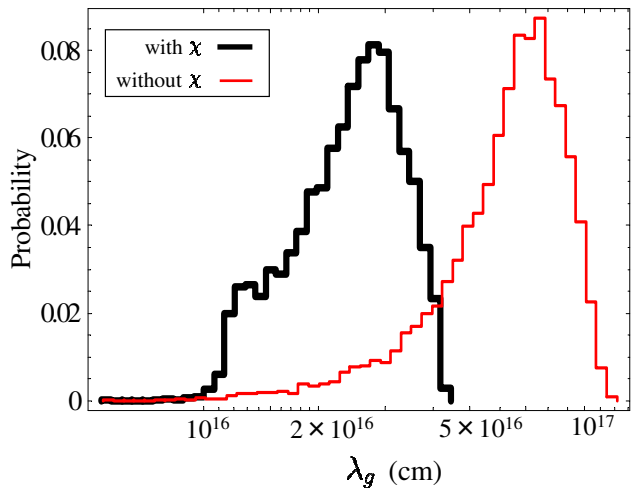} }
\vspace*{8pt}
 \caption{\label{lambda_mc}
Histograms showing the lower bounds on $\lambda_g$ obtained from Monte Carlo simulations with $10^4\so$ equal-mass IMBH binaries with circular orbits of $\chi= 0.2$ in $\omega$ Centauri with random orientation for the orbital angular momenta.
We show here the results using DPF.
The (black) thick and (red) thin histogram each represents the one with and without taking $\chi$ as a variable parameter, respectively. 
(This figure is taken from Ref.~{\cite{yagiDPF}}.) }
\end{figure}

In Fig.~{\ref{lambda_mc}}, we show the probability distribution of the lower bound on $\lambda_g$ with Monte Carlo simulations.
We assume $10^4\so$ equal-mass binaries of $\chi=0.2$ with circular orbits in $\omega$ Centauri. 
One sees that the constraint is now slightly weaker than the one with the sky-averaged analysis.
However, this is still much stronger than the one from the binary pulsar observations. 
These probability distributions shift to larger $\lambda_g$ by roughly $\sqrt{7}=2.6$ when we use adv.~DPF.

\subsubsection{Prospects for Other Space-Borne GW Interferometers}

 \begin{table}
\tbl{The proposed bounds on $\lambda_g$ (in units of $10^{21}$cm) with LISA, eLISA and ASTROD-GW. We assume spin-aligned BH/BH binaries at 3Gpc with a single interferometer and perform pattern-averaged Fisher analysis. We exclude both $\beta$ and $\sigma$ into parameters. (This should mimic the precessing case.) We consider binaries with circular orbits and do not include $I_e$ into parameters.}
%%\begin{ruledtabular}
%\begin{center}
{\begin{tabular}{c||ccc}  \hline\hline
 BH masses & LISA & eLISA  & ASTROD-GW \\ \hline
$(10^7+10^7)M_\odot$ & 6.941 & 1.914 & 12.79  \\
$(10^7+10^6)M_\odot$ & 3.946 &  1.300 & 5.851  \\ 
$(10^6+10^6)M_\odot$ & 4.631 & 2.010 & 5.411  \\ 
$(10^6+10^5)M_\odot$ & 2.743 & 1.330 & 2.492  \\ \hline\hline
\end{tabular} \label{table-massive-elisa}}
%\end{center}
%%\end{ruledtabular}
\end{table}

In this subsection, we explain the proposed bounds on $\lambda_g$ using eLISA and ASTROD-GW.
The bounds using these detectors are shown in Table~\ref{table-massive-elisa}.
As for the BD case, we show the results with LISA for comparison.
Here, we do not include spins into parameters.
The bounds for the precessing case should be similar, as can be understood by comparing Tables~\ref{table-corr} and~\ref{table-massive-elisa}. 
eLISA can place the bounds that are weaker than the LISA ones by a factor of a few. 
This is because eLISA is less sensitive in the lower frequency part.
Among the GW interferometers considered in this paper, ASTROD-GW performs the best in constraining massive gravity theories.
This is because it is the most sensitive in the lower frequency range. 
It can place even stronger constraint if it detects GW signals from a binary with larger BH masses.

%----------------------------
\subsection{Other Theories}

Up to here, we focused on BD and MG theories, but GW observations will allow us to test other theories. One example is quadratic gravity~\cite{yunes-stein} in which general quadratic curvature terms coupled to a scalar field is added to Einstein-Hilbert term at the level of action, together with the kinetic and the potential terms of the scalar field. This theory includes Einstein-Dilaton-Gauss-Bonnet (EDGB)~\cite{EDGB} theory in the even parity sector and dynamical Chern-Simons (CS)~\cite{jackiw,CSreview} gravity in the odd parity sector as specific examples, which are motivated from e.g. superstring theory~{\cite{polchinski1,polchinski2}}. The gravitational and scalar radiation energy flux in quadratic gravity are calculated in Ref.~\cite{quadratic} within the post-Newtonian approach. In the even parity sector, the scalar dipole radiation exists, just like in BD theory, and this gives -1PN dissipative correction to the gravitational waveform phase relative to GR. This dominates the 0PN conservative correction coming from the scalar interaction. In Ref.~{\cite{kent-LMXB}}, the current strongest bound on EDGB theory is obtained from the orbital decay rate of the low-mass X-ray binary A0620-00~{\cite{A0620-00}}, and possible future constraints using GW interferometers are also estimated. Ground-based detectors are not so useful in constraining the theory, but eLISA should be able to place comparable bound and DECIGO/BBO can place 100 times stronger constraint compared to the current bound.

\begin{figure}[htb]
\begin{center}
\begin{tabular}{l}
\includegraphics[width=6.5cm,clip=true]{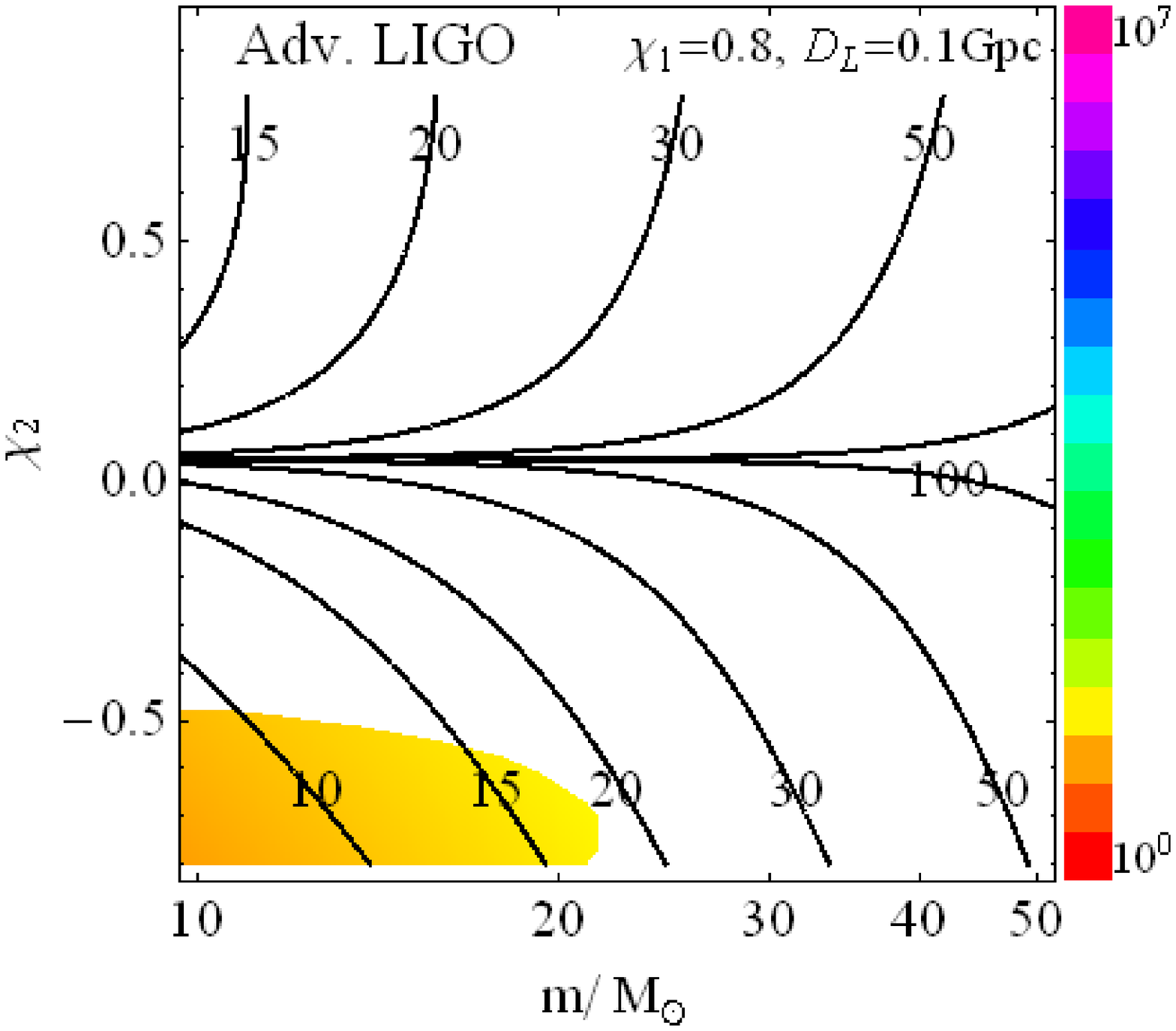}  
\includegraphics[width=6.5cm,clip=true]{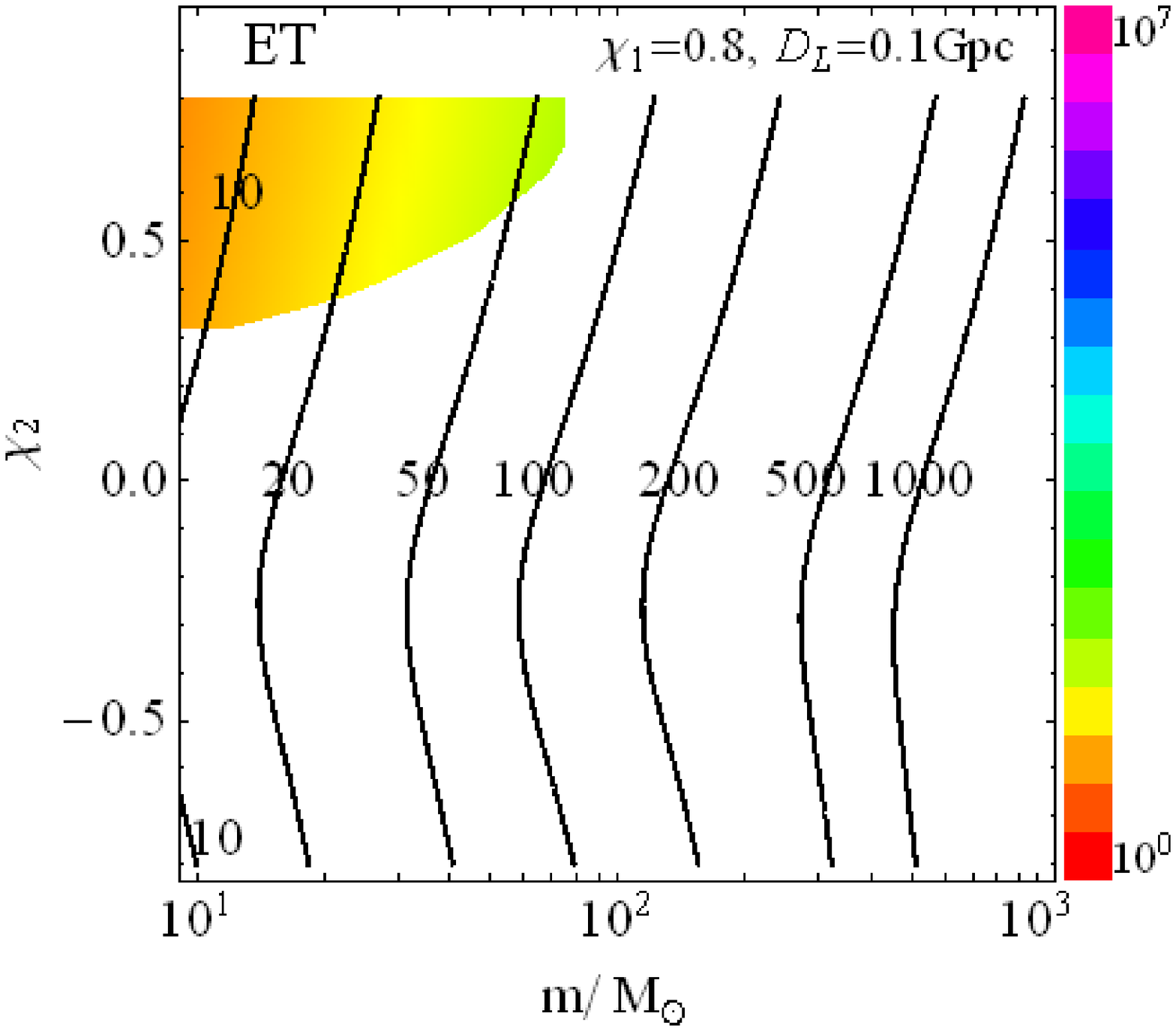} \\
\includegraphics[width=6.5cm,clip=true]{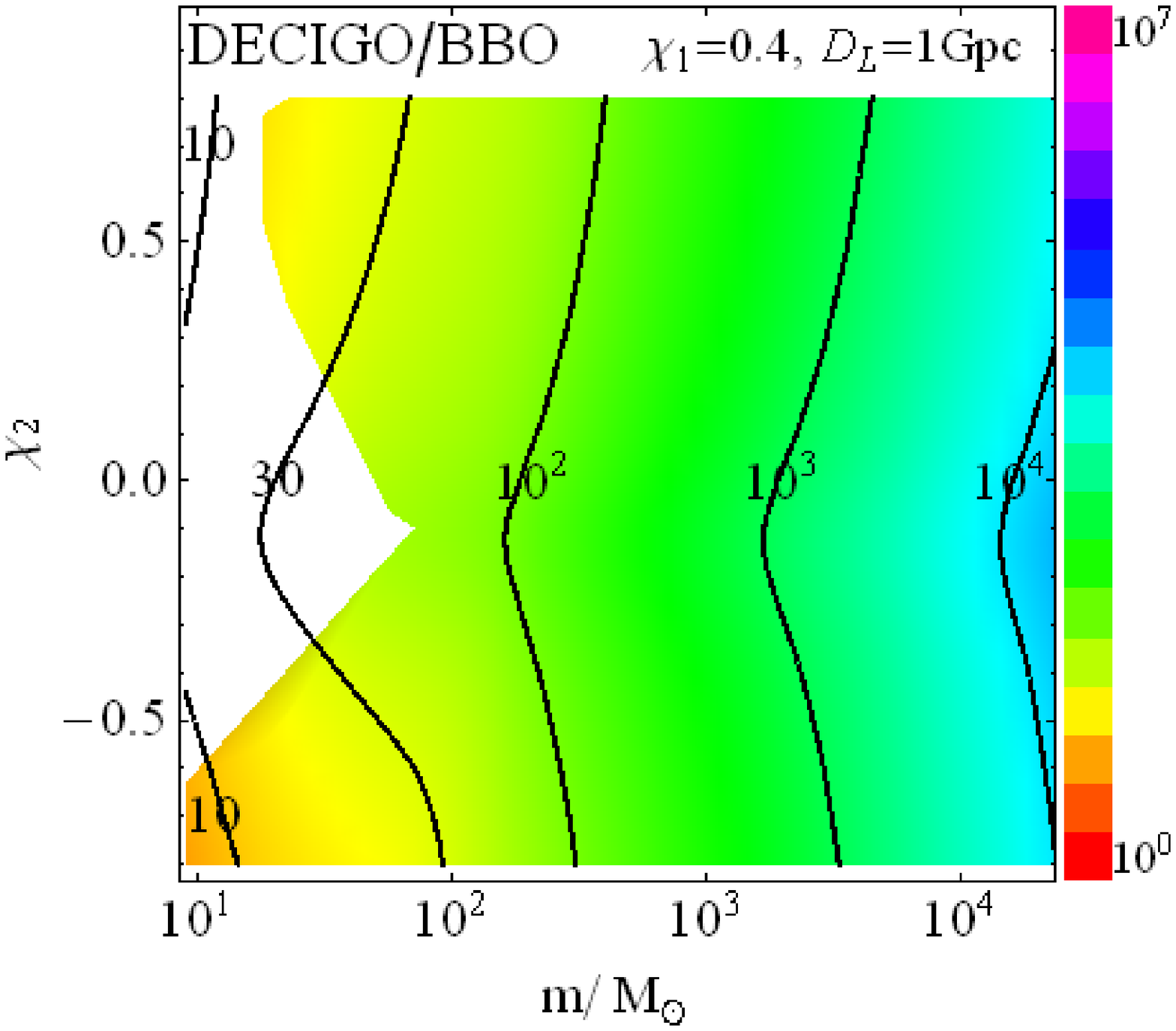}  
\includegraphics[width=6.5cm,clip=true]{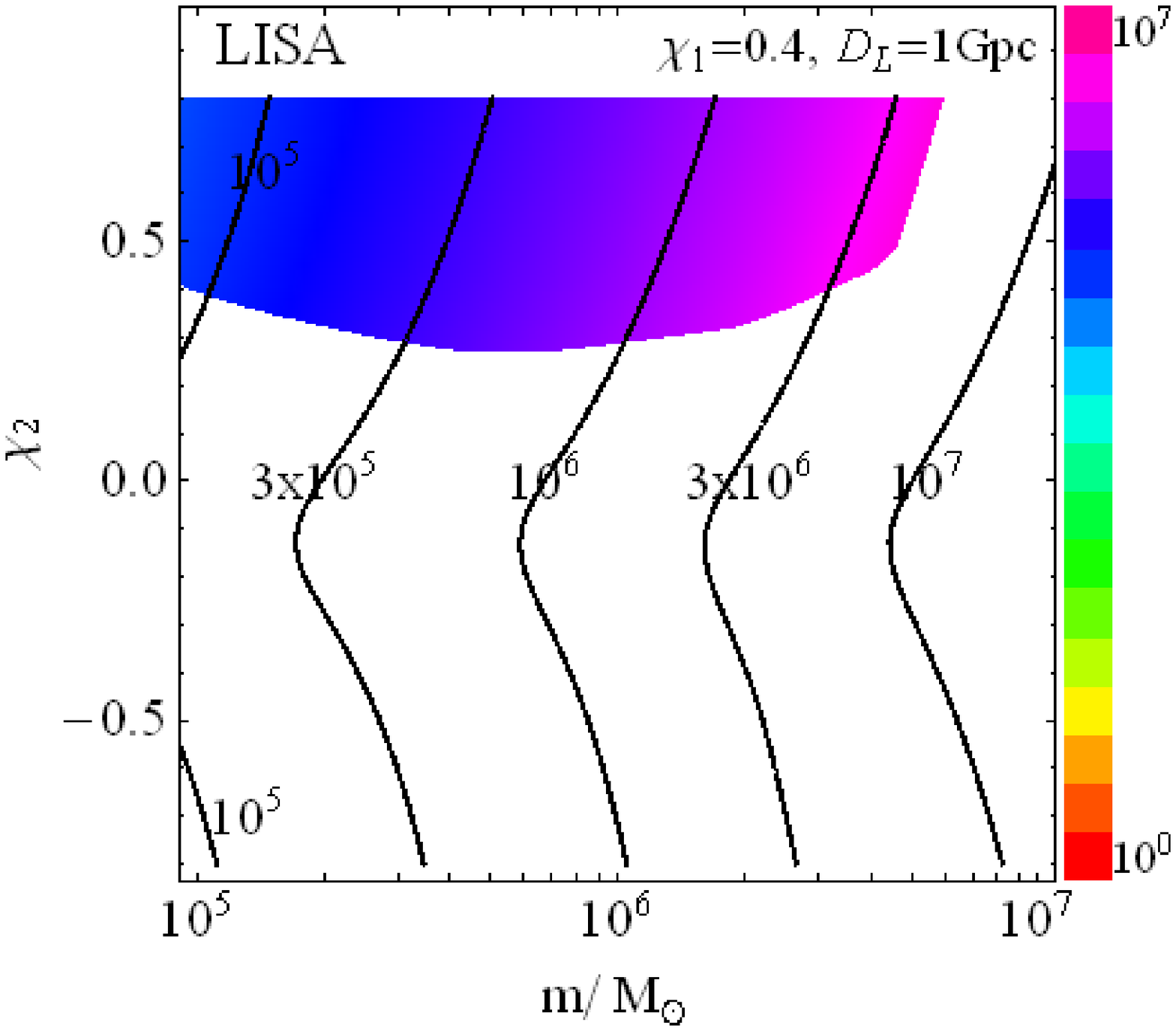} 
\end{tabular}
\caption{\label{fig:CS_constraints} 
Projected $1\sigma$ upper bounds on the characteristic length scale of dynamical CS gravity $\xi^{1/4}$ in km with the second-generation ground-based detectors (top left), ET (top right), DECIGO/BBO (bottom left) and LISA (bottom right) for spinning BH binaries with $m_1/m_2=2$. The fixed values for the dimensionless spin parameter of the primary $\chi_1$ and the luminosity distance $D_L$ are shown at the top right of each panel. The constraints using  the second-generation ground-based detectors have been obtained by assuming that the spins are known \emph{a priori}. This roughly models projected constraints with precessing BH binary observations and may be correct within an error of 30$\%$. For other detectors, binaries are assumed to be spin-aligned (or anti-aligned). The colored contours show the regions of parameter space where the constraints on $\xi^{1/4}$ also satisfy the small coupling approximation (the approximation that we used to derive the gravitaional waveform) at the fiducial luminosity distances chosen. This figure is taken from Ref.~{\cite{kent-CSGW}}.}
\end{center}
\end{figure}

\begin{figure}[htb]
\begin{center}
 \includegraphics[width=8.5cm]{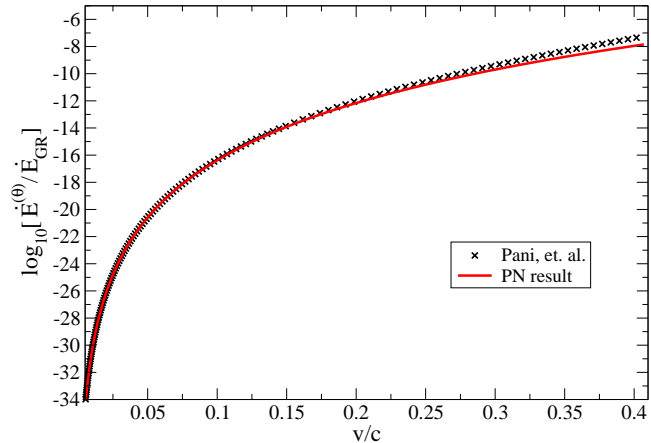}
 \caption{\label{fig:comparison} 
 The scalar radiation energy flux $\dot{E}^{(\vartheta )}$ relative to the GR gravitational one $\dot{E}_\mrm{GR}$ against $v/c$ for a non-spinning BH binary in dynamical CS gravity. The crosses are the ones obtained numerically in Ref.~\cite{pani-DCS-EMRI} using BH perturbation while the solid curve is the analytic result obtained in Ref.~\cite{quadratic} using PN formalism.
Observe that at low velocities, in the regime where the PN approximation is valid, the two curves agree. (This figure is taken from Ref.~{\cite{quadratic}}.)}
 \end{center}
\end{figure}

As for the odd parity sector of quadratic gravity (dynamical CS gravity), both scalar radiation and corrections to the gravitational radiation give 2PN dissipative correction to the gravitational waveform phase~{\cite{quadratic}}. Also in this theory, there is a quadrupole moment deformation in the BH solution valid to quadratic order in spin~\cite{kent-CSBH}, which gives 2PN conservative correction. Another conservative correction at the same PN order also arises from the scalar interaction. Combining all of these 2PN CS effects, we construct gravitational waveform from BH binaries and performed Fisher analysis~{\cite{kent-CSGW}} to obtain the constraints on the characteristic length scale of the theory $\xi^{1/4}$. (The theory reduces to GR in the limit $\xi \to 0$.) We present proposed constraints on the theory with GW observations in Fig.~\ref{fig:CS_constraints}. We found that ground-based detectors and DECIGO/BBO have ability to constrain the theory by 6--7 orders of magnitude stronger than the solar system bound~{\cite{alihaimoud-chen}} of $\xi^{1/4} < \mathcal{O}(10^8)$km. On the other hand, LISA is not as good as these detectors, but still, it should be able to place 2--3 orders of magnitude stronger constraint than the solar system experiments, which is roughly consistent with the results in Refs.~{\cite{sopuerta-yunes-DCS-EMRI,canizares}}. (The difference comes from the fact that Ref.~\cite{kent-CSGW} considers 2PN dissipative and conservative effects at quadratic order in spin while Refs.~{\cite{sopuerta-yunes-DCS-EMRI,canizares}} consider 4.5PN conservative effect at linear order in spin.) If the BHs are not spinning, then the corrections to the scalar and gravitational radiation flux would be of 7PN and 6PN orders respectively. This is first calculated numerically by Pani \textit{et al}.~\cite{pani-DCS-EMRI} using BH perturbation. Later, the scalar radiation correction is confirmed analytically under PN analysis (as shown in Fig.~\ref{fig:comparison}) and the origin of 6PN correction for the gravitational radiation is also discussed~{\cite{quadratic}}. Quasi-normal modes in this theory for non-spinning BHs have been calculated by Molina \textit{et al}.~\cite{molina} and the corrections might be detected by future GW observations.

For other theories, gravitational waves from compact binaries allow us to probe the size of the extra dimension~\cite{mcwilliams,kent-RSII} and the time variation of the gravitational constant $\dot{G}$ \ {\cite{yunes-Gdot}}. GWs from NS oscillations can be used to constrain not only scalar-tensor theories as mentioned above, but also other theories such as tensor-vector-scalar theory~{\cite{sotani-fp-TeVeS,sotani-w-TeVeS,sotani-torsional-TeVeS}}. Direct detections of additional polarizations~\cite{nishizawa-pol1,nishizawa-pol2} would be smoking-guns for the deviation from GR.

%%%%%%%%%%%%%%%%%%%%%%%%%%%%%%%%%%%%%%%%%%%%%%
\section{Conclusions}
\label{sec:conclusions}

Observations complementary to the ground-based GW interferometers can be performed with the space-borne ones. As for the prototype of DECIGO, DPF is aimed to be launched in 2016--2017. The main goals of this satellite are to test the important technologies that are crucial to the space GW mission, and to carry out observations of GWs and Earth gravity. Although the event rate is not so promising, DPF is sensitive enough to detect GWs from galactic IMBH binaries. In this review article, we first showed what information can be achieved if these GWs  are detected. Since the sensitivity frequency range of DPF is lower than the ground-based detectors, DPF can detect inspiral signals earlier than the latter. Therefore, DPF may be able to give an alert to the ground-based detectors so that they can be prepared for observing merger and ringdown phases. Also, DPF can determine binary parameters such as masses and spins to several $\%$ accuracies. On the other hand, the measurement accuracy of the distance and the angular resolution are not so high. 

One very important aspect of GW observations is that they can test alternative theories of gravity in the \emph{strong}-field regime. In this article, we considered extending 2 of the most important characteristics in GR: (i) there are only tensor degrees of freedom and (ii) the graviton is massless (or GWs propagate at the speed of light). We consider BD and MG theories as representatives of the modifications of (i) and (ii), respectively. The former is the simplest example of the scalar-tensor theory. There exists a scalar dipole radiation which changes the evolution of the binary system. This gives a correction to the GW phase at ``-1PN'' order relative to GR. It has a negative PN order correction because scalar radiation  is dipolar while GWs are quadrupolar. This correction is suppressed by the inverse of the BD parameter that describes the strength of the coupling between the scalar and the matter fields. We carried out the Fisher analysis and found that DECIGO/BBO can place 4--5 orders of magnitude stronger constraint compared to the solar system experiment. This constraint is still 1--2 orders of magnitude more stringent than the one from future solar system mission ASTROD~{\cite{ASTROD-I}}. DECIGO/BBO has more advantages than LISA because (i) the number of GW cycles are larger which allows us to perform more accurate test, and (ii) the velocity of the binary constituents is smaller which gives larger dipolar correction. We also found that eLISA and ASTROD-GW would place similar bounds as LISA if we fix the SNRs to be the same. We note that scalar-tensor theories might also be constrained from GWs associated with NS oscillations~{\cite{sotani-fp-ST,sotani-w-ST}}.

As for MG theories, the propagation speed of GWs now becomes smaller than the speed of light, which gives 1PN correction to GW phase. We carried out similar analysis mentioned above and found that in this case, LISA performs better than DECIGO/BBO. This is because the correction to the propagation speed becomes larger for lower frequency GWs. LISA can place 4 orders of magnitude stronger constraint than the solar system bound. DPF may be able to place comparable bound to the solar system one and 2 orders of magnitude stronger constraint than the binary pulsar observations.  Space-borne GW interferometers would supply valuable information on astrophysics and gravity. eLISA would place slightly weaker constraint than LISA, while ASTROD-GW can place the strongest bounds among all the GW interferometers considered in this paper.

One might worry that the magnification effect of the gravitational lensing on GWs~{\cite{holzwald,holzlinder}} might spoil the results shown in this article. However, this effect only affects the amplitude of GWs, and hence, as discussed e.g. in Ref.~{\cite{cutlerholz}}, this increases the measurement errors on the luminosity distance and the angular resolution of the binary. However, since the amplitude and the phase are almost uncorrelated, this effect would not affect the measurement accuracies of the parameters that appear in the phase.

%%%%%%%%%%%%%%%%%%%%%%%%%%%%%%%%%%%%%%%%%%%%%%%%%%%%%%%
\section*{Acknowledgments}

The author would like to thank Bala Iyer, Wei-Tou Ni and K.~G.~Arun for organizing the 5th ASTROD Symposium, and for inviting him to give a talk in the workshop and to write this review article. The author also thanks Masaki Ando for introducing some of his works to Prof. Ni and covering my travel expenses to India.

%%%%%%%%%%%%%%%%%%%%%%%%%%%%%%%%%%%%%%%%%%%%%%%%%%%%%%%%
\appendix

%%%%%%%%%%%%%%%%%%%%%%%%%%%%%%%%%%%%%%%%%%%%%%%%%
\section{Lengthy Formulas}

%%%%%%%%%%%%%%%%%%%%%%%%%%%%%%%%%%%%%%%%%%%%%%%
\subsection{Spin-Aligned Binaries}
\label{app-spin-aligned}

When expressing the observed waveform for the spin-aligned case, we need to express $\theta_{\mathrm{S}}(t)$, $\phi_{\mathrm{S}}(t)$ and $\psi_{\mathrm{S}}(t)$ in terms of the angles $\bar{\theta}_\mrm{S}$, $\bar{\phi}_\mrm{S}$, $\bar{\theta}_\mrm{L}$ and $\bar{\phi}_\mrm{L}$.
The first two are expressed as~\cite{cutler1998,bertibuonanno}
%
%\begin{widetext}
\begin{eqnarray}
\cos \theta_{\mathrm{S}}(t)&=&\frac{1}{2}\cos \bar{\theta_{\mathrm{S}}}
                                          -\frac{\sqrt{3}}{2}\sin \bar{\theta_{\mathrm{S}}} \cos[\bar{\phi}(t)-\bar{\phi_{\mathrm{S}}}],  \\
\phi_{\mathrm{S}}(t)&=&\alpha_1+\frac{\pi}{12}+\tan^{-1}
                              \left( \frac{\sqrt{3}\cos{\bar{\theta_{\mathrm{S}}}}
                              +\sin \bar{\theta_{\mathrm{S}}}\cos [\bar{\phi}(t)-\bar{\phi_{\mathrm{S}}}]}
                              {2\sin \bar{\theta_{\mathrm{S}}}\sin [\bar{\phi}(t)-\bar{\phi_{\mathrm{S}}}]} \right).
\end{eqnarray}
%\end{widetext}
For the polarization angle $\psi_{\mathrm{S}}(t)$ (see Eq.~(\ref{tanpsi})), first $\hat{\bm{z}}\cdot\hat{\bm{N}}=\cos\theta_{\mathrm{S}}$.
Next when we neglect the spin precession effects, $\hat{\bm{L}}$ is constant and $\hat{\bm{L}}\cdot\hat{\bm{z}}$, $\hat{\bm{L}}\cdot\hat{\bm{N}}$, and $\hat{\bm{N}}\cdot(\hat{\bm{L}}\times\hat{\bm{z}})$ are given as~\cite{cutler1998,bertibuonanno}
%
%\begin{widetext}
\begin{eqnarray}
\hat{\bm{L}}\cdot\hat{\bm{z}}&=&\frac{1}{2}\cos\bar{\theta}_{\mathrm{L}}
                   -\frac{\sqrt{3}}{2}\sin\bar{\theta}_{\mathrm{L}} \cos[\bar{\phi}(t)-\bar{\phi}_{\mathrm{L}}], \label{lz} \\
\hat{\bm{L}}\cdot\hat{\bm{N}}&=&\cos\bar{\theta}_{\mathrm{L}}\cos\bar{\theta}_{\mathrm{S}}
                  +\sin\bar{\theta}_{\mathrm{L}}\sin\bar{\theta}_{\mathrm{S}}\cos(\bar{\phi}_{\mathrm{L}}-\bar{\phi}_{\mathrm{S}}), \label{ln} \\
\hat{\bm{N}}\cdot(\hat{\bm{L}}\times\hat{\bm{z}})&=&\frac{1}{2}\sin\bar{\theta}_{\mathrm{L}}\sin\bar{\theta}_{\mathrm{S}}
                                             \sin(\bar{\phi}_{\mathrm{L}}-\bar{\phi}_{\mathrm{S}}) \notag \\
                                             &&-\frac{\sqrt{3}}{2}\cos\bar{\phi}(t)(
                                             \cos\bar{\theta}_{\mathrm{L}}\sin\bar{\theta}_{\mathrm{S}}\sin \bar{\phi}_{\mathrm{S}} 
                                             -\cos\bar{\theta}_{\mathrm{S}}\sin\bar{\theta}_{\mathrm{L}}\sin \bar{\phi}_{\mathrm{L}} ) \notag \\
                                             &&-\frac{\sqrt{3}}{2}\sin\bar{\phi}(t)(
                                             \cos\bar{\theta}_{\mathrm{S}}\sin\bar{\theta}_{\mathrm{L}}\cos \bar{\phi}_{\mathrm{L}} 
                                             -\cos\bar{\theta}_{\mathrm{L}}\sin\bar{\theta}_{\mathrm{S}}\cos \bar{\phi}_{\mathrm{S}} ). \label{nlz}
\end{eqnarray}
%\end{widetext}

%%%%%%%%%%%%%%%%%%%%%%%%%%%%%%%%%%%%%%%%%%%%%%
\subsection{Precessing Binaries}
\label{app-prec}

The precession equations for circular orbit binaries are~\cite{apostolatos} 
%
%\begin{widetext}
\begin{eqnarray}
\dot{\bm{L}}&=&\frac{1}{a^3}\left[ \frac{4m_1+3m_2}{2m_1}\bm{S}_1+\frac{4m_2+3m_1}{2m_2}\bm{S}_2 \right] 
                           \times\bm{L} \notag \\
                          && -\frac{3}{2}\frac{1}{a^3}[(\bm{S}_2\cdot\hat{\bm{L}})\bm{S}_1
                            +(\bm{S}_1\cdot\hat{\bm{L}})\bm{S}_2]\times\hat{\bm{L}}-\frac{32}{5}\frac{\mu^2}{a}
                            \left( \frac{M_t}{a} \right)^{5/2} \hat{\bm{L}},  \label{Lprec} \\
\dot{\bm{S}}_1&=&\frac{1}{a^3}\left[ \frac{4m_1+3m_2}{2m_1}(\mu \sqrt{M_t a}\hat{\bm{L}}) \right]\times \bm{S}_1
                              +\frac{1}{a^3}\left[ \frac{1}{2}\bm{S}_2-\frac{3}{2}(\bm{S}_2\cdot\hat{\bm{L}})\hat{\bm{L}} \right]
                              \times\bm{S}_1, \label{S1prec} \\ 
\dot{\bm{S}}_2&=&\frac{1}{a^3}\left[ \frac{4m_2+3m_1}{2m_2}(\mu \sqrt{M_t a}\hat{\bm{L}}) \right]\times \bm{S}_2
                              +\frac{1}{a^3}\left[ \frac{1}{2}\bm{S}_1-\frac{3}{2}(\bm{S}_1\cdot\hat{\bm{L}})\hat{\bm{L}} \right]
                              \times\bm{S}_2. \label{S2prec} 
\end{eqnarray}
%\end{widetext}
The first term of each equation represents the spin-orbit interactions and the second term represents the spin-spin interactions.
The last term of Eq.~(\ref{Lprec}) is the angular momentum loss due to the radiation reaction.
This changes the total angular momentum $\bm{J}\equiv \bm{L}+\bm{S}_1+\bm{S}_2$ as
\begin{equation}
\dot{\bm{J}}=-\frac{32}{5}\frac{\mu^2}{a}\left( \frac{M_t}{a} \right)^{5/2} \hat{\bm{L}}.
\end{equation}

From Eq.~(\ref{L-prec}), the quantities $(\hat{\bm{L}}\cdot\hat{\bm{N}})$, $(\hat{\bm{L}}\cdot\hat{\bm{z}})$ and $[\hat{\bm{N}}\cdot(\hat{\bm{L}}\times\hat{\bm{z}})]$, which are needed to compute the polarization angle $\psi_{\mathrm{S}}(t)$ in the beam-pattern coefficients $F_{\alpha}^+$ and $F_{\alpha}^{\times}$, are expressed as~\cite{vecchio}
%
%\begin{widetext}
\begin{eqnarray}
\hat{\bm{L}}\cdot\hat{\bm{z}}&=&(\hat{\bm{J}}_0\cdot\hat{\bm{z}})\cos\lambda_{\mathrm{L}}
                     +\frac{1-2(\hat{\bm{J}}_0\cdot\hat{\bm{z}})\cos\bar{\theta}_{\mathrm{J}}}
                     {2\sin\bar{\theta}_{\mathrm{J}}}\sin\lambda_{\mathrm{L}}\cos\alpha \notag \\
                     & &+\frac{(\hat{\bm{J}}_0\times\hat{\bar{\bm{z}}})\cdot\hat{\bm{z}}}{\sin\bar{\theta}_{\mathrm{J}}}
                     \sin\lambda_{\mathrm{L}}\sin\alpha, \label{55} \\
\hat{\bm{L}}\cdot\hat{\bm{N}}&=&(\hat{\bm{J}}_0\cdot\hat{\bm{N}})\cos\lambda_{\mathrm{L}}
                     +\frac{\cos\bar{\theta}_{\mathrm{S}}-(\hat{\bm{J}}_0\cdot\hat{\bm{N}})\cos\bar{\theta}_{\mathrm{J}}}
                     {\sin\bar{\theta}_{\mathrm{J}}}\sin\lambda_{\mathrm{L}}\cos\alpha \notag \\
                     & &+\frac{(\hat{\bm{J}}_0\times\hat{\bar{\bm{z}}})\cdot\hat{\bm{N}}}{\sin\bar{\theta}_{\mathrm{J}}}
                     \sin\lambda_{\mathrm{L}}\sin\alpha, \label{54} \\
\hat{\bm{N}}\cdot(\hat{\bm{L}}\times\hat{\bm{z}})&=
                     &\hat{\bm{N}}\cdot (\hat{\bm{J}}_0\times\hat{\bm{z}})\cos\lambda_{\mathrm{L}}
                     +\frac{\hat{\bm{N}}\cdot(\hat{\bar{\bm{z}}}\times\hat{\bm{z}})
                     -\hat{\bm{N}}\cdot(\hat{\bm{J}}_0\times\hat{\bm{z}})\cos\bar{\theta}_{\mathrm{J}}}
                     {\sin\bar{\theta}_{\mathrm{J}}}\sin\lambda_{\mathrm{L}}\cos\alpha \notag \\
                     & &+\frac{\hat{\bm{N}}\cdot (\hat{\bm{J}}_0\times\hat{\bar{\bm{z}}})\times\hat{\bm{z}}}
                     {\sin\bar{\theta}_{\mathrm{J}}}
                     \sin\lambda_{\mathrm{L}}\sin\alpha, \label{56}
\end{eqnarray}
%\end{widetext}
where
%
%\begin{widetext}
\begin{eqnarray}
\hat{\bm{J}}_0\cdot\hat{\bm{z}}&=&\frac{1}{2}\cos\bar{\theta_{\mathrm{J}}}
                   -\frac{\sqrt{3}}{2}\sin\bar{\theta_{\mathrm{J}}} \cos[\bar{\phi}(t)-\bar{\phi_{\mathrm{J}}}], \\
\hat{\bm{J}}_0\cdot\hat{\bm{N}}&=&\cos\bar{\theta_{\mathrm{J}}}\cos\bar{\theta_{\mathrm{S}}}
                    +\sin\bar{\theta_{\mathrm{J}}}\sin\bar{\theta_{\mathrm{S}}}\cos(\bar{\phi_{\mathrm{J}}}-\bar{\phi_{\mathrm{S}}}), \label{Jn} \\
\hat{\bm{N}}\cdot (\hat{\bm{J}}_0\times\hat{\bm{z}})&=&\frac{1}{2}\sin\bar{\theta_{\mathrm{J}}}\sin\bar{\theta_{\mathrm{S}}}
                                             \sin(\bar{\phi_{\mathrm{J}}}-\bar{\phi_{\mathrm{S}}}) \notag \\
                                             &&-\frac{\sqrt{3}}{2}\cos\bar{\phi(t)}(
                                             \cos\bar{\theta_{\mathrm{J}}}\sin\bar{\theta_{\mathrm{S}}}\sin \bar{\phi_{\mathrm{S}}} 
                                             -\cos\bar{\theta_{\mathrm{S}}}\sin\bar{\theta_{\mathrm{J}}}\sin \bar{\phi_{\mathrm{J}}} ) \notag \\
                                             &&-\frac{\sqrt{3}}{2}\sin\bar{\phi(t)}(
                                             \cos\bar{\theta_{\mathrm{S}}}\sin\bar{\theta_{\mathrm{J}}}\cos \bar{\phi_{\mathrm{J}}} 
                                             -\cos\bar{\theta_{\mathrm{J}}}\sin\bar{\theta_{\mathrm{S}}}\cos \bar{\phi_{\mathrm{S}}} ), \label{nJz} \\
(\hat{\bm{J}}_0\times\hat{\bar{\bm{z}}})\cdot\hat{\bm{N}}&=&\sin\bar{\theta}_{\mathrm{S}}\sin\bar{\theta}_{\mathrm{J}}
                             \sin (\bar{\phi}_{\mathrm{J}}-\bar{\phi}_{\mathrm{S}}), \\
\hat{\bm{N}}\cdot(\hat{\bar{\bm{z}}}\times\hat{\bm{z}})&=&
                       \frac{\sqrt{3}}{2}\sin\bar{\theta}_{\mathrm{S}}\sin(\bar{\phi}(t)-\bar{\phi}_{\mathrm{S}}), \\
(\hat{\bm{J}}_0\times\hat{\bar{\bm{z}}})\cdot\hat{\bm{z}}&=&
                        \frac{\sqrt{3}}{2}\sin\bar{\theta}_{\mathrm{J}}\sin(\bar{\phi}(t)-\bar{\phi}_{\mathrm{J}}), \\
\hat{\bm{N}}\cdot (\hat{\bm{J}}_0\times\hat{\bar{\bm{z}}})\times\hat{\bm{z}}&=&
                  -\frac{1}{2}\sin\bar{\theta}_{\mathrm{J}}[ \sqrt{3}\cos\bar{\theta}_{\mathrm{S}}\cos\{ \bar{\phi}(t)-\bar{\phi}_{\mathrm{J}} \}                  +\sin\bar{\theta}_{\mathrm{S}}\cos(\bar{\phi}_{\mathrm{J}}-\bar{\phi}_{\mathrm{S}})]\,. \nn \\
\end{eqnarray}
When spins are zero, $\sin\lambda_{\mathrm{L}}=0$, $\bar{\theta}_{\mathrm{J}}=\bar{\theta}_{\mathrm{L}}$ and $\bar{\phi}_{\mathrm{J}}=\bar{\phi}_{\mathrm{L}}$.
Then, Eqs.~(\ref{55}),~(\ref{54}) and~(\ref{56}) each reduces to Eqs.~(\ref{lz}),~(\ref{ln}) and~(\ref{nlz}), respectively.

$\hat{\bm{L}}\cdot\hat{\bm{u}}$ and $\hat{\bm{N}}\cdot(\hat{\bm{L}}\times\hat{\bm{u}})$ in Eq.~(\ref{thomas-new}) are given as~\cite{vecchio}
\begin{eqnarray}
\hat{\bm{L}}\cdot\hat{\bm{u}}&=&-\hat{\bm{N}}\cdot(\hat{\bm{L}}\times\hat{\bm{J}}_0) \\ \notag
                                                       &=&\hat{\bm{N}}\cdot(\hat{\bm{J}}_0\times\hat{\bm{z}})
                              \frac{\sin\lambda_{\mathrm{L}}}{\sin\theta_{\mathrm{J}}}\cos\alpha
                              -(\cos\theta_{\mathrm{S}}-\cos\theta_{\mathrm{J}}(\hat{\bm{J}}_0\cdot\hat{\bm{N}}))
                              \frac{\sin\lambda_{L}}{\sin\theta_{\mathrm{J}}}\sin\alpha, \\
\hat{\bm{N}}\cdot(\hat{\bm{L}}\times\hat{\bm{u}}) &= &\cos \lambda_{\mathrm{L}}
                               -(\hat{\bm{L}}\cdot\hat{\bm{N}})\cdot(\hat{\bm{J}}_0\cdot\hat{\bm{N}}),
\end{eqnarray}
where $\hat{\bm{N}}\cdot(\hat{\bm{J}}_0\times\hat{\bm{z}})$ and
$\hat{\bm{J}}_0\cdot\hat{\bm{N}}$ are given as Eqs.~(\ref{nJz}) and~(\ref{Jn}), respectively.
%\end{widetext}

%%%%%%%%%%%%%%%%%%%%%%%%%%%%%%%%%%%%%%%%%%
\subsection{GW Energy-Momentum Tensor in BD Theory}
\label{app-tmunu}

The energy-momentum tensor for GWs, $t^{\mu\nu}$, in BD theory is given as~\cite{will1977}
\ba
16\pi t^{\mu\nu}&=&-\frac{1}{2}{\chi_{\alpha\beta}}^{,\mu}\chi^{\alpha\beta,\nu}-\chi_{\alpha\beta}\chi^{\alpha\beta,\mu\nu}
                            -\chi_{\alpha\beta}(\chi^{\mu\nu,\alpha\beta}-2\chi^{\alpha(\nu,\mu)\beta}) \nn \\
                          & &+\frac{1}{2}\chi\chi^{,\mu\nu}+\frac{1}{4}\chi^{,\mu}\chi^{,\nu}+\frac{3}{2}\chi\Box\chi^{\mu\nu} 
                            +\chi_{,\alpha}\chi^{\mu\nu,\alpha}-\chi_{,\alpha}\chi^{\alpha(\nu,\mu)} \nn \\
                         &  &-{\chi^{\mu}}_{\alpha,\beta}\chi^{\nu\alpha,\beta}+{\chi^{\mu}}_{\alpha,\beta}\chi^{\nu\beta,\alpha}
                     -{\chi_{,\alpha}}^{,(\mu}\chi^{,\nu)\alpha}+\frac{1}{2}\chi^{\mu\nu}\Box\chi-2\chi_{\alpha}^{(\mu}\Box\chi^{\nu)\alpha} \nn \\
                         &  &+\frac{1}{2}\eta^{\mu\nu}\left( \frac{3}{2}\chi_{\alpha\beta,\gamma}\chi^{\alpha\beta,\gamma}
                            -\frac{3}{4}\chi_{,\alpha}\chi^{,\alpha}+2\chi_{\alpha\beta}\Box\chi^{\alpha\beta}
                            -\chi_{\alpha\beta,\gamma}\chi^{\alpha\gamma,\beta}-\chi\Box\chi+\chi_{\alpha\beta}\chi^{\alpha\beta} \right) \nn \\
                         &   &+4\varphi\Box\chi^{\mu\nu}+\Box\varphi\chi^{\mu\nu}-2{\varphi_{,\alpha}}^{,(\mu}\chi^{\nu)\alpha}   
                            -2{\varphi_{,\alpha}}(\chi^{\alpha(\mu,\nu)}-\chi^{\mu\nu,\alpha}) 
                            +\eta^{\mu\nu}\varphi_{,\alpha\beta}\chi^{\alpha\beta} \nn \\ 
& &+(2\omega_\mrm{BD}-1)\varphi^{,\mu}\varphi^{,\nu}-4\varphi\varphi^{,\mu\nu}
                             +\eta^{\mu\nu}\left[4\varphi\Box\varphi+\left(\frac{5}{2}-\omega_\mrm{BD}\right)\varphi_{,\alpha}\varphi^{,\alpha} \right] \nn \\
                           &  & +O(\chi^3,\chi^2\varphi,\chi\varphi^2,\varphi^3) \label{tmunu-bd}
\ea
with $\varphi\equiv\tilde{\phi}/\phi_0$.

%\begin{thebibliography}{000} %for 3 digits
%\begin{thebibliography}{00}  %for 2 digits
%\begin{thebibliography}{0}    %for 1 digit
\bibliographystyle{ws-ijmpd}
\bibliography{ref}

\end{document}